\documentclass[12pt]{article}
\usepackage{amsmath,amssymb}
\usepackage[dvipdfmx]{graphicx}
\usepackage{bm}
\usepackage{color}
\usepackage{comment}
\def\mydate{9 September, 2014}
\def\ignore#1{{}}

\makeatletter
\@addtoreset{equation}{section}
\makeatother

\newcommand{\siml}{%
\hspace{0.3em}\raisebox{0.4ex}{$<$}\hspace{-0.75em}\raisebox{-.7ex}{$\sim$}\hspace{0.3em}}

\newcommand{\bea}{\begin{eqnarray}}
\newcommand{\eea}{\end{eqnarray}}

\newcommand{\beeq}{\begin{equation}}
\newcommand{\eneq}{\end{equation}}
\newcommand{\beqn}{\begin{eqnarray}}
\newcommand{\eeqn}{\end{eqnarray}}

\def\dd{\partial}
\def\la{\raise.16ex\hbox{$\langle$}\lower.16ex\hbox{}  }
\def\ra{\raise.16ex\hbox{$\rangle$}\lower.16ex\hbox{} }
\def\go{\rightarrow}

\def\onehalf{ \hbox{$\frac{1}{2}$} }

\def\onefourth{ \hbox{$\frac{1}{4}$} }

\def\Tr{{\rm Tr \,}}
\def\tr{{\rm tr \,}}
\def\eff{{\rm eff}}

\def\cL{{\cal L}}

\def\cM{{\cal M}}

\def\EM{{\rm EM}}

\def\KK{{\rm KK}}

\def\brane{{\rm brane}}

\def\vect{{\rm vec}}
\def\sp{{\rm sp}}

\def\psibar{ \psi \kern-.65em\raise.6em\hbox{$-$} }
\def\psibarl{ \psi \kern-.65em\raise.6em\hbox{$-$} \lower.6em\hbox{} }

\newcommand{\nlight}{n_F^{\rm light}}
\newcommand{\nheavy}{n_F^{\rm heavy}}

\begin{document}
\allowdisplaybreaks[1]

\thispagestyle{empty}

%\begin{titlepage}
%%%%% PREPRINT NUMBERS %%%%%%
{\small \noindent \mydate    \hfill OU-HET 807, KIAS-P14008}
%{\small \noindent \myred{updated YH}    \hfill}

\vspace{2.5cm}

%%%%%%%%%%%%%%%%%%% TITLE %%%%%%%%%%%%%%%%%%
\baselineskip=35pt plus 1pt minus 1pt

\begin{center}
{\LARGE \bf  Dark matter}
\vskip 10pt
{\LARGE \bf 
in the SO(5)$\times $U(1) gauge-Higgs unification
}
\end{center}
%%%%%%%%%%%%%%%% AUTHORS %%%%%%%%%%%%%%%%%%%%%%%

\vspace{1.5cm}
\baselineskip=22pt plus 1pt minus 1pt

\begin{center}
{\bf
Shuichiro Funatsu$^*$, Hisaki Hatanaka$^\dagger$, Yutaka Hosotani$^*$\\
Yuta Orikasa$^{\dagger,\ddagger}$ and Takuya Shimotani$^*$
}

\vskip 5pt

$^*${\small \it Department of Physics, 
Osaka University, 
Toyonaka, Osaka 560-0043, 
Japan} \\

$^\dagger${\small \it School of Physics, KIAS, Seoul 130-722, Republic of Korea}\\

and

\vskip 7pt

$^\ddagger${\small \it Department of Physics and Astronomy, 
Seoul National University,\\ 
Seoul 151-742, Republic of Korea}\\

\end{center}

%%%%%%%%%%% ABSTRACT %%%%%%%%%%%%%%%

\vskip 2.cm
\baselineskip=20pt plus 1pt minus 1pt

\begin{abstract}
In the $SO(5) \times U(1)$ gauge-Higgs unification  the lightest, neutral component  of $n_F$
$SO(5)$-spinor fermions (dark fermions), which are relevant for having 
the observed unstable  Higgs boson,  becomes  the dark matter of the  universe.
We show that the relic abundance of the dark matter determined by WMAP and Planck data
is reproduced,  below the bound placed by the direct detection experiment by LUX, by a model 
with one light and three heavier  ($n_F=4$)  dark fermions with the lightest one 
of a mass  from 2.3$\,$TeV  to 3.1$\,$TeV.
The corresponding Aharonov-Bohm phase $\theta_H$ in the fifth dimension 
ranges from 0.097 to 0.074.
The case of $n_F=3$  ($n_F = 5, 6$) dark fermions yields the relic abundance  smaller (larger) than 
the observed limit.
\end{abstract}

%\end{titlepage}

%\tableofcontents

\newpage

\baselineskip=20pt plus 1pt minus 1pt

\section{Introduction}

The Higgs boson of a mass around 125.5$\,$GeV was discovered at LHC.\cite{Aad:2012tfa, Chatrchyan:2012ufa} 
It is not clear, however,  whether or not the particle discovered is precisely 
the Higgs boson specified in the standard model (SM).
Physics beyond the standard model may be hiding, showing up at the upgraded LHC.
Couplings of the Higgs boson to other particles may slightly deviate from those in SM,
and new particles may be produced, say, in the  4 - 7 TeV range.
SM lacks a principle governing dynamics of the Higgs boson.  
Further SM has no clue to explain the dark matter (DM) in the universe.

In the gauge-Higgs unification (GHU) the Higgs boson is unified with gauge bosons.
The 4D Higgs boson appears as a part of the extra-dimensional component 
of gauge fields so that its dynamics are governed by the gauge principle.\cite{YH1}-\cite{Lim2013} 
It has been shown that in the $SO(5) \times U(1)$ GHU
in the Randall-Sundrum warped space the low energy physics appears almost
the same as that in SM, consistent with all LHC data.\cite{ACP}-\cite{LHCsignals}
Contributions of Kaluza-Klein (KK)
excited modes to the $H \go \gamma \gamma$ decay, for example, turn out very small.\cite{FHHOS2013}
Higgs couplings to gauge bosons, quarks and leptons at the tree level are suppressed 
by a common factor $\cos \theta_H$ where $\theta_H$ is the Aharonov-Bohm phase
in the extra dimension.\cite{SH1}-\cite{Hasegawa:2012sy} 
All of the precision measurements, the tree-unitary constraint, 
and the $Z'$ search indicate that $\theta_H < 0.2$.\cite{ACP, Haba:2009a}
The $SO(5) \times U(1)$ GHU predicts new structure at higher energies.
The masses of the 1st KK modes of $Z$ and $\gamma$ are predicted to be $3 \sim 7\,$TeV
for $\theta_H = 0.1 \sim 0.2$.  The Higgs cubic and quartic self-coupling should be smaller than
those in the SM by 10\% - 20\%.\cite{LHCsignals}  Many other signals of GHU have been
investigated.\cite{Lim2007b}-\cite{Adachi:2014wva}

Another important issue is the dark matter.\cite{Kolb}  
Supersymmetric theory, the leading model of physics beyond the SM, predicts the lightest 
supersymmetric particle as a dark matter candidate.\cite{JKG, EFO}
The lightest KK particle in  universal extra dimension models \cite{Servant:2002aq}-\cite{Split-UED2}, 
the lightest T-odd particle in the little Higgs models \cite{Perelstein:2006bq, Hooper:2006xe},
a fermionic composite state in the composite Higgs models \cite{DiazCruz:2007be}-\cite{Chala:2012af}, 
and axions \cite{Preskill:1982cy}-\cite{Covi:1999ty} can be identified as  dark matter.
In the Higgs portal scenario the Higgs boson couples to  dark matter in the hidden 
sector \cite{Silveira:1985rk}-\cite{Okada:2014oda}, and 
the dynamical dark matter scenario has been proposed.\cite{DynamicalDM}
Is there a dark matter candidate in the $SO(5) \times U(1)$ gauge-Higgs unification  model?
Can it explain the relic abundance reduced from the WMAP/Planck  data 
and other observations, within the constraints from direct detection searches?
A few scenarios for dark matter in GHU have been proposed.\cite{PPSS, Carena2009, HKT, Haba2009}
In this paper we would like to show that the realistic $SO(5) \times U(1)$ gauge-Higgs unification model 
contains a natural candidate for dark matter.  

In the minimal $SO(5) \times U(1)$ gauge-Higgs unification model, in which 
only quark-lepton vector multiplets and associated brane fermions are introduced 
in the fermion sector, the effective potential is minimized at $\theta_H = \onehalf \pi$, 
which in turn implies that the Higgs boson becomes stable, 
contradicting with the observation.\cite{HOOS, HTU1, HKT}
To have an unstable Higgs boson, it is necessary to introduce fermion multiplets in the spinor
representation of $SO(5)$ which do not appear at low energies.\cite{FHHOS2013}
Indeed, the presence of these fermions, with the
gauge fields and top quark multiplet,  naturally  leads  to $0 < \theta_H < \onehalf \pi$, 
yielding predictions consistent with the observation.  
One remarkable property is that independent of the details of these $SO(5)$-spinor fermions
there appears the universality relations among $\theta_H$, the masses of KK $Z$/photon,
and the Higgs self couplings.

We show that the lightest, neutral component of the $SO(5)$-spinor fermions is absolutely stable, and
becomes the dark matter of the universe. 
For this reason the $SO(5)$-spinor fermion is called as a dark fermion in the present paper.
It is heavy with a mass around $2 \sim 4 \,$TeV, but its couplings to the Higgs boson are small.
From its relic abundance  the number and structure of the dark fermion multiplets are inferred.
It is curious that the Higgs dynamics are intimately related to the dark matter in the gauge-Higgs unification.

The paper is organized as follows.  
In Section 2  the $SO(5) \times U(1)$  model is introduced.  
In Section 3 it is shown that the neutral components of dark fermions become the dark matter, and
the relic abundance is evaluated.
In Section 4 the spin-independent cross section of the dark matter candidate with nucleons
is evaluated, and the compatibility with the constraint coming from the direct detection
experiments, XENON100 and LUX \cite{XENON, LUX},  is examined.  
It will be found that  the model with $n_F=4$ nondegenerate dark fermions
 with the lightest one of a mass  2.3$\,$TeV$\sim$3.1$\,$TeV 
 explains the relic abundance of the dark matter determined 
  from the WMAP/Planck data below the bound placed by the direct detection observation of LUX.
 Section 5 is devoted to the conclusion and discussions.
 In the appendices wave functions and couplings of dark fermions and relevant gauge bosons 
 are summarized.

 \section{Model}

The model of the $SO(5)\times U(1)$ GHU is defined 
in the Randall-Sundrum (RS) warped space with a metric
\begin{equation}
ds^2=G_{MN}dx^M dx^N=e^{-2\sigma(y)} \eta_{\mu\nu}dx^\mu dx^\nu+dy^2 ,
\end{equation}
where $\eta_{\mu\nu} = \text{diag}(-1,1,1,1)$, $\sigma(y)=\sigma(y+2L)=\sigma(-y)$, 
and $\sigma(y)=k|y|$ for $|y| \le L$.
The Planck and TeV branes are located at $y = 0$ and $y = L$, respectively.
The bulk region $0 < y < L$ is anti-de Sitter (AdS) spacetime 
with a cosmological constant $\Lambda = -6k^2$.
The warp factor $z_L \equiv e^{kL}$ is large, $z_L \gg1$, and the Kaluza-Klein   mass scale is given by 
$m_{KK} = \pi k/(z_L - 1) \sim \pi kz_L^{-1}$.

The model consists of $SO(5)\times U(1)_X$ gauge fields $(A_M , B_M )$, 
quark-lepton multiplets  $\Psi_a$, $SO(5)$-spinor fermions (dark fermions) $\Psi_{F_i}$, 
brane fermions $\hat{\chi}_{\alpha R}$,  and brane scalar $\hat \Phi$.\cite{HNU, FHHOS2013}
The bulk part of the action is given by
\beqn
&&\hskip -1.cm
S_{\text{bulk}}=\int d^5x\sqrt{-G} \, \Bigl[-\tr 
\Bigl(\frac{1}{\;4\;}F^{(A) MN} F^{(A)}_{MN}+\frac{1}{2\xi}
(f^{(A)}_{\text{gf}})^2+\mathcal{L}^{(A)}_{\text{gh}}\Bigr) \cr
\noalign{\kern 10pt}
&&\hskip 2.5cm
-\Bigl(\frac{1}{\;4\;}F^{(B) MN} F^{(B)}_{MN}+\frac{1}{2\xi}
(f^{(B)}_{\text{gf}})^2+\mathcal{L}^{(B)}_{\text{gh}}\Bigr) \cr
\noalign{\kern 10pt}
&&\hskip 2.5cm
+\sum_{a} \bar{\Psi}_a\mathcal{D}(c_a)\Psi_a
+ \sum_{i=1}^{n_F} \bar{\Psi}_{F_i} \mathcal{D}(c_{F_i})\Psi_{F_i} \Bigr],\cr
\noalign{\kern 10pt}
&&\hskip -1.cm
\mathcal{D}(c)= \Gamma^A {e_A}^M
\Big(\partial_M+\frac{1}{8}\omega_{MBC}[\Gamma^B,\Gamma^C]
-ig_AA_M-ig_BQ_{X}B_M\Big)-c\sigma'(y) .
\label{action1}
\eeqn
The gauge fixing and ghost terms are denoted as functionals with subscripts gf and gh, respectively. 
$F^{(A)}_{M N}=\partial_M A_N-\partial_N A_M-ig_A\bigl[A_M,A_N \bigr]$, and 
$F^{(B)}_{M N}=\partial_M B_N-\partial_N B_M$. 
The color $SU(3)_C$ gluon fields and their interactions have been suppressed in the present paper.
The $SO(5)$ gauge fields $A_M$ are decomposed as
\begin{equation}
A_M=\sum^3_{a_L=1}A_M^{a_L}T^{a_L}+\sum^3_{a_R=1}A_M^{a_R}T^{a_R}
+\sum^4_{\hat{a}=1}A_M^{\hat{a}}T^{\hat{a}} ,
\end{equation}
where $T^{a_L, a_R}  (a_L , a_R = 1, 2, 3)$ and $T^{\hat{a}}  (\hat{a} = 1, 2, 3, 4)$ 
are the generators of $SO(4) \simeq SU(2)_L \times SU(2)_R$ and $SO(5)/SO(4)$, 
respectively.

In the fermion part $\bar{\Psi}=i\Psi^\dagger \Gamma^0$ and $\Gamma^M$ 
matrices are given by
\beqn
&&\hskip -1.cm
\Gamma^{\mu}=\gamma^\mu = 
\begin{pmatrix} &\sigma^{\mu} \cr \bar\sigma^{\mu}& \end{pmatrix} ~,~~
\Gamma^5= \gamma^5 = \begin{pmatrix} 1& \cr &-1\end{pmatrix} ~, \cr
\noalign{\kern 10pt}
&&\hskip -1.cm
\sigma^{\mu}=(1, \, \vec{\sigma}) ~,~~
\bar \sigma ^{\mu}= (-1,\, \vec{\sigma}) ~.
\label{DiracM}
\eeqn
The quark-lepton multiplets $\Psi_a$ are introduced in the vector representation of $SO(5)$. 
In contrast, $n_F$ dark fermions $\Psi_{F_i}$ are introduced in the spinor representation.
The $c$ term in Eq. (\ref{action1}) gives a bulk kink mass, 
where $\sigma'(y)=k\epsilon(y)$ is a periodic step function with a magnitude $k$. 
The dimensionless parameter $c$ plays an important role in controlling profiles of 
fermion wave functions.

The orbifold boundary conditions at $y_0 = 0$ and $y_1 = L$ are given by
\beqn
&&\hskip -1.cm
\begin{pmatrix}A_{\mu}\\A_y\end{pmatrix} (x,y_j-y)=
P_{\vect} \begin{pmatrix}A_{\mu}\\-A_y\end{pmatrix} (x,y_j+y) P_{\vect}^{-1},  \cr
\noalign{\kern 10pt}
&&\hskip -1.cm
\begin{pmatrix}B_{\mu}\\B_y\end{pmatrix} (x,y_j-y)=
\begin{pmatrix}B_{\mu}\\-B_y\end{pmatrix}(x,y_j+y), \cr
\noalign{\kern 10pt}
&&\hskip -1.cm
\Psi_a(x,y_j-y)=P_{\vect} \Gamma^5 \Psi_a(x,y_j+y),\cr
\noalign{\kern 10pt}
&&\hskip -1.cm
\Psi_{F_i} (x, y_j -y)=
\eta_{F_i} (-1)^j P_{\sp} \Gamma^5\Psi_{F_i} (x, y_j + y),  
~~~ \eta_{F_i} = \pm 1 ,  \cr
%(-1)^{j+1} P_{\sp} \Gamma^5\Psi_{F_i} (x^\mu, y_j + y), 
%\end{cases}   \cr
\noalign{\kern 10pt}
&&\hskip -1.cm
P_{\vect}=\text{diag} \, ( -1,-1, -1,-1, +1) , \quad 
P_{\sp} =\text{diag} \, ( +1,+1, -1,-1 ) .
\label{BC1}    
\eeqn
The $SO(5)\times U(1)_X$ symmetry is reduced to 
$SO(4)\times U(1)_X \simeq SU(2)_L\times SU(2)_R\times U(1)_X$ 
by the orbifold boundary conditions. 
Various orbifold boundary conditions fall into a finite number of 
equivalence classes of boundary conditions.\cite{HHHK, HHK}
The physical symmetry of the true vacuum in each equivalence class of boundary 
conditions is dynamically determined at the quantum level by the Hosotani mechanism.
Recently dynamics for selecting boundary conditions have been proposed as well.\cite{Yamamoto2014}
The Hosotani mechanism has been explored and established, not only in perturbation theory, 
but also on the lattice nonperturbatively.\cite{Cossu}

The brane action $S_\brane$ contains brane fermions $\hat \chi_{\alpha R} (x)$, 
brane scalar $\hat \Phi (x)$, $A_\mu(x, y=0)$ and $\Psi_a(x, y=0)$.
It manifestly preserves gauge-invariance in $SO(4) \times U(1)_X$. 
$\hat \Phi$ develops non-vanishing expectation value $\la \hat \Phi \ra \gg m_\KK$,
which results in spontaneous breaking of  $SO(4) \times U(1)_X$ into 
$SU(2)_L \times U(1)_Y$ and in making all exotic fermions heavy.

The 4D Higgs field, which is a bidoublet  in $SU(2)_L \times SU(2)_R$, 
appears as a zero mode in the $SO(5)/SO(4)$ part of the fifth dimensional 
component of the vector potential $A^{\hat{a}}_y(x,y)$ 
with custodial symmetry.\cite{ACP, Pomarol2009, Sakamura2014}
Without loss of generality one can set $\langle A^{\hat{a}}_y\rangle 
\propto \delta^{a4}$ when the EW symmetry is spontaneously broken. 
The zero modes of $A^{\hat{a}}_y$ (a = 1,2,3) are absorbed by $W$ and $Z$ bosons. 
The 4D neutral Higgs field $H(x)$ is a fluctuation mode of 
the Wilson line phase $\theta_H$ which is an Aharonov-Bohm phase in the fifth dimension;
\beqn
&&\hskip -1.cm
A^{\hat{4}}_y (x,y)= \big\{ \theta_Hf_H +H(x) \big\}  u_H(y)+\cdots ~, \cr
\noalign{\kern 10pt}
&&\hskip -1.cm
\exp \Big\{ \frac{i}{2} \theta_H \cdot 2 \sqrt{2}T^{\hat{4}} \Big\} 
= \exp  \bigg\{ ig_A	\int^L_0 dy  \la A_y \ra \bigg\}   ~, \cr
\noalign{\kern 10pt}
&&\hskip -1.cm
f_H=\frac{2}{g_A} \sqrt{\frac{k}{z_L^2-1}}
=\frac{2}{g_w} \sqrt{\frac{k}{L(z_L^2-1)}} ~.
\label{WilsonPhase1}
\eeqn
Here the wave function of the 4D Higgs boson is given by 
$u_H(y) = [2k/(z_L^2 - 1)]^{1/2}e^{2ky}$ for $0 \leq y \leq L$ and 
$u_H(-y) = u_H(y) = u_H(y + 2L)$. $g_w = g_A/\sqrt{L}$ 
is the dimensionless 4D $SU(2)_L$ coupling.

For each generation two vector multiplets $\Psi_1$ and $\Psi_2$ for quarks and 
two vector multiplets $\Psi_3$ and $\Psi_4$ for leptons are introduced.
In contrast, the dark fermion $\Psi_{F_i}$ belongs to the spinor representation
of $SO(5)$, having  four components
\beeq
\Psi_{F_i} = \begin{pmatrix}
\psi^i_{l1}\\ \psi^i_{l2}\\  \psi^i_{r1} \\  \psi^i_{r2} \end{pmatrix} .
\eneq
$\psi^i_l$ and $\psi^i_r$  are  $SU(2)_L$ and  $SU(2)_R$ doublets,  respectively. 
They mix with each other for $\theta_H \not= 0$.  
%Mass eigenstates are denoted as $F_{l,r}^{i+} $ and $F_{l,r}^{i0}$ .
%$F_i^+$ is a linear combination of $\psi_{i1}$ and $\psi_{i3}$,
%whereas $F^0$ is a linear combination of $\psi_{i2}$ and $\psi_{i4}$.
The electric charge is given by $Q_\EM=T^{3_L}+T^{3_R}+Q_X$. 
We take $Q_X= \onehalf$ for $\Psi_{F_i} $ so that it contains charge 1 and 0
components.

The KK decomposition of $\Psi_{F_i}$ fields are summarized in Appendix B.
With the boundary condition (\ref{BC1}) $\Psi_{F_i}(x,z)$ does not have zero modes,
and is expanded in the KK modes $F^{+ (n)}_{i} (x)$ and $F^{0 (n)}_{i} (x)$
($n=1,2,3, \cdots$) as in (\ref{KKF1}).
The mass spectrum is determined by (\ref{SpectrumF1}).
With $\eta_{F_i} = +1$ in  the boundary condition for  $\Psi_{F_i} $ 
in \eqref{BC1} and for small $\theta_H$ 
the odd KK number modes $F^{+ (n)}_{i}, F^{0 (n)}_{i}$ ($n$: odd)
are mostly $SU(2)_R$ doublets, containing  $SU(2)_L$ components slightly. 
The even KK number modes $F^{+ (n)}_{i}, F^{0 (n)}_{i}$ ($n$: even)
are mostly $SU(2)_L$ doublets.
Consequenltly the  first KK modes $F^{+ (1)}_{i}, F^{0 (1)}_{i}$ 
couple to the $SU(2)_L$ gauge bosons ($W$ and $Z$) very weakly.
On the other hand, with $\eta_{F_i} = -1$, 
$F^{+ (n)}_{i}, F^{0 (n)}_{i}$ ($n$: odd) are mostly $SU(2)_L$ doublets,
and the  first KK modes $F^{+ (1)}_{i}, F^{0 (1)}_{i}$ 
couple to $W$ and $Z$ with the standard weak coupling strengths.

The dark fermion number is conserved so that the lightest mode
of the dark fermions becomes stable.
At the tree level the  first KK modes $F^{+ (1)}_{i}$ and $F^{0 (1)}_{i}$ are degenerate.
Their mass is about 1.5$\,$TeV  to 4$\,$TeV. 
The charged component $F^{+ (1)}_{i}$ receives  radiative corrections by photon 
and becomes heavier than the neutral component $F^{0 (1)}_{i}$.
Their mass difference is estimated to be about $20$ GeV for a cutoff scale 
$\Lambda= 100$ TeV.  $F^{+ (1)}_{i}$ eventually decays into $F^{0 (1)}_{i}$ and SM particles.

The lightest modes of $F^{0 (1)}_{i}$'s are absolutely stable, and become 
DM of the universe.  
%the dark matter (DM) of the universe.  
 In the following discussions we denote $F^{0 (1)}_{i}$ simply by $F^{0}_{i}$.
We shall see below that  the observed relic abundance of the DM
and the bound from direct detection search of DM particles put severe constraint
on the value of $\theta_H$ and the number and degeneracy of dark fermions.

\section{Relic density}\label{sec:Abundance}
\newcommand{\etaF}{{\eta_F}}

By considering annihilations and decays of dark fermions in the early universe,
one can evaluate the relic density of the dark fermion. We mostly follow the arguments
 in Refs. \cite{Kolb}, \cite{Servant:2002aq} and \cite{Griest}.
The Boltzmann equation for $F^0_i$ is given by
\begin{eqnarray}
&&\hskip -1.cm
\frac{d n_{(F^0_i)}}{dt} 
= - 3 H n_{(F^0_i)} 
- \sum_{X,X'} [\langle \sigma(\bar{F}^0_i F^0_i \to X X') v 
\rangle 
(n_{(F^0_i)} n_{(\bar{F}^0_i)} - n_{(F^0_i)}^{\rm eq} n_{(\bar{F^0_i})}^{\rm eq} )]
\cr
\noalign{\kern 10pt}
&& \hskip -.5cm
- \sum_{X,X'} \left[ \langle \sigma(F_i^- F_i^0 \to X X')v\rangle
(n_{(F^0_i)} n_{(F^-_i)} - n_{(F^0_i)}^{\rm eq} n_{(F^-_i)}^{\rm eq} )
\right]
\cr
\noalign{\kern 10pt}
&& \hskip -.5cm
- \sum_j \left[ \langle \sigma(\bar{F}_i^0 F_i^0 \to F_j^+ F_j^-)v \rangle (n_{(F^0_i)} n_{(\bar{F}^0_i)} - n_{(F^0_i)}^{\rm eq} n_{(\bar{F^0_i})}^{\rm eq} )\right]
\cr
\noalign{\kern 10pt}
&& \hskip -.5cm
-\sum_{X,X'} \Big\{
\langle \sigma(F^0_i X\to F^+_i X') v\rangle n_{(F^0_i)} n_{(X)}
 - \langle \sigma v(F^+_i X'\to F^0_i X)\rangle n_{(F^+_i)} n_{(X')}  \Big\} .
\label{eq:Boltzmann0}
\end{eqnarray}
Similar relations are obtained for $\bar{F}^0_i$ and $F^\pm_i$.
Here $H$ is the Hubble constant,
$n_{(F)}$ denotes the number density of $F$,
and $X$ represents a SM field.
The number density of $F$ in the thermal equilibrium is given by
$n_{(x)}^{\rm eq} = g_x (m_x T/2\pi)^{3/2} \exp(-m_x/T)$ where $g_x$ and $m_x$ are the number of the degrees of freedom and mass of $x$, respectively.
If $F^\pm$ is heavier than $F^0$, a term describing  $F^+ \to F^0 f \bar{f}'$ decay should 
be added on the right-hand side of \eqref{eq:Boltzmann0};
\begin{eqnarray}
 - \left(n_{(F^+_i)} - n_{(F^+_i)}^{\rm eq}\right)\Gamma(F^+_i \to F^0_i f \bar{f}'),
 \label{eq:F+decay}
\end{eqnarray}
where $f,f'$ are fermions in the SM and $\Gamma$ denotes a decay width.

The effective interactions relevant to annihilations of dark fermions are given by
\begin{eqnarray}
{\cal L}_{\rm eff} 
&\supset& 
Z_{\mu} 
\biggl\{
\sum_{i=1}^{n_F} \bar{F}^0_i \gamma^\mu \frac{g_w}{\cos\theta_W} (V_F + \gamma_5 A_F) F^0_i
+ \sum_{i=1}^{n_F} \bar{F}^+_i \gamma^\mu \frac{g_w}{\cos\theta_W} (V_{F+} + \gamma_5 A_{F+}) F^+_i
\cr
\noalign{\kern 5pt}
&&\hskip 3.cm
+ \sum_{f} \bar{f} \gamma^\mu \frac{g_w}{\cos\theta_W} (v_f + \gamma_5 a_f) f  \biggr\}   \cr
\noalign{\kern 10pt}
&&
+ 
\!\!\!\!\!
\sum_{V=Z^{(1)},Z_R^{(1)}} 
\!\!\!\!\!
V_\mu
\biggl\{
\sum_{i=1}^{n_F} \bar{F}^0_i \gamma^\mu g_w (V^{(V)}_F + \gamma_5 A^{(V)}_F) F^0_i
+\sum_{i=1}^{n_F} \bar{F}^+_i \gamma^\mu g_w (V^{(V)}_{F+} + \gamma_5 A^{(V)}_{F+}) F^+_i
\nonumber\\
&&
\hskip 3.cm
 + \sum_{f} \bar{f} \gamma^\mu g_w (v^{(V)}_f + \gamma_5 a^{(V)}_f) f
\biggr\}
\nonumber\\
\noalign{\kern 10pt}
&&
+ \sum_{V=\gamma,\gamma^{(1)}} 
\!\!\!\!\!
V_\mu
\biggl\{
\sum_{i=1}^{n_F} \bar{F}^+_i \gamma^\mu e (V^{(V)}_{F+} + \gamma_5 A^{(V)}_{F+}) F^+_i
+ \sum_{f} \bar{f} \gamma^\mu e (v^{(V)}_f + \gamma_5 a^{(V)}_f) f
\biggr\}
\nonumber\\
\noalign{\kern 10pt}
&&
- H \sum_{i=1}^{n_F} Y_{F_i} (\bar{F}^0_i F^0_i  + \bar{F}^+_iF^+_i )
- H \sum_f y_f \bar{f} f
\nonumber\\
\noalign{\kern 10pt}
&&
+ 
\!\!\!\!\sum_{ V=\gamma,\gamma^{(1)},Z,Z^{(1)}, Z_R^{(1)}}
\!\!\!\!\!\!\!\!\!\!
i g_{VW^+W^-}
(\eta^{\mu\rho}\eta^{\nu\sigma} - \eta^{\mu\sigma}\eta^{\nu\rho})
\nonumber\\
&&\hskip 3.cm
\times  \left\{
   W_\rho^- V_\sigma \partial_\mu W_\nu^+
+  V_\rho W_\sigma^+ \partial_\mu W_\nu^-
+  W_\rho^+ W_\sigma^- \partial_\mu V_\nu \right\},
\end{eqnarray}
and by charged currents in Eq.~\eqref{eq:charged-couplings}.
Here $H$ denotes the Higgs boson, and $f$ refers to a fermion in the SM (quarks, leptons and neutrinos).
%{\bf [Definition of $\gamma^\mu$ is given in Sec. 2.]}
%\textcolor{blue}{$\gamma^\mu$ ($\mu=0,1,2,3$) are 4-dimensional gamma matrices satisfying $(\gamma^0)^2 = -(\gamma^{i})^2 = 1$ ($i=1,2,3$) and $\gamma_5 \equiv i\gamma^0\gamma^1\gamma^2\gamma^3$.}

%
For the decays \eqref{eq:F+decay} the corresponding interaction terms in the effective Lagrangian are
\begin{eqnarray}
{\cal L}_{\rm eff} &\supset& 
\sum_{V=W,W^{(1)},W_R^{(1)}} V_\mu^- \, \frac{g_w}{\sqrt{2}}
\biggl\{
\sum_{i=1}^{n_F}  {\bar F^0}_i \gamma^\mu (V^{(V)}_F + \gamma_5 A^{(V)}_F)F^+_i
\cr
\noalign{\kern 10pt}
&&\hskip 1.cm
+\sum_{\{f,f'\}} U^{(V)\rm CKM}_{f f'}   \bar{f}' \gamma^\mu (v^{(V)}_f + \gamma_5 a^{(V)}_f ) f
\biggr\}
+ (h.c.),
\label{eq:charged-couplings}
\end{eqnarray}
where $f$ and $f'$ refer to up-type quark (neutrino) and down-type quark (charged lepton), respectively. A CKM-like mixing matrix $U^{(V)\rm CKM}$ is a unit matrix for leptons and is assumed to approximately coincide to the CKM-matrix for $V=W$.
For the spinor fermion $F$, the right- and left-handed couplings $g^{V}_{FR/L} \equiv g_w (V_F^{(V)} \pm A_F^{(V)})/2$ are given in the Appendix \ref{apdx-WFF-couplings}, and for the SM fermions the couplings can be found in Ref.~\cite{LHCsignals}.
In particular, $W_R$ boson is found to have no couplings to the SM fermions.

\subsection{Decays v.s. conversions of charged dark fermions}

\begin{figure}[tbp]
\centerline{\includegraphics[width=5cm]{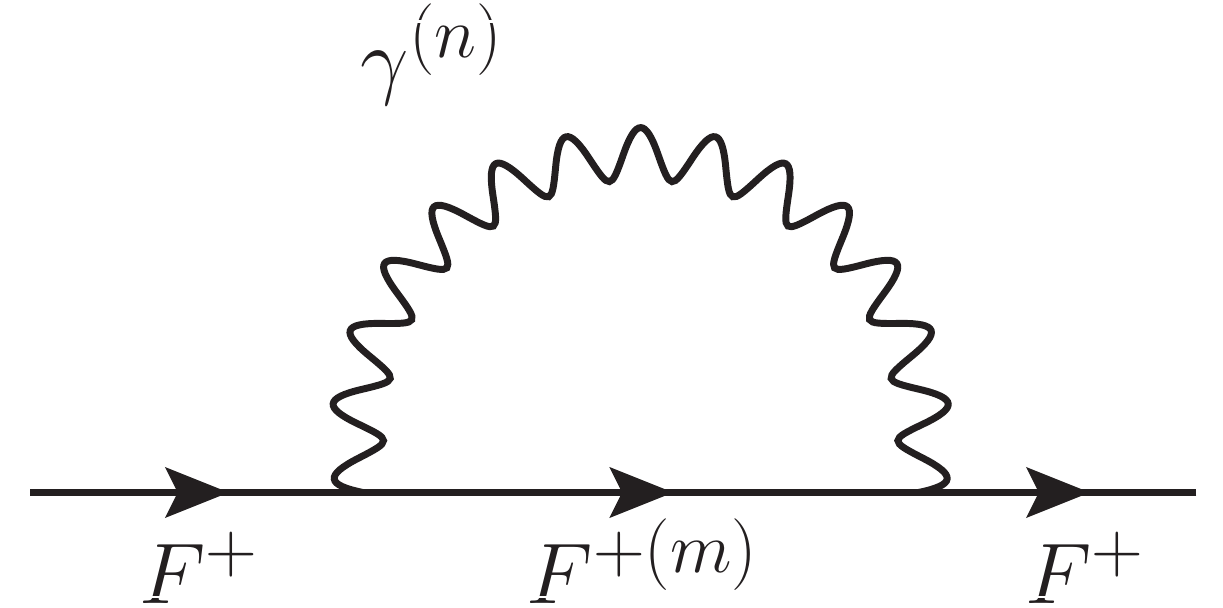}}
\caption{Diagrams contributing to the fermion mass difference $\Delta m_F = \delta m_{F^+} - \delta m_{F^0}$.}\label{fig:fermion-mass}
\end{figure}

At the quantum level, masses of $F^\pm$ and $F^0$ receive finite corrections $\delta m_{F^+}$ and 
$\delta m_{F^0}$, respectively, and the degeneracy is lifted by one-loop corrections involving the photon and 
KK photons, which appear only in $\delta m_{F^+}$ as depicted in Fig.~\ref{fig:fermion-mass}.
The mass difference  between $F^{\pm}$ and $F^0$, $\delta m_{F^\pm} - \delta m_{F^0}$, can be evaluated in an analogous way as in the universal extra dimension \cite{Cheng:2002iz}, and in the case of the warped extra dimension 
it is estimated by
\begin{eqnarray}
\delta m_{F^\pm} - \delta m_{F^0}
 &\sim& m_F \frac{\alpha_{\rm EM}}{4\pi}\cdot K,
\label{eq:mass-diff}
\end{eqnarray}
where $\alpha_{\rm EM}$ is the electromagnetic fine-structure constant. 
In UED $K = \ln(\Lambda^2/\mu^2)$ where $\Lambda$ and $\mu$ is the cut-off scale and a renormalization scale, respectively, and $\Lambda/\mu \sim {\cal O}(10)$. 
In the RS space-time only the first few KK excited states of each fields enter the quantum corrections.
In particular the coupling of right-handed $F^{\pm(1)}$ to $\gamma^{(1)}$ is several times as large as the electromagnetic coupling. It follows that $K \sim {\cal O}(10)$. Similarly, quantum corrections due to higher-KK modes to the gauge couplings also become small, and a large cut-off scale is allowed.\cite{Randall2001}

\begin{figure}[htbp]
\centerline{\includegraphics[width=4cm]{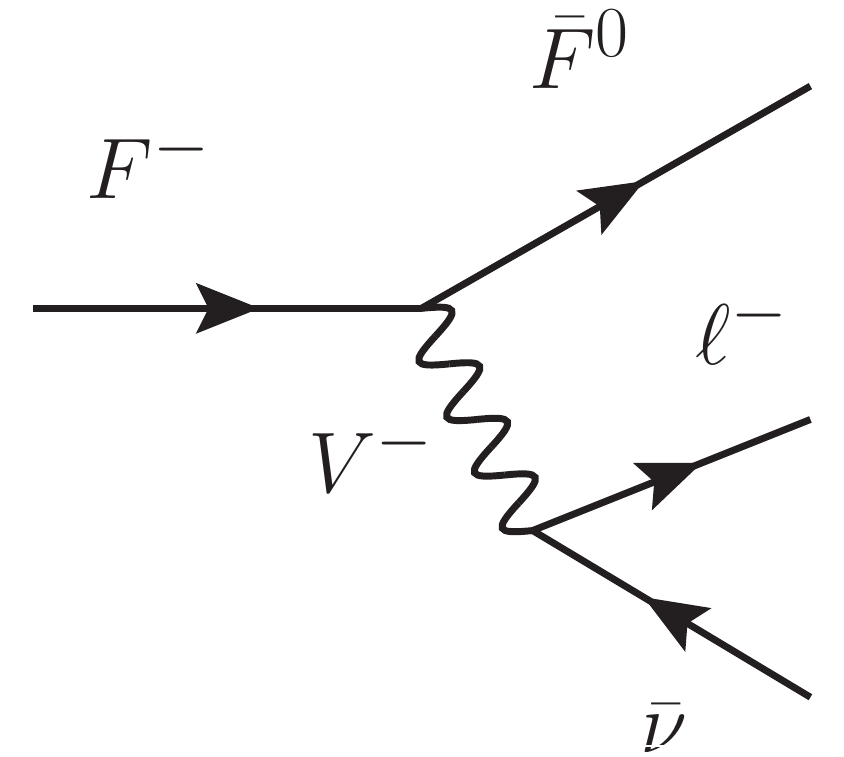}}
\caption{Charged dark fermion decay. 
$\ell^-$ and $\bar{\nu}$ can be replaced with down-type quarks and up-type anti-quarks, respectively.}\label{fig:f-decay}
\end{figure}

A charged dark fermion decays to a neutral dark fermion and a charged vector bosons, hence to charged leptons and neutrinos, or light down-type quarks and up-type antiquarks. (See Fig.~\ref{fig:f-decay})

In the $SO(5)\times U(1)$ GHU model, we have three charged vector bosons at low energies: 
$W$, the 1st KK excited state of $W$, and the lowest KK mode of the $W_R$ boson.
A charged dark fermion $F^+$ decays to $F^0$ mainly by emitting a $W$ boson, 
because $W^{(1)}$ is heavy and interacts with $F^+$ and $F^0$ very weakly, and 
$W_R^{(1)}$ cannot decay to the SM fermions. 
If the mass difference between the charged and neutral dark fermions,
$\Delta m_F \equiv m_{F^\pm} - m_{F^0} \simeq
\delta m_{F^\pm} - \delta m_{F^0}$,  is much smaller than $m_W$, the decay rate is given by
\begin{eqnarray}
\lefteqn{
\Gamma(F^- \to \bar{F}^0 \ell \bar{\nu})
}\nonumber\\
 &=& \frac{G_F^2}{192\pi^3} m_{F^-}^5
\left[ (g'^W_{\bar{l}\nu L})^2 +  (g'^W_{\bar{l}\nu R})^2 \right]
%\nonumber\\
%\noalign{\kern 10pt} &&\quad  \times
\biggl\{
\left[ (g'^W_{F L})^2 +  (g'^W_{F R})^2 \right] 
f\left(\frac{m_{F^0}^2}{m_{F^-}^2}\right)
- 4 g'^W_{FL} g'^W_{FR} g\left(\frac{m_{F^0}^2}{m_{F^-}^2}\right)
\biggr\}
\nonumber\\
\noalign{\kern 10pt}
&=& \frac{G_F^2}{192 \pi^3} \Delta m_F^5 
\left[ (g'^W_{\bar{l}\nu L})^2 +  (g'^W_{\bar{l}\nu R})^2 \right]
%\nonumber\\ \noalign{\kern 10pt} && \quad \times
\biggl\{
\frac{64}{5}[(g'^W_{F L})^2 +  (g'^W_{F R})^2 - g'^W_{F L} g'^W_{F R}]
+ {\cal O}\left(\frac{\Delta m_F^6}{m_F^6}\right)
\biggr\} ,
\label{eq:f-decay}
\end{eqnarray}
where $g'^{V}_{\bar{f}f L/R} \equiv g^{V}_{\bar{f}f L/R}/g_w$,
$g'^{V}_{F L/R} \equiv g^{V}_{F L/R}/g_w$ and
\begin{eqnarray}
f(x) &=& 1-8x + 8x^3 - x^4 - 12x^2\ln x ~,
\nonumber\\
g(x) &=& 1+9x-9x^2 - x^3 + 6x(1+x)\ln x ~.
\end{eqnarray}
In the second equality in \eqref{eq:f-decay}, we have assumed $\Delta m_F \ll m_{F^\pm}, m_{F^0}$ 
and have invoked approximations
\begin{eqnarray}
f((1-x)^2) &=& \tfrac{64}{5}x^5 - \tfrac{96}{5}x^6 + {\cal O}(x^7),
\quad
g((1-x)^2) = \tfrac{16}{5}x^5 - \tfrac{8}{5}x^6 + {\cal O}(x^7).
\end{eqnarray}
Hence the lifetime of $F^-$ is given by
\begin{eqnarray}
\tau_{F^\pm} 
&\simeq&
\tau_\mu \left( \frac{m_\mu}{\Delta m_F}\right)^5 
\frac{5}{64}
\left[ (g'^W_{FL})^2 +  (g'^W_{FR})^2
- g'^W_{FL} g'^W_{FR}\right]^{-1}, 
\end{eqnarray}
where $\tau_\mu = 2.2\times10^{-6}\,\text{sec}$ and $m_\mu = 105\,\text{MeV}$ are 
the lifetime and mass of the muon, respectively. $(g'^L_{W\bar{l}\nu},g'^R_{W\bar{l}\nu})  =(1,0)$ is used.
In order that the $F^\pm$ lifetime is much shorter than the typical time scale of the weakly interacting massive particle (WIMP)-DM formation, i.e. $\tau_{F\pm} \ll 10^{-10}\,\text{sec}$, the mass difference of dark fermions must be the order of $10\,\text{GeV}$ or larger.
The mass difference \eqref{eq:mass-diff} will satisfy this condition for $m_F \gtrsim 2$ TeV with $K \sim {\cal O}(10)$.
Hereafter we assume that these conditions are satisfied and $F^\pm$ decays sufficiently quickly. 
We also note that if  charged fermions $F^+$ do not decay sufficiently fast,  they would remain 
after the DM freeze-out and would subsequently decay to $F^0$, resulting in doubling 
the relic DM density.

\begin{figure}[htbp]
\centerline{%
\includegraphics[width=5cm]{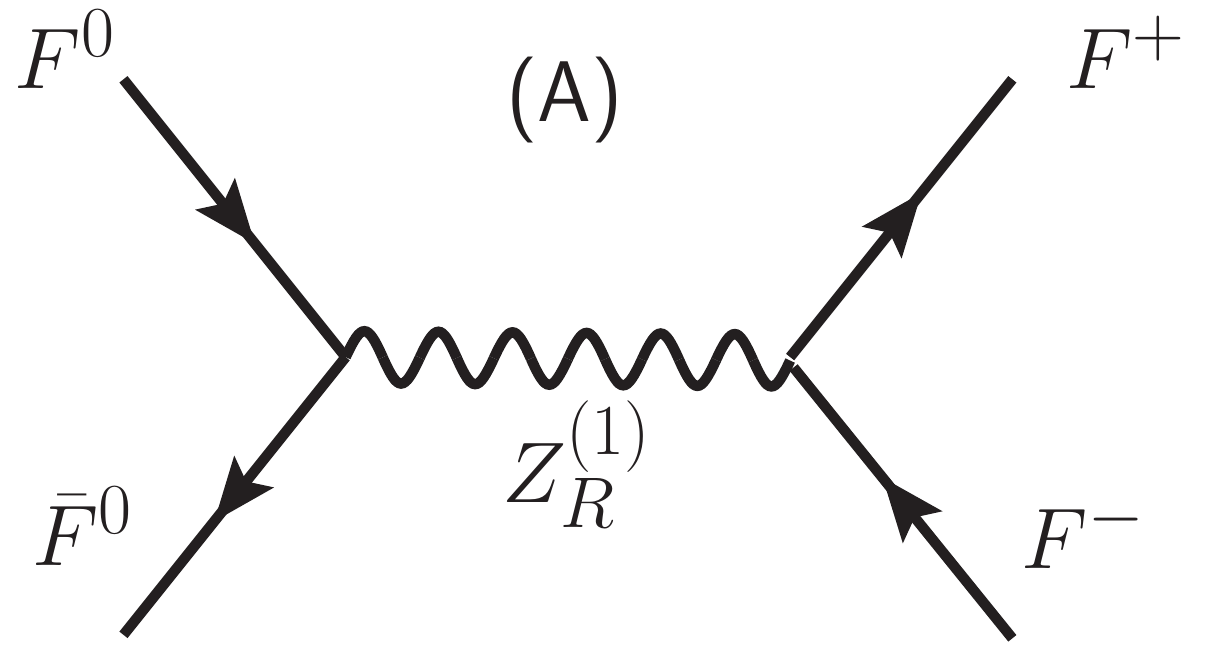}
\includegraphics[width=3cm]{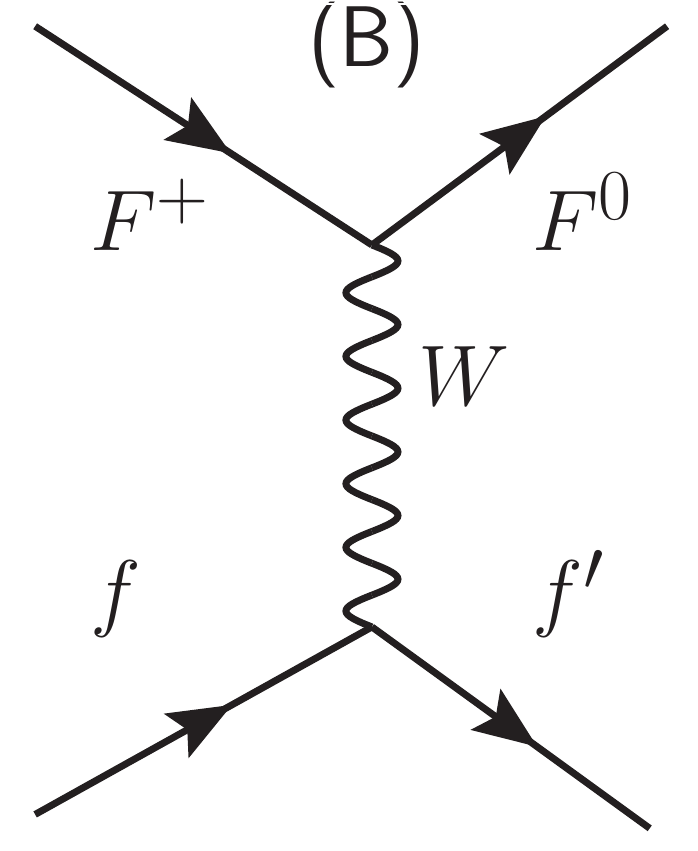}}
\caption{Processes of $F^0 \leftrightarrow F^+$ conversions. 
(A) $F^0 + \bar{F}^0 \leftrightarrow F^+ + F^-$ mediated by $Z_R^{(1)}$.
(B) $F^0 + f \leftrightarrow F^+ + f'$ by exchanging the $W$ boson, where $f$ and $f'$ are SM fermions.}\label{fig:conversion}
\end{figure}

In the right-hand side of the Boltzmann equation \eqref{eq:Boltzmann0},
the last two terms correspond to $F^0 \leftrightarrow F^+$ conversion
detpicted in Fig.~\ref{fig:conversion}.

The process depicted as (A) in Fig.~\ref{fig:conversion},
in particular $F^+ F^-$ pair production through this process is kinematically allowed
since $m_F \gg \Delta m_F$.
Although the process depicted as (B) in Fig.~\ref{fig:conversion} is suppressed by the
small $F\bar{F}W$ coupling which is order of $10^{-3}$,
this conversion process can dominate  due to the large ratio of 
$n_{(X)}^{\rm eq}/n^{\rm eq}_{(F)} \sim 
(T/m_F)^{3/2} \exp(m_F/T) \sim 10^{10}$ for $T/m_F \sim 30$ \cite{Griest}.

Thus we have $n_{(F^\pm)}^{\rm eq} \sim n_{(F^0)}^{\rm eq}$ before the freeze-out,
and after the freeze-out $F^+$ decay to $F^0$. The relic density of the dark fermion in the present universe is given by the sum of the charged and neutral dark fermions at the freeze-out. In the followings we calculate the number density of  all dark fermions.

%%%%%%%%%%%%%%%%%%%%%%%%%%%%%%%%
\subsection{Pair annihilations and relic density of dark fermions}\label{sec:annihilation}

\begin{figure}[htbp]
\centerline{%
\includegraphics[width=4cm]{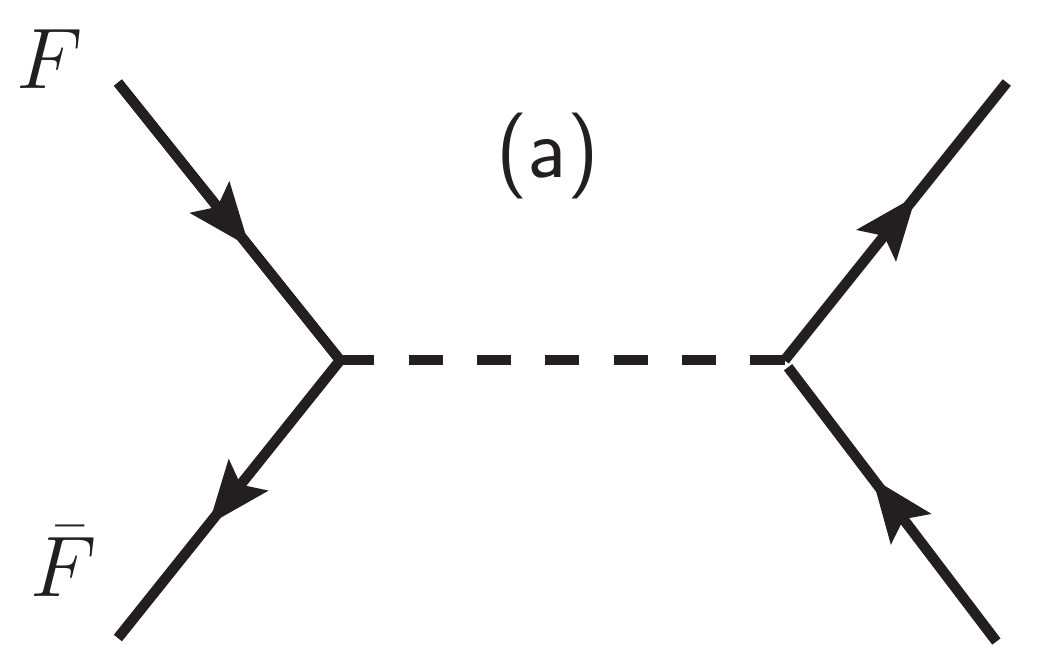}\quad
\includegraphics[width=4cm]{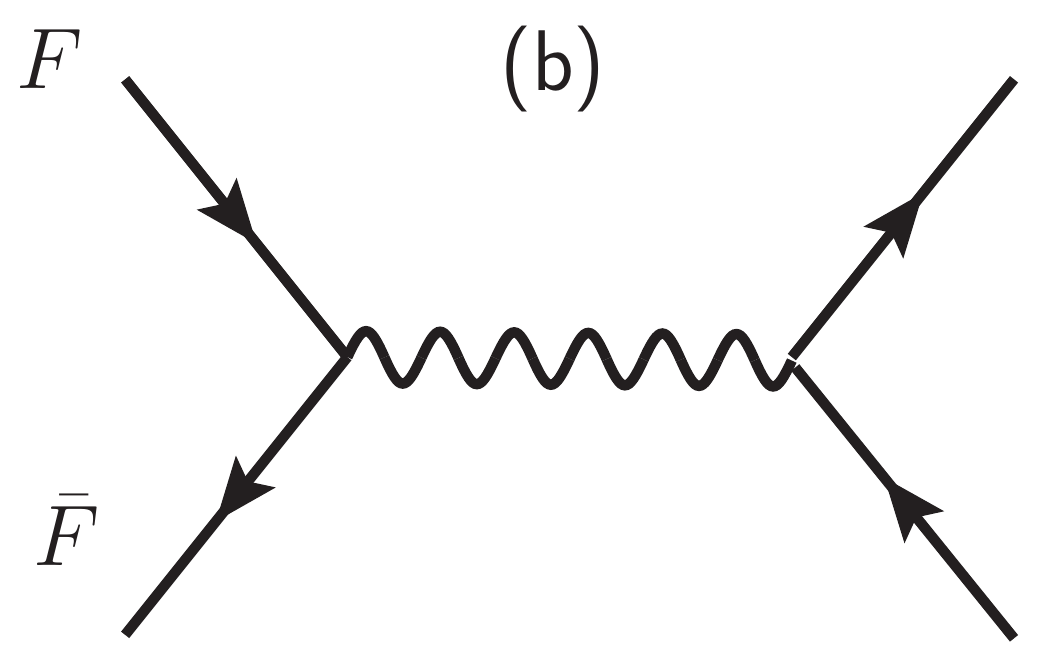}\quad
}
\vspace{0.5cm}
\centerline{\includegraphics[height=3.5cm]{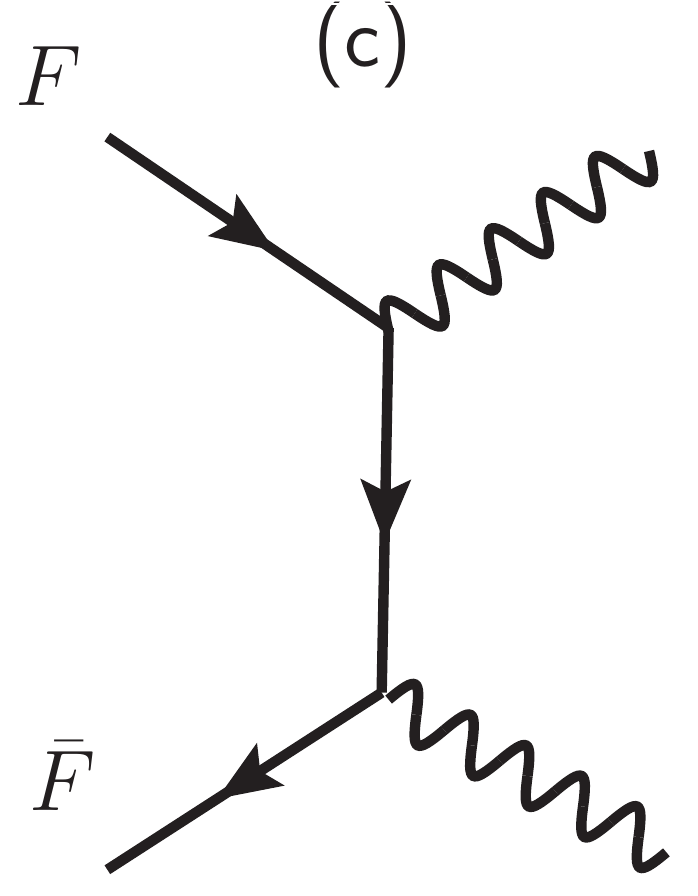}\quad
\includegraphics[height=3.5cm]{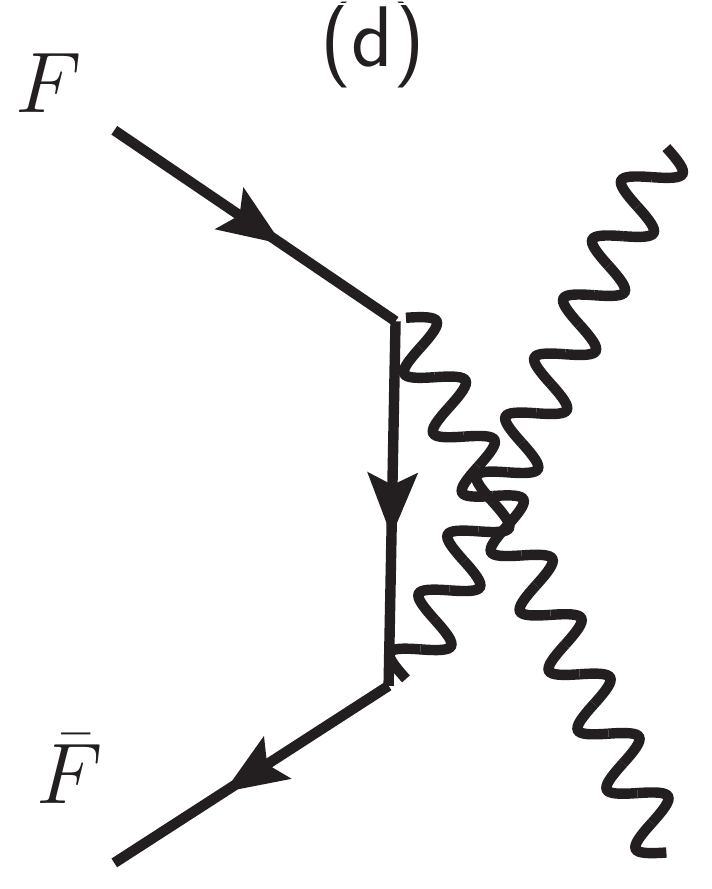}\quad
\raisebox{0.25cm}{\includegraphics[width=4cm]{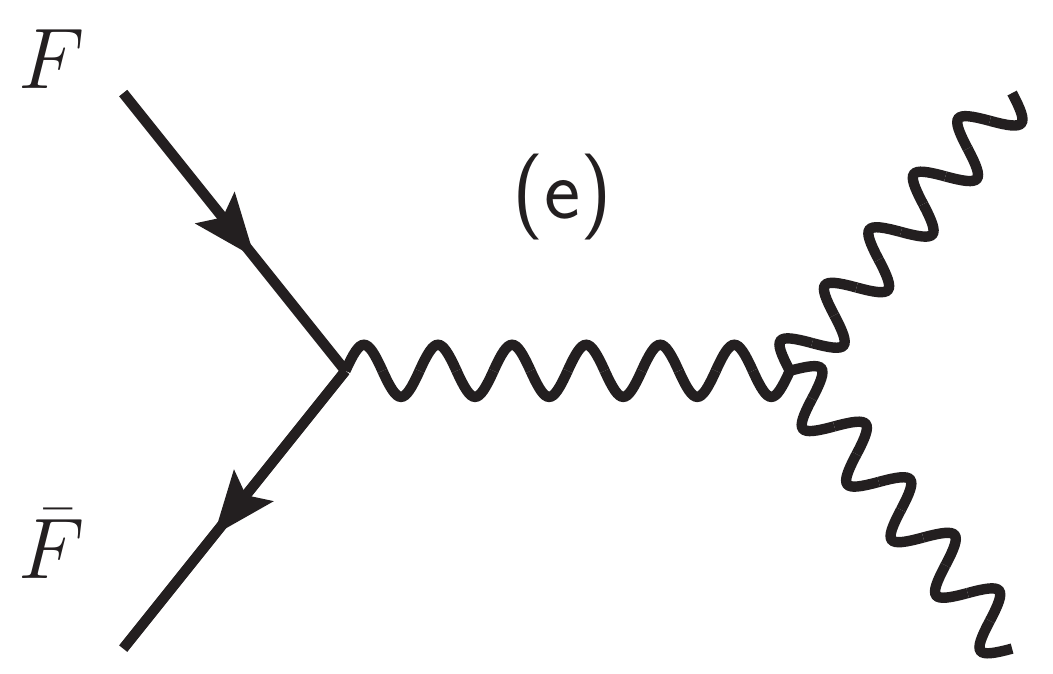}}
}
\caption{$F\bar{F}$ annihilation diagrams. 
(a) s-channlel annihilation to a fermion-pair through the Higgs boson
(b) s-channel, to fermions through a vector boson
(c)(d) t- and u-channel annihilations to two vector bosons
(e) s-channel annihilation to two vector bosons
}\label{fig:ann-diagrams}
\end{figure}

The annihilation processes and corresponding diagrams of the dark fermions are tabulated 
in Table. \ref{tbl:process}  and Fig. \ref{fig:ann-diagrams}.
We note that the masses of the first excited states of SM fermions [bosons] are about $m_{KK}$ [$0.8 m_{KK}$]. Mass of dark fermions is smaller than the half of $m_{KK}$,
so that the the final states of the annihilation of dark fermions involve only SM particles.
\begin{table}[tbp]
\caption{Pair annihilation processes of  dark fermions ($F = F^0,\,F^+$).
(a-i)-(a-v) are annihilation processes of neutral and charged dark fermions,
whereas (ac-i)-(ac-iv) are those of charged dark fermions.
(co-i)-(co-v) are for co-annihilation of the neutral and charged dark fermions.
In the intermediate states `$n$' denotes the KK-excitation level ($n\ne0$).
In the final states $q$, $l$ and $\nu$ denotes quarks, charged leptons and neutrinos in the SM.
Corresponding diagrams are shown in Fig.~\ref{fig:ann-diagrams}.}\label{tbl:process}
\begin{center}
\begin{tabular}{ccc}
\hline\hline
& process & diagrams \\
\hline
annihilation \\
(a-i) & $F\bar{F}\to (S=H,H^{(n)}) \to q\bar{q}$, $l\bar{l}$ & (a) \\
(a-ii) & $F\bar{F}\to (V=Z,Z^{(n)},Z_R^{(n)}) \to q\bar{q}$, $l\bar{l}$, $\nu\bar{\nu}$ & (b) \\
(a-iii) & $F\bar{F}\to ZZ$, t- and u- channels & (c), (d) \\
(a-iv) & $F\bar{F}\to W^+W^-$ t-channel & (c)\\
(a-v) & $F\bar{F}\to(V=Z,Z^{(n)},Z_R^{(n)})\to W^+W^-$ & (e) \\
(ac-i) & $F^+F^- \to \gamma\gamma$, t- and u-channels & (c) (d) \\
(ac-ii) & $F^+F^- \to Z\gamma$, t- and u-channels & (c) (d) \\
(ac-iii) & $F^+F^- \to (V=\gamma, \gamma^{(n)}) \to q\bar{q}, l\bar{l}$ & (b) \\
(ac-iv) & $F^+F^- \to (V=\gamma, \gamma^{(n)}) \to W^+ W^-$ & (e) \\
\hline
co-annihilation \\
(co-i)  & $F^+ \bar{F}^0 \to (V = W^+, W^{+(n)},W_R^{+(n)})\to q\bar{q}'$, $\nu\bar{l}$ & (b) \\
(co-ii) & $F^+ \bar{F}^0 \to (V = W^+, W^{+(n)},W_R^{+(n)})\to W^+Z$ &  (e) \\
(co-iii)& $F^+ \bar{F}^0 \to (V = W^+, W^{+(n)},W_R^{+(n)}) \to W^+ \gamma$ & (e) \\
(co-iv) & $F^+\bar{F}^0 \to W^+ Z$, t- and u-channels & (c), (d) \\
(co-v) & $F^+\bar{F}^0  \to W^+ \gamma$, t- and u-channels & (c), (d) \\
\hline\hline
\end{tabular}
\end{center}
\end{table}
%%%%%%%%%%%%%%%%

We consider the case where $\theta_H$ is small ($z_L \lesssim 10^5$ or $\theta_H \lesssim 0.15$).
In such a case, dark fermion is heavy and 
some of annihilation amplitudes are processes are suppressed by $\sin^2\theta_H$.
We find that for most of the processes annihilation cross-sections 
are too small to explain the current relic density.
In particular, we find that $\bar{F}FW$, $\bar{F}FZ,\bar{F}FZ^{(n)}$ and $Z_R^{(n)}WW$ 
couplings are  suppressed by $\sin^2\theta_H$ factor. 
(See Appendix \ref{sec:apdx-fermion-boson-couplings} and \ref{sec:apdx-boson-boson-couplings}).
One finds that the  process (a-i) is suppressed by the small Higgs Yukawa couplings of $F\bar{F}$ and
processes (a-ii) with $V=Z$ and $Z^{(n)}$ are suppressed by the small $Z^{(n)}F\bar{F}$ couplings.
The processes (a-iii) and (a-iv) are suppressed by the small $W^-F^+\bar{F}$ and $ZF\bar{F}$ couplings.
All processes of (a-v) are suppressed by the small $Z^{(n)}F\bar{F}$ coupling and 
small $Z_R^{(n)} W^+ W^-$ couplings.
Thus one finds that only the process (a-ii) with $V=Z_R^{(1)}$ is unsuppressed and could be enhanced by both
the Breit-Wigner resonance\cite{UED-Kakizaki3} of $Z_R^{(1)}$ and the large right-handed couplings of $Z_R^{(1)}$ 
to quarks and leptons. 
%The process (vii) is suppressed by the small $ZF\bar{F}$ coupling.

For the annihilation of charged dark fermions,
we see that the process (ac-i) is not suppressed by couplings. 
However, the annihilation cross section
\begin{eqnarray}
\sigma (F^+_i F^-_i \to \gamma\gamma) \cdot v
&=& \frac{e^4}{8\pi m_F^2} + {\cal O}(v^2) ~, 
\end{eqnarray}
where $v$ is the relative velocity of initial particles,  is numerically small and negligible 
with $m_F \gtrsim 2\,\text{TeV}$.
The process (ac-ii) is suppressed by $F\bar{F}Z$ couplings.
The cross section in the process (ac-iii) with $V=\gamma$ is estimated as
\begin{eqnarray}
\sum_{f}\sigma (F^+_i F^-_i \to \gamma \to f\bar{f})\cdot v 
&=& 8 \cdot \frac{e^4}{16\pi m_F^2} + {\cal O}(v^2) ~.
\end{eqnarray}
The process (ac-iii) with $V=\gamma^{(1)}$ can be enhanced by both large right-handed coupling 
of fermions and Breit-Wigner resonances.
The process (ac-iv) is suppressed by the small $\gamma^{(1)}W^+ W^-$ coupling.

As for coannihilation, we have tabulated  possible processes in Table. \ref{tbl:process}.
We find that the process (co-i) with $V=W^+,W^{+(n)}$ is suppressed by small 
$F^+\bar{F}^0 W^-$ couplings and the process (co-i) with $V=W_R$ is forbidden 
because of vanishing $W_R\bar{f}f$ couplings.
The process (co-ii) with $V=W,W^{(n)}$ is suppressed by small $\bar{F}FW$ couplings,
and (co-ii) with $V=W_R$ is suppressed  by the small $W_R-W-Z$ couplings.
The process (co-iii) with $V=W, W^{+(n)}$ is suppressed by small $F\bar{F}W^{(n)}$ couplings.
The process (co-iii) with $V = W_R$ is forbidden by the vanishing $W_R W \gamma$ coupling 
which ensures the ortho-normality of the KK gauge bosons.
The processes (co-iv) and (co-v) are suppressed by small $F\bar{F}Z$ and $F^+\bar{F}^0 W^-$ 
couplings.  Hence we found that  all of the co-annihilation processes are either vanishing or strongly
suppressed.

Thus we find that relevant processes for  dark fermion annihilation are
the following s-channel processes
\begin{eqnarray}
\begin{matrix}
&F^0 \bar{F}^0 &\to& Z_R^{(1)} &\to& q\bar{q},\, l\bar{l},\, \nu\bar{\nu},
\\
&F^+ F^- &\to& \gamma,\, \gamma^{(1)} &\to& q\bar{q},\,  l\bar{l},
\\
&F^+ F^- &\to& Z_R^{(1)} &\to& q\bar{q},\, l\bar{l},\, \nu\bar{\nu},
\label{eq:relevant-processes}
\end{matrix}
\end{eqnarray}
and all other annihilation and co-annihilation processes are negligible.

In the followings, we calculate the relic density of the dark fermions
using annihilation cross sections of the processes  given in \eqref{eq:relevant-processes}.
For charged dark fermions,
the annihilation cross section of $F_i^+ F_i^-$ to the SM fermions is given by
\begin{eqnarray}
\lefteqn{
\sum_{f} \sigma(F_i^+ F_i^- \to \{\gamma,\,\gamma^{(1)},\,Z_R^{(1)} \}\to \bar{f}f)
}\nonumber\\
&=& 
8 \cdot \frac{e^4}{16\pi \beta s^2} \Big(s + 4 m_F^2 + \frac{1}{3}s \beta^2\Big)
%\frac{1}{2 n_F}
\nonumber\\
\noalign{\kern 10pt}
&&
\hskip -1.cm
+\frac{1}{64\pi\beta} \biggl[
\frac{s}{(s-m^2_{Z_R^{(1)}})^2 + m_{Z_R^{(1)}}^2 \Gamma_{Z_R^{(1)}}^2}
g_w^4 \bigg(\sum_{f} 
\bigg[ \Big(g_{fL}^{Z_R^{(1)} }\Big)^2 
+ \Big(g_{fR}^{Z_R^{(1)}}\Big)^2 \bigg] \bigg)
\nonumber\\
\noalign{\kern 10pt}
&&\hskip 1.cm
\times
\biggl\{
 \left(1 + \frac{\beta^2}{3}\right) \bigg[ \Big(g_{F^+ L}^{Z_R^{(1)}}\Big)^2 
 + \Big(g_{F^+ R}^{Z_R^{(1)}} \Big)^2 \bigg]
 + 8 \frac{m_F^2}{s} g_{F^+ L}^{Z_R^{(1)}} g_{F^+ R}^{Z_R^{(1)}}  \biggr\}
\nonumber\\
\noalign{\kern 10pt}
&& 
\hskip -1.cm
+\frac{s}{(s-m_{\gamma^{(1)}}^2)^2 + m_{\gamma^{(1)}}^2 \Gamma_{\gamma^{(1)}}^2}
e^4 \left(\sum_{f} \left[ \left(g_{fL}^{\gamma^{(1)}}\right)^2 + \left(g_{fR}^{\gamma^{(1)}}\right)^2 \right] \right)
\nonumber\\
\noalign{\kern 10pt}
&&\hskip 1.cm
\times
\biggl\{
 \left(1 + \frac{\beta^2}{3}\right) \left[ \left(g_{F^+ L}^{\gamma^{(1)}}\right)^2 + \left(g_{F^+ R}^{\gamma^{(1)}} \right)^2 \right]
 + 8 \frac{m_F^2}{s} g_{F^+ L}^{\gamma^{(1)}} g_{F^+ R}^{\gamma^{(1)}}  \biggr\}
\nonumber\\
\noalign{\kern 10pt}
&&
\hskip -1.cm
+ 2 \cdot \frac{(s-m_{Z_R^{(1)}}^2)(s-m_{\gamma^{(1)}}^2) + m_{Z_R^{(1)}} m_{\gamma^{(1)}} \Gamma_{Z_R^{(1)}}\Gamma_{\gamma^{(1)}}}{[(s-m_{Z_R^{(1)}}^2)^2 
   + m_{Z_R^{(1)}}^2 \Gamma_{Z_R^{(1)}}^2] [(s - m^2_{\gamma^{(1)}})^2 + m_{\gamma^{(1)}}^2 \Gamma_{\gamma^{(1)}}^2]}
\cdot s
\nonumber\\
\noalign{\kern 10pt}
&&
\hskip 0.cm
\times
g_w^2 e^2 \left(\sum_{f} \left[ g _{fL}^{Z_R^{(1)}} g_{fL}^{\gamma^{(1)}}
 + g_{fR}^{Z_R^{(1)}} g_{fR}^{\gamma^{(1)}} \right] \right)
\nonumber\\
\noalign{\kern 10pt}
&&
\hskip -1.cm
\times
\biggl\{
 \Big(1 + \frac{\beta^2}{3}\Big) \bigg[ 
   g_{F^+ L}^{Z_R^{(1)}} g_{F^+ L}^{\gamma^{(1)}}  
 + g_{F^+ R}^{\gamma^{(1)}} g_{F^+ R}^{Z_R^{(1)}} \bigg]
 + 4 \frac{m_F^2}{s} \left[
   g_{F^+ L}^{Z_R^{(1)}} g_{F^+ R}^{\gamma^{(1)}} 
 + g_{F^+ L}^{\gamma^{(1)}} g_{F^+ R}^{Z_R^{(1)}} 
 \right] \biggr\}
\biggr],
\label{eq:annihilation}
\nonumber\\
\end{eqnarray}
where 
$g_{FL/FR}^{V} \equiv V^{(V)}_{F} \mp A^{(V)}_{F}$, 
$g_{fL/fR}^{V} \equiv v^{(V)}_f \mp a^{(V)}_f$
($V = Z_R^{(1)},\,\gamma^{(1)}$) and the couplings are summarized in 
Sec.~\ref{sec:vector-coupling}.
$\beta \equiv \sqrt{1 - 4 m_F^2/s}$ and $s$ is the invariant mass of $F\bar{F}$.
We have neglected $\gamma$-$\gamma^{(1)}$ and $\gamma$-$Z_R^{(1)}$ interference terms. 
$F_i^0\bar{F}_i^0$ annihilation cross section 
$\sum_f \sigma( F^0\bar{F}^0 \to Z_R^{(1)} \to \bar{f} f)$ 
is obtained from \eqref{eq:annihilation} by replacing $f_{L+/R+}^{(V)}$ with 
$f_{L/R}^{(V)} \equiv V_{F}^{(V)} \pm A_{F}^{(V)}$
and ignoring $e^2$ and $e^4$ terms.
%$1/2n_F$ counts the probability of Dirac fermions with $n_F$-species meeting their antiparticles.
$\Gamma_{Z_R^{(1)}}$ and $\Gamma_{\gamma^{(1)}}$ are the total decay rate of $Z_R^{(1)}$ and $\gamma^{(1)}$ bosons, and $\Gamma_{Z_R^{(1)}}$ is estimated to be
\begin{eqnarray}
&&\hskip -1.cm
\Gamma_{Z_R^{(1)}} =
\sum_{f} {N_{c,f}} \frac{m_{Z_R^{(1)}}}{24\pi} 
g_w^2 \gamma \Big(g_{fL}^{Z_R^{(1)}}, g_{fR}^{Z_R^{(1)}}, m_f^2/m_{Z_R^{(1)}}^2\Big)
\nonumber\\
\noalign{\kern 10pt}
&& \hskip -.5cm
+ \sum_{F} \frac{m_{Z_R^{(1)}}}{24\pi} g_w^2
\biggl[
\gamma\Big(g_{F^0 L}^{Z_R^{(1)}}, g_{F^0 R}^{Z_R^{(1)}}, m_F^2/m_{Z_R^{(1)}}^2\Big)
 + \gamma\Big(g_{F^+ L}^{Z_R^{(1)}}, g_{F^+ R}^{Z_R^{(1)}}, m_F^2/m_{Z_R^{(1)}}^2\Big)
\biggr] ~,
\nonumber\\
\noalign{\kern 10pt}
&&\hskip -1.cm
\gamma(g_L,g_R,x) \equiv \sqrt{1-4x} [g_L^2 + g_R^2 - x (g_L^2 + g_R^2 - 6 g_L g_R)]~.
\label{eq:ZR-decayrate}
\end{eqnarray}$\Gamma_{\gamma^{(1)}}$ is obtained in an analogous way.
Here $N_{c,f}=3$ ($1$) when $f$ is a quark (charged lepton or neutrino). 
$m_f$ is the mass of the SM fermion.
%$f_{L+}$ and $f_{R+}$ are left- and right-handed couplings of $F^\pm$ to the $Z_R^{(1)}$ boson.
We note that the $F^\pm$ contributions in \eqref{eq:ZR-decayrate} are rather large.

Let $n_{0}$ [$n_{+}$] be the number-density of $F^0_i$ and $\bar{F}^0_i$ 
[$F^+_i$ and $F^-_i$] ($i=1,\dots,n_F$),
and $\sigma_0$ [$\sigma_+$] be the annihilation cross section of $F^0_i$ [$F^+_i$].
Then the evolution of the total number density of the DM, 
is given by $n \equiv 2n_F (n_0 + n_+)$,
and the time-evolution of $n$ is governed by the Boltzmann equation
\begin{eqnarray}
\frac{d n}{dt} &=& - 3 H n - 2 n_F \langle \sigma_0  v \rangle (n_0^2 - n_{0,\rm eq}^2)
- 2 n_F \langle \sigma_+ v \rangle (n_+^2 - n_{+,\rm eq}^2),
\end{eqnarray}
where $n_{0/+,\rm eq}$ is the number-density in the thermal equilibrium 
and  approximated by
$n_{0/+,\rm eq} \approx g_{0/+} (m_{F^{0/\pm}} T/2\pi)^{3/2} \exp(-m_{F^{0/\pm}}/T)$ with $g_{0/+}=2$ being the number of degrees of freedom of $F^0_i$ and $F^+_i$.
Using the relations $n_{0,+}/n_{0,+\rm eq} = n/n_{\rm eq}$ and
$n_{0,\rm eq}/n_{\rm eq} = n_{+,\rm eq} /n_{\rm eq} = 1/4n_F$,
we obtain
\begin{eqnarray}
\frac{dn}{dt} &=& -3 H n 
- \langle \sigma_{\rm eff} v \rangle (n^2 - n_{\rm eq}^2),
\quad
\sigma_{\rm eff} v \equiv \frac{\sigma_0 v + \sigma_+ v}{8n_F}.
\label{eq:Boltzmann}
\end{eqnarray}
We introduce $Y_{({\rm eq})} \equiv n_{({\rm eq})}/S$ where $S = 2\pi^2 g_* T^3/45$ is the entropy density.
$g_*$ is the degree of freedom at the freeze-out temperature $T_f$ and we take $g_* = 92$. Conservation of entropy per co-moving volume, $S a_{sf}^3 = \text{constant}$ ($a_{sf}$ is the scale factor of the expanding universe), reads
$dn/dt + 3Hn = S dY/dt$.
The Hubble constant is given by $H^2 = 4\pi^3 g_* T^4/(45M_{Pl}^2)$
and $t = 1/2H$ in the radiation-dominant era.
$M_{Pl}$ is the Planck mass.
Hence we rewrite the Boltzmann equation as
\begin{eqnarray}
\frac{dY}{dx} = \frac{\langle \sigma_{\rm eff} v\rangle}{H} \frac{1}{x} S (Y^2 - Y_{\text{eq}}^2),
\label{eq:Boltzmann1}
\end{eqnarray}
where $x \equiv T/m_F$ and $T$ is the temperature of the universe.
$\langle \sigma v\rangle = \langle \sigma v \rangle(x)$ is the thermal-averaged cross section discussed later. 
$n_{\text{eq}}$ is the density in the thermal equilibrium, and becomes
\begin{eqnarray}
n_{\text{eq}} &=& g_{\rm eff} \left( \frac{m_F T}{2\pi}\right)^{3/2} e^{-m_F/T}
\label{eq:density-equilibrium}
\end{eqnarray}
($g_{\rm eff}=2\cdot 4 n_F$ is the degree of freedom of the dark fermions)
in the non-relativistic limit.
Defining $\Delta \equiv Y - Y_{\text{eq}}$ and $\Delta'\equiv d\Delta/dx$, 
$Y_{\rm eq}' \equiv dY_{\rm eq}/dx$, we rewrite \eqref{eq:Boltzmann1} as
\begin{eqnarray}
\Delta' &=& - Y'_{\text{eq}} + f(x) \Delta (2 Y_{\text{eq}} + \Delta),
\quad
f(x) = \sqrt{\frac{\pi g_*}{45}} m_F M_{Pl} \langle \sigma v\rangle,
\label{eq:Boltzman2}
\end{eqnarray}
which is written at early times ($x \gg x_f \equiv T_f/m_F$, $|\Delta'| \ll |Y'_{\text{eq}}|$) as
\begin{eqnarray}
\Delta &=& \frac{Y'_{\text{eq}}}{f(x) (2Y_{\text{eq}} + \Delta)}.
\label{eq:early-time}
\end{eqnarray}
At late times ($T \ll T_f$),
$Y_{\text{eq}} \ll Y \sim \Delta$ and $|\Delta'| \gg |Y'_{\rm eq}|$,
hence \eqref{eq:Boltzman2} reads
\begin{eqnarray}
\Delta^{-2} \Delta' = f(x).
\label{eq:rate-time}
\end{eqnarray}
Integrating \eqref{eq:rate-time} with $x$ from zero to $x_f\equiv T_f/m_F$,
we obtain
\begin{eqnarray}
Y_{0}^{-1} \simeq \Delta_{0}^{-1} = \int_0^{x_f} f(x) dx
= \sqrt{\frac{\pi g_*}{45}} M_{Pl} m_F J_f,
\quad
J_f \equiv \int_{0}^{x_f} \langle \sigma_{\rm eff} v \rangle(x) dx,
\label{Jf_definition}
\end{eqnarray}
where we have used $\Delta(x_f) \gg \Delta(x=0)$.
Thus the relic density of the dark fermions at the present time is given by 
\begin{eqnarray}
\Omega_{DM} h^2 &=& \frac{\rho_{DM}}{\rho_c} h^2 = \frac{m_F S_0 Y_0 h^2}{\rho_c}
=
 \frac{1.04 \times 10^9}{M_{Pl}} \frac{1}{\sqrt{g_*}} \frac{1}{J_f},
\label{eq:relic_density}
\end{eqnarray}
where $\rho_{DM} = m_F S_0 Y_0$ and 
$\rho_c = 3 H_0^2 M_{Pl}^2/8\pi = 1.054\times10^{-5}\text{GeV}\text{cm}^{-3}$ have been 
made  use of.  $S_0=2889.2\text{cm}^{-3}$ is the entropy-density of the present universe.

The freeze-out temperature is determined by solving the condition
\begin{eqnarray}
\Delta(x_f) &=& c Y_{\text{eq}}(x_f),
\label{eq:freeze-out}
\end{eqnarray}
with $\Delta$ in the early-time.
$c$ is an numerical factor of order unity and determined by matching the late-time and early-time solutions. Hereafter we take $c=1/2$.
Eq. \eqref{eq:freeze-out} with \eqref{eq:early-time} reads
the following transcendental relation
\begin{eqnarray}
x_f^{-1} 
&=& \ln \left(
c(c+2)  \sqrt{\frac{45}{8}} \frac{g_{\rm eff}}{2\pi^3} \frac{m_F M_{Pl} x_f^{1/2} \langle \sigma_{\rm eff} v \rangle}{g_*^{1/2}}\right),
\end{eqnarray}
which can be solved by numerical iteration.

The precise form of the velocity-averaged cross section $\langle \sigma v \rangle$ is given 
in Ref.~\cite{Srednicki:1988ce}. When $\sigma v$ is expanded in $v^2$ as
\begin{eqnarray}
\sigma v &=& a + b v^2 + \cdots = a + b [(s-4m_F^2)/m_F^2] + \cdots,
\label{eq:v-expand}
\end{eqnarray}
we obtain
\begin{eqnarray}
\langle \sigma v \rangle 
&=& 4\pi \left( \frac{m_F}{4\pi T}\right)^{3/2} 
\int_{0}^{\infty} dv \, v^2 e^{-m_F v^2/4T} \sigma v
\nonumber\\
&=& a + 6bT/m_F + \cdots.
\end{eqnarray}
In the present case $x_f \sim 1/30$ and therefore
only the first term in the $v^2$ expansion in Eq.~\eqref{eq:v-expand} is kept in the following analysis.

\subsection{Relic density of degenerate dark fermions}
First we consider the case in which all dark fermions are degenerate.
In the numerical study of this paper, we have adopted 
$\alpha_{EM} \equiv e^2/4\pi = 1/128$,
$\sin^2\theta_W = 0.2312$,
$m_Z=91.1876$ GeV and $m_{\rm top} = 171.17$ GeV.\cite{PDG2012}
In Table.~\ref{tbl:Degenerate}, we have summarized 
values of $\theta_H$, the bulk mass parameters of the top quark $c_{\rm top}$ and the dark fermion $c_F$, and mass of the dark fermion $m_F$ for particular values of $(z_L, n_F)$.
$\theta_H$, $c_{\rm top}$ and $c_F$ are chosen so that we obtain $126\text{GeV}$ Higgs mass
\cite{FHHOS2013,LHCsignals}.
\begin{table}[tbp]
\caption{$\theta_H$, $c_{\rm top}$, $c_F$ and $m_F$ for $z_L$ and $n_F= 3,4,5$ and $6$,
in the case where dark fermions are degenerate.}\label{tbl:Degenerate}
\begin{center}
\begin{tabular}{c|c|cccc}
\hline\hline
$n_F$ & $z_L$ &
$\theta_H$ & $c_{\rm top}$ &
$c_F$ & $m_F$ 
\\
& & & & & [TeV]  % &  & 
\\ 
\hline
$3$ & $10^8$ &
    $0.360$ & $0.357$ &
    $0.385$ & $0.670$ %& $-0.106$ 
\\
    & $10^6$ &
    $0.177$ & $0.296$ &
    $0.309$ & $1.54$ %& $-0.071$ 
\\
    & $10^5$ &
    $0.117$ & $0.227$ &
    $0.235$ & $2.54$ %& $-0.064$  
\\
    & $2\times10^4$ &
    $0.0859$ & $0.137$ &
     $0.127$ & $3.88$ %& $-0.089$ 
    \\
\hline
$4$ & $10^8$ & 
    $0.355$ & $0.357$ &
    $0.423$ & $0.567$ %& $-0.082$ 
\\
    & $10^6$ &
    $0.174$ & $0.292$ &
    $0.374$ & $1.27$ %& $-0.049$ 
\\
    & $10^5$ &
    $0.115$ & $0.227$ &
    $0.332$ & $2.03$ %& $-0.038$ 
\\
    &$3\times10^4$ &
    $0.0917$ &  $0.168$ &
    $0.299$ & $2.66$ %& $-0.034$ 
\\
    & $10^4$ &
    $0.0737$ & $0.0366$ & 
    $0.256$ & $3.46$ %& $-0.033$
\\
\hline
$6$ & $10^8$ & 
    $0.348$ & $0.356$ &
    $0.461$ & $0.455$ %& $-0.060$ 
\\
    & $10^6$ &
    $0.171$ & $0.292$ &
    $0.434$ & $1.00$ %& $-0.034$ 
\\
    & $10^5$ &
    $0.113$ & $0.227$ &
    $0.414$ & $1.57$ %& $-0.024$ 
\\
    & $10^4$ &
    $0.0724$ & $0.0365$ &
    $0.379$ & $2.57$ %& $-0.017$ 
\\
\hline\hline
\end{tabular}
\end{center}
\end{table}

\begin{figure}[tbp]
\centerline{\includegraphics[width=9cm]{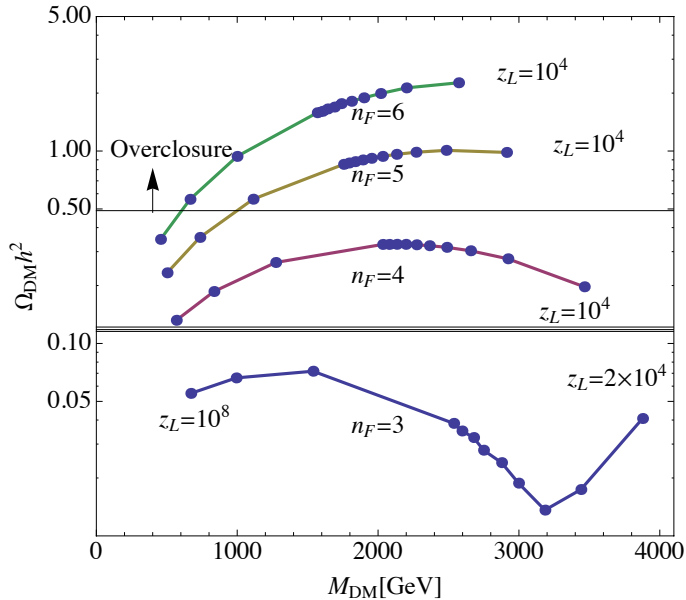}}
\caption{%
Relic density of  neutral dark fermions in the case of $n_f$ degenerate  dark fermion 
multiplets ($n_F=3,4,5, 6$).
Data points are, from right to left,  $z_L=10^4$ ($2\times 10^4$) to $10^5$ with a step of $10^4$,  $10^6$, $10^7$ and $10^8$ for $n_F=4,5,6$ ($n_F=3$).
The current observed limit of $\Omega_{\rm DM} h^2$ and the lower bound of the over-closure of 
the universe are  indicated as horizontal lines.}\label{fig:density-nomix}
\end{figure}

In Fig.~\ref{fig:density-nomix}  the relic density of the dark fermions for $n_F=3,4,5$ and 6 is plotted.
In the plot, the best value [68\% confidence level (CL) limits] of the  relic density
of the cold dark matter observed by Planck \cite{Ade:2013zuv}:
\begin{eqnarray}
\Omega_{\rm CDM} h^2 &=& 0.11805\quad[0.1186 \pm 0.0031],
\label{eq:cdm-observed}
\end{eqnarray}
has been also shown.
Here Hubble's expansion-rate $H_0 \equiv 100h\,\text{km}\,\text{s}^{-1}\text{Mpc}^{-1}$,
$100h = 67.11$ [$67.4\pm1.4$].
In our previous work \cite{LHCsignals}, we have constrained $z_L$ by
$z_L \lesssim 10^6$ because no evidence of the neutral boson resonances in LHC have been seen.
For $z_L \lesssim 10^6$, we found that no parameter regions can explain the current DM density.
For $n_F=3$ we obtain $\Omega_{\rm DM} h^2 \lesssim 0.08$ for any value of $z_L$.
For $n_F=4$ and $z_L \le 10^6$, we have $\Omega_{\rm DM} h^2 \gtrsim 0.2$.
For $n_F=5$ and $6$, predicted densities are larger than the limit on the closure universe.

We remark that for $n_F=3$ the relic density becomes very small at $z_L \sim 3\times10^4$
due to the fact that the masses of the 1st KK vector bosons are very close to  twice  the mass of dark fermions, and the enhancement due to the Breit-Wigner resonance happens. 
A similar mechanism occurs in some of the universal extra-dimension models\cite{UED-Kakizaki1,UED-Kakizaki2,Split-UED1,Split-UED2}. 

%%%%%%%%%%%%%%%%%%%%%%%%%%%%%%%%%%%%%%%%%%%%%%%%%%
\subsection{Current mixing} 

So far it has been supposed that $n_F$ multiplets of  $SO(5)$-spinor fermions $\Psi_{F_i}$
are degenerate. 
There is an intriguing scenario that some of them are heavier than others, only
the lightest $F^{0(1)}_i$'s becoming the dark matter.  
A typical mass of $F^{0(1)}_i$ is $1 \sim 3 \,$TeV.  
We show that the mass difference   of $O( 200)\,$GeV and small mixing could fulfill this job.

\begin{figure}[htbp]
\begin{center}
\includegraphics[height=3cm]{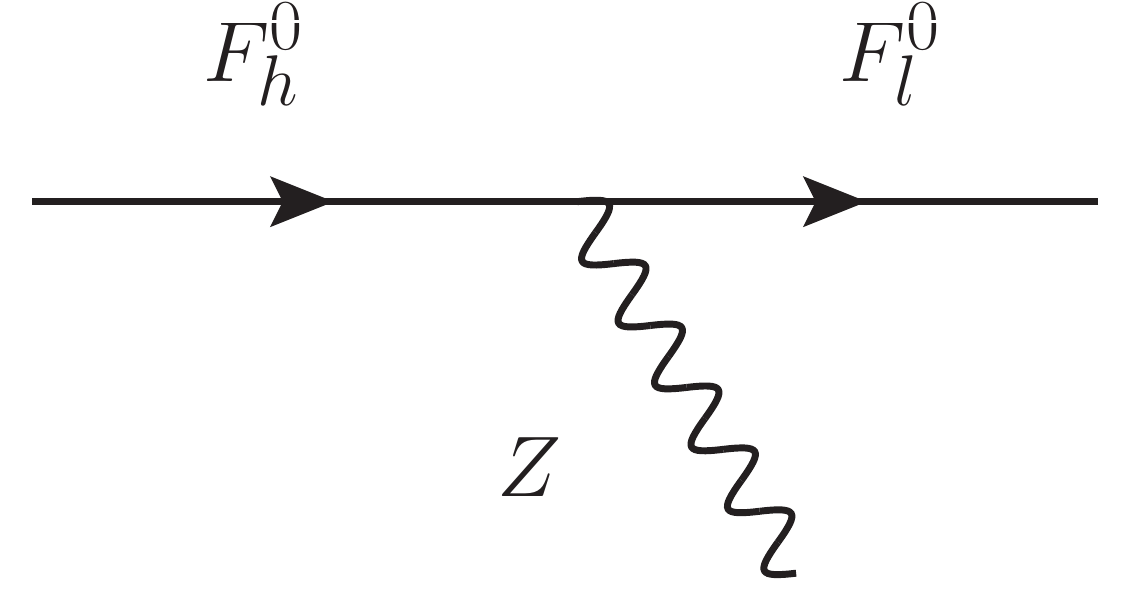}
\quad
\includegraphics[height=3cm]{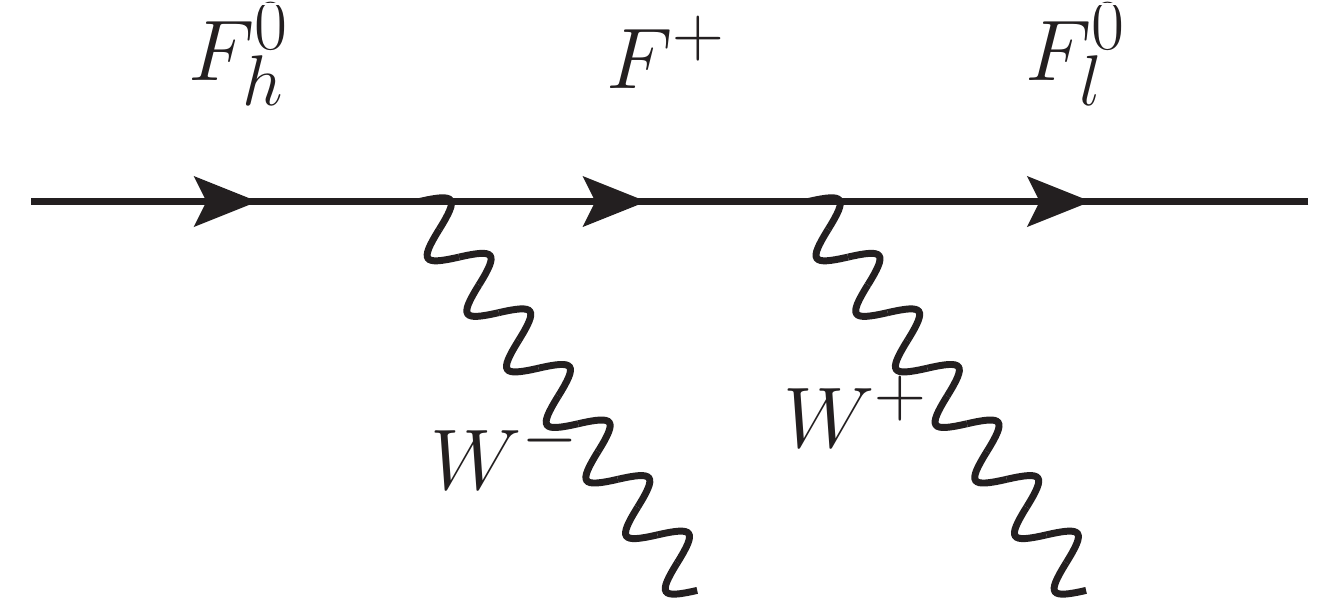}
\caption{$F_h^0$ decay to $F_l^0$ by emitting one $Z$ boson or  two $W$ bosons.}
\label{fig:FtoFWW}
\end{center}
\end{figure}

Let us denote the lightest particles of heavy and light $SO(5)$-spinor fermions 
by $(F^+_h, F^0_h)$ and $(F^+_l, F^0_l)$, respectively.  Charged $F^+_l$ and 
$F^+_h$ are heavier than the corresponding neutral ones, and are supposed to
decay sufficiently fast.  $F^0_h$ also needs to decay sufficiently fast in order
for the scenario to work.  
$F^0_h$  can decay either as $\go F_l^0 + Z$ or 
as  $\go F^+_l + W^- \go F^0_l + W^+ + W^-$ as shown  in Fig.~\ref{fig:FtoFWW}.   
For this process the off-diagonal neutral or charged current is necessary.   
We examine in this subsection how the off-diagonal  currents are generated.

To be concrete, let us suppose that there are only two $SO(5)$-spinor fermion
multiplets, $\Psi_{F_h}$ and $\Psi_{F_l}$, which are gauge-eigenstates.   
We suppose that $\Psi_{F_l}$ obeys the boundary condition $\eta_{F_l} = +1$ 
in (\ref{BC1}), whereas $\Psi_{F_h}$ satisfies the flipped boundary condition $\eta_{F_h} = -1$.
It is easy to confirm that their KK spectrum is given by  (\ref{SpectrumF1}) for both
$\Psi_{F_h}$ and $\Psi_{F_l}$.
The lowest mode $(F_h^{+ (1)}, F_h^{0 (1)})$ is mostly
an $SU(2)_L$ doublet, whereas   $(F_l^{+ (1)}, F_l^{0 (1)})$
is mostly an $SU(2)_R$ doublet.

Let us denote gauge (mass) eigenstates of the lightest modes of $\Psi_{F_h}, \Psi_{F_l}$ 
by $ \hat F^+_h, \hat F^0_h, \hat F^+_l, \hat F^0_l$   ($  F^+_h,  F^0_h, F^+_l,  F^0_l$).
The most general form of bulk mass terms for $\Psi_{F_h}$ and $\Psi_{F_l}$ is 
\beeq
\cL_F^{\rm 5D \, mass} = - \sigma'(y) \big\{ c_{F_h} \bar \Psi_{F_h} \Psi_{F_h} 
+ c_{F_l} \bar \Psi_{F_l} \Psi_{F_l} \big\} 
- \tilde\Delta \big\{ \bar \Psi_{F_h} \Psi_{F_l} + \bar \Psi_{F_l} \Psi_{F_h} \big\} ~.
\label{action2}
\eneq
We note that $\bar \Psi_{F_h} \Psi_{F_h}$ and $\bar \Psi_{F_l} \Psi_{F_l}$ are  odd 
under parity $y \go -y$, whereas $\bar \Psi_{F_h} \Psi_{F_l}$ is even.
The $\tilde \Delta$ term induces mass mixing among $\hat F^+_h$ and $\hat F^+_l$,
and among $\hat F^0_h$ and $\hat F^0_l$.  
$c_{F_h}$ and $c_{F_l}$ generate masses $\hat m_h$ and $\hat m_l$ 
for $(\hat F^+_h, \hat F^0_h)$ and  $(\hat F^+_l, \hat F^0_l)$. 
We suppose that $c_{F_h} < c_{F_l}$ so that $\hat m_h > \hat m_l$.
As described in Sec.~3.1, charged states acquire radiative corrections (\ref{eq:mass-diff}),
$a \, \hat m_h$ ($a \, \hat m_l$) for $\hat F^+_h$ ($\hat F^+_l$) where $a$ is 
$O(10^{-3} \sim  10^{-2})$.

Hence the mass matrices are given by
\beqn
&&\hskip -1.cm
\cL_F^{\rm 4D \,   mass} = - (\bar {\hat F}^+_h , \bar {\hat F}^+_l) \, \cM_+
\begin{pmatrix} {\hat F}^+_h \cr {\hat F}^+_l \end{pmatrix}
-  (\bar {\hat F}^0_h , \bar {\hat F}^0_l) \, \cM_0  
\begin{pmatrix} {\hat F}^0_h \cr {\hat F}^0_l \end{pmatrix} , \cr
\noalign{\kern 10pt}
&&\hskip -1.cm
\cM_+  = \begin{pmatrix} (1+a) \hat m_h & \Delta \cr \Delta & (1+a) \hat m_l \end{pmatrix}
%\cM_{F^+}  = \begin{pmatrix} (1+a) \hat m_h & \Delta \cr \Delta & (1+a) \hat m_l \end{pmatrix}
~~,~~
\cM_0  = \begin{pmatrix}  \hat m_h & \Delta \cr \Delta &  \hat m_l \end{pmatrix} .
\label{massF2}
\eeqn
We suppose that $\Delta \ll \hat m_h , \hat m_l$. 
We diagonalize the two matrices to obtain
\beqn
&&\hskip -1.cm
\cL_F^{\rm 4D \,   mass} = - m_{F_h^+} \bar F_h^+ F_h^+ - m_{F_l^+} \bar F_l^+ F_l^+
- m_{F_h^0} \bar F_h^0 F_h^0 - m_{F_l^0} \bar F_l^0 F_l^0 ~, \cr
\noalign{\kern 10pt}
&&\hskip -1.cm
\begin{pmatrix} F_h^+ \cr F_l^+ \end{pmatrix} 
= V \big( \onehalf \alpha_+ \big) \begin{pmatrix} \hat F_h^+ \cr \hat F_l^+ \end{pmatrix} ,~~
%= V\Big(\frac{ \alpha_+}{2} \Big) \begin{pmatrix} \hat F_h^+ \cr \hat F_l^+ \end{pmatrix} ,~~
\begin{pmatrix} F_h^0 \cr F_l^0 \end{pmatrix} 
= V\big(\onehalf \alpha_0 \big) \begin{pmatrix} \hat F_h^0 \cr \hat F_l^0 \end{pmatrix} , \cr
%= V\Big(\frac{ \alpha_0}{2} \Big) \begin{pmatrix} \hat F_h^0 \cr \hat F_l^0 \end{pmatrix} , \cr
\noalign{\kern 10pt}
&&\hskip -1.cm
\begin{pmatrix} m_{F_h^+} \cr m_{F_l^+} \end{pmatrix} 
= \onehalf (1+a) (\hat m_h + \hat m_l ) \pm 
\sqrt{ \onefourth (1+a)^2 (\hat m_h - \hat m_l)^2 +\Delta^2 } ~, \cr
\noalign{\kern 10pt}
&&\hskip -1.cm
V(\alpha) = \begin{pmatrix} \cos  \alpha &  \sin  \alpha \cr
- \sin  \alpha &  \cos  \alpha \end{pmatrix} , ~~
\tan \alpha_+ = \frac{2 \Delta}{(1+a) (\hat m_h - \hat m_l)} ~.
\label{massF3}
\eeqn
The masses $(m_{F_h^0}, m_{F_l^0})$ and angle $\alpha_0$ are obtained from 
$(m_{F_h^+}, m_{F_l^+})$ and $\alpha_+$ by taking $a \go 0$.

The couplings to $Z$ (the neutral currents) are given originally by
\beeq
Z_\mu \sum_{F_j = F_h^+, F_l^+, F_h^0, F_l^0}
\Big\{ g^Z_{F_j L}\bar {\hat F}_{jL} \gamma^\mu {\hat F}_{jL}
+ g^Z_{F_j R}\bar {\hat F}_{jR} \gamma^\mu {\hat F}_{jR} \Big\} ~.
\label{FZcoupling1}
\eneq
Similarly the couplings to $W$ (the charged currents) are given by
\beeq
W_\mu \sum_{j=h,l} \Big\{  g^W_{F_j L}\bar {\hat F}_{jL}^+ \gamma^\mu {\hat F}_{jL}^0
+ g^W_{F_j R}\bar {\hat F}_{j R}^+ \gamma^\mu {\hat F}_{j R}^0 \Big\} + (h.c.).
\label{FWcoupling1}
\eneq
We recall that  $(F_h^+, F_h^0)$ is mostly an $SU(2)_L$ doublet, whereas 
$(F_l^+, F_l^0)$ is mostly an $SU(2)_R$ doublet 
with the  boundary conditions imposed on $\Psi_{F_h}$ and $\Psi_{F_l}$.
Therefore   $g^Z_{F_h^0 L} \gg g^Z_{F_l^0 L}$ and $g^W_{F_h L} \gg g^W_{F_l L}$ etc..
In terms of mass eigenstates the neutral current becomes
\beqn
&&\hskip -1.cm
(\bar F_{hL}^0 , \bar F_{lL}^0 ) \bigg\{ \frac{g^Z_{F_h^0 L} + g^Z_{F_l^0 L}}{2}
+ \frac{g^Z_{F_h^0 L} - g^Z_{F_l^0 L}}{2} \, U( \alpha_0) \bigg\}\gamma^\mu 
\begin{pmatrix} F_{hL}^0 \cr  F_{lL}^0 \end{pmatrix} \cr
\noalign{\kern 10pt}
&&\hskip -1.cm
+ (\bar F_{hL}^+ , \bar F_{lL}^+ ) \bigg\{ \frac{g^Z_{F_h^+ L} + g^Z_{F_l^+ L}}{2}
+ \frac{g^Z_{F_h^+ L} - g^Z_{F_l^+ L}}{2} \, U( \alpha_+) \bigg\}\gamma^\mu 
\begin{pmatrix} F_{hL}^+ \cr  F_{lL}^+ \end{pmatrix} \cr
\noalign{\kern 10pt}
&&\hskip +1.cm
+ ( L \go R ) ~,
\label{FZcoupling2}
\eeqn
where
\beqn
U(\alpha) = \begin{pmatrix} \cos \alpha & - \sin \alpha \cr - \sin\alpha & - \cos\alpha \end{pmatrix} .
\eeqn
The charged current is
\beqn
&&\hskip -1.cm
(\bar F_{hL}^+ , \bar F_{lL}^+ ) 
\bigg\{ \frac{g^W_{F_h L} + g^W_{F_l L}}{2} V\Big(\frac{\alpha_+ - \alpha_0}{2} \Big)
+ \frac{g^W_{F_h L} - g^W_{F_l L}}{2} \, U\Big( \frac{\alpha_+ + \alpha_0}{2} \Big) \bigg\}
\gamma^\mu 
\begin{pmatrix} F_{hL}^0 \cr  F_{lL}^0 \end{pmatrix} \cr
\noalign{\kern 10pt}
&&\hskip 1.cm
+ (L \go R) ~.
\label{FWcoupling2}
\eeqn
We recognize that off-diagonal neutral and charged currents are generated for dark fermions
obeying the distinct boundary conditions.

For small $\theta_H$, heavy dark fermions have much larger couplings to $W$ and $Z$
than light dark fermions.
Let us suppose that $\Delta  \ll \hat m_h - \hat m_l$ so that 
$\onehalf \alpha_0\sim \Delta/(\hat m_h - \hat m_l)  \ll 1$ and $\alpha_+ \sim \alpha_0/(1+a) \ll 1$.
The $Z$ coupling of  $F_{lL/R}^0$ is  
$\sim g^Z_{F_l^0 L/R} +  g^Z_{F_h^0 L/R} (\onehalf \alpha_0)^2$.  We assume that 
$(\onehalf \alpha_0)^2 \ll {\rm sup} ( | g^Z_{F_l^0 L} / g^Z_{F_h^0 L}| , | g^Z_{F_l^0 R} / g^Z_{F_h^0 R}|)$ 
so that  the estimate of the cross section for the direct detection experiments discussed
in the next section remains valid.

The couplings for $F_{hL}^0 \go F_{lL}^0 +Z$ and for $F_{hL}^0 \go F_{lL}^+ + W^-$ are
approximately  $ - \onehalf g^Z_{F_h^0 L} \alpha_0$ and $- \onehalf g^W_{F_h L} \alpha_0$,
respectively.  With a moderate 
$\onehalf \alpha_0  \sim \frac{1}{3}  \, {\rm sup} ( | g^Z_{F_l^0 L} / g^Z_{F_h^0 L}|^{1/2}, 
| g^Z_{F_l^0 R} / g^Z_{F_h^0 R}|^{1/2} )$, 
$F_{hL}^0$ decays sufficiently fast.  Only the light dark fermion $F_{lL}^0$
becomes a candidate of dark matter.

%%%%%%%%%%%%%%%%%%%%%%%%%%%%%%%%%%%%%%%%%%%%%%%%%%%%%%%%%%%%%%%%%%%%%%%%%%%%%%%%%%%%%%
\subsection{Relic density of non-degenerate dark fermions}

Let us examine the case with non-degenerate dark fermions.
We separate the $n_F$ dark fermions $(F_i^+,  F_i^0)$  ($i=1,\cdots,n_F$) into 
$\nlight$ light fermions $(F_l^+,  F_l^0)$ (with bulk mass $c_{F_h}$) and 
$\nheavy$ heavy fermions $(F_h^+,  F_h^0)$ (with $c_{F_h}$). 
Here $\Delta c_F \equiv c_{F_l} - c_{F_h} > 0$.
$c_{F_l}$ and $c_{F_h}$ are chosen so as to keep the values of $\theta_H$ and $m_H$ unchanged.
In Table.~\ref{tbl:diff_nf4},  the values of $c_{F_l}$, $\Delta c_F$ and the corresponding 
fermion masses are tabulated. The changes in the  couplings of $\nlight$ fermions 
to vector bosons from those in the degenerate case are found to be small.
\begin{table}[tbp]
\caption{%
Parameters in the non-degenerate case of dark fermions, $(\nlight,\nheavy)$.
Bulk mass parameter $c_{F_l}$ and  the  masses $m_{F_h}$ and $m_{F_l}$
of $F_{h}$ and $F_{l}$ are tabulated for various $\Delta c_F \equiv c_{F_l} - c_{F_h}$ (see text)
and $z_L$. Even small $\Delta c_F$ gives rise to  large mass difference.
}\label{tbl:diff_nf4}
\vskip 8pt
\begin{center}
\begin{tabular}{cc|ccc|ccc}
\hline\hline
$\Delta c_F$ &&
 $0.04$ &         &        &
 $0.06$ &         &        \\
$(\nlight,\nheavy)$ & $z_L$ & 
$c_{F_l}$           & $m_{F_h}$ & $m_{F_l}$ &
$c_{F_l}$           & $m_{F_h}$ & $m_{F_l}$ \\
       &&
       & [TeV] & [TeV] & 
       & [TeV] & [TeV] \\
\hline
(1,3)&
$10^6$ 
      & $0.404$ & $1.32$ & $1.13$ 
      & $0.418$ & $1.34$ & $1.06$
\\
&$10^5$ 
       & $0.362$ & $2.09$ & $1.86$
       & $0.377$ & $2.12$ & $1.77$ 
\\
&$3\times10^4$
      & $0.329$ & $2.72$ & $2.46$ 
      & $0.344$ & $2.76$ & $2.36$
\\
&$10^4$ 
       & $0.286$ & $3.54$ & $3.24$ 
       & $0.240$ & $3.58$ & $3.14$ \\
\hline
(2,2)
&$10^5$ 
       & $0.352$ & $2.15$ & $1.92$ 
       & $0.361$ & $2.21$ & $1.86$
       \\
&$10^4$ 
      & $0.276$ & $3.61$ & $3.32$ 
      & $0.285$ & $3.69$ & $3.25$
      \\
\hline
$(3,1)$ 
&$10^5$
            & $0.342$ & $2.21$ & $1.98$ 
            & $0.346$ & $2.30$ & $1.95$ \\
&$10^4$
            & $0.266$ & $3.68$ & $3.39$ 
            & $0.270$ & $3.80$ & $3.36$ \\
\hline\hline
\end{tabular}
\end{center}
\end{table}

\begin{figure}[htbp]
\centerline{
\includegraphics[width=3cm]{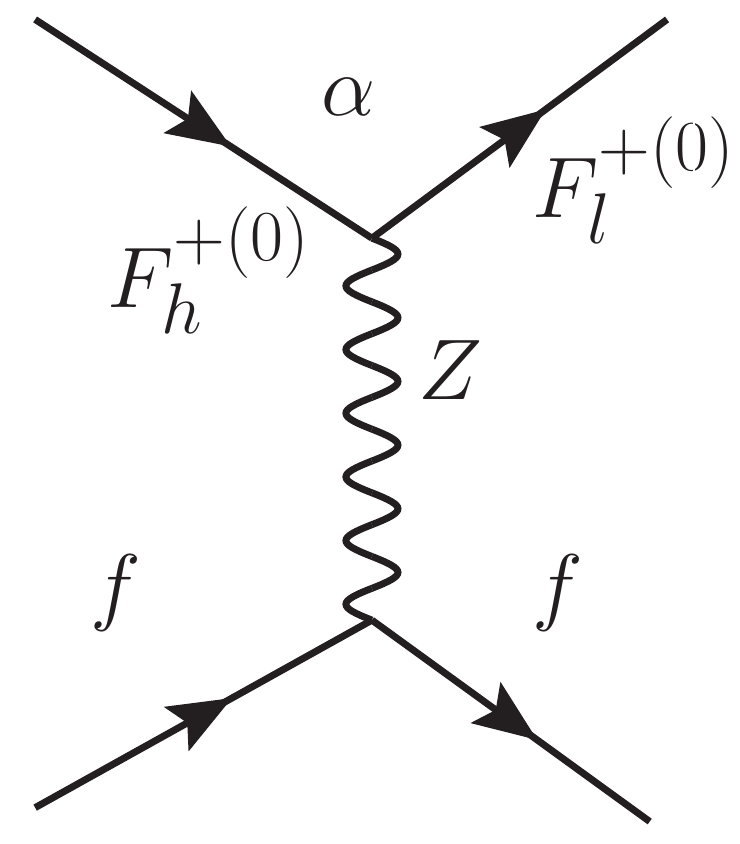}%
\includegraphics[width=3cm]{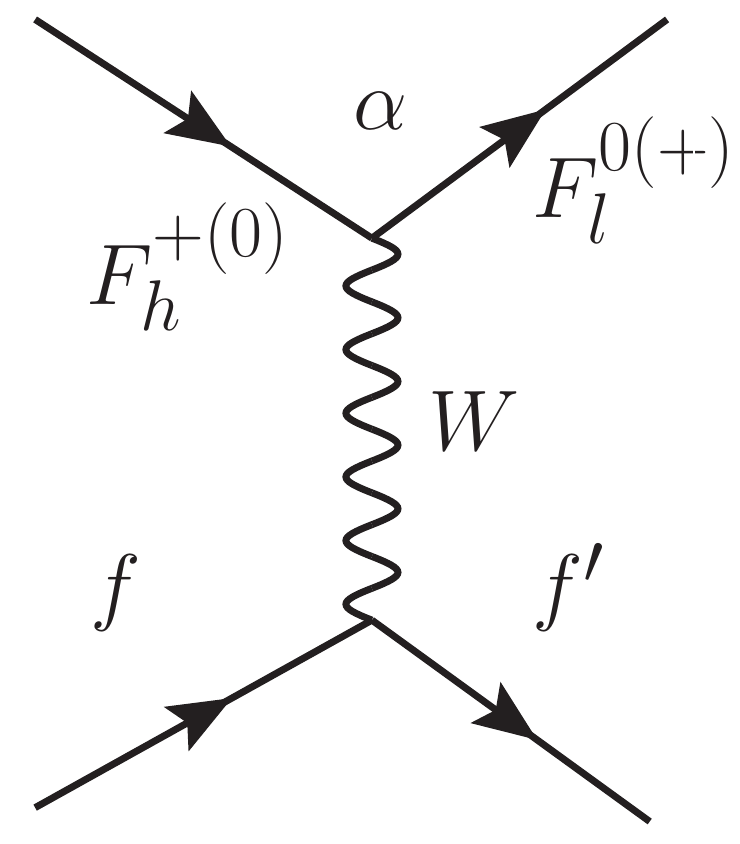}}
\caption{Dominant processes of $F_h \leftrightarrow F_l$ conversion. $f$ and $f'$ are the SM fermions.
$\alpha$ denotes the suppression of the $FFW$, $FFZ$ vertex factor by the mixing.}\label{fig:hl-conversion}
\end{figure}

At the temperature $T \gtrsim m_{F_h} - m_{F_l}$, the heavy-light conversion process depicted in Fig.~\ref{fig:hl-conversion} dominates,  and both $F_h$ and $F_l$ obey the Boltzmann distribution.
When $m_{F_h} - m_{F_l} \gtrsim T_f = {\cal O}(100\,\text{GeV})$,
the number density of $F_h$ becomes much smaller than that of $F_l$. 

In contrast to $F_l$, $F_h$ obey the boundary condition $\eta_{F_h}=-1$ and
its couplings to $W$ and $Z$ are not suppressed, whereas its coupling to $Z_R$ is suppressed. 
Thus the dominant annihilation processes of $F_h$ are s-channel processes of
$F\bar{F}$ annihilation to the SM fermions through $Z^{(1)}$ 
and $\gamma^{(1)}$ [(a-ii) with $V=Z^{(1)}$ and (ac-iv) with $V=\gamma^{(1)}$ in Table.~\ref{tbl:process}]
and  co-annihilation through $W^{(1)}$ [(co-i) with $V=W^{(1)}$ in Table.~\ref{tbl:process}]. 
The time evolutions of the total dark fermion density is given by
\begin{eqnarray}
\frac{dn}{dt} &=& - 3 H n
- 2 \nlight \langle \sigma_{l0} v \rangle (n_{l0}^2 - n_{l0,\rm eq}^2)
- 2 \nlight \langle \sigma_{l+} v \rangle (n_{l+}^2 - n_{l+,\rm eq}^2)
\nonumber\\&&
\phantom{-3 H n} 
- 2 \nheavy \langle \sigma_{h0} v \rangle (n_{h0}^2 - n_{h0,\rm eq}^2)
- 2 \nheavy \langle \sigma_{h+} v \rangle (n_{h+}^2 - n_{h+,\rm eq}^2)
\nonumber\\&&
\phantom{-3 H n}
- 4 \nheavy \langle \sigma_{hc} v \langle (n_{h0}n_{h+} - n_{h0}^{\rm eq} n_{h+}^{\rm eq})
\nonumber\\
\noalign{\kern 10pt}
&=& - 3 H n
 - 2 \nlight 
 \left( \frac{n_{l}^{\rm eq}}{n_{\rm eq}}\right)^2
 [\langle \sigma_{l0}v \rangle + \langle \sigma_{l+}v\rangle]
 (n^2 - n_{\rm eq}^2)
\nonumber\\&&
\phantom{-3 H n}
 - 2 \nheavy 
 \left( \frac{n_{h}^{\rm eq}}{n_{\rm eq}}\right)^2
 [\langle \sigma_{h0}v \rangle + \langle \sigma_{h+}v \rangle 
 + 2 \langle \sigma_{hc}v \rangle]
 (n^2 - n_{\rm eq}^2)
\nonumber\\
\noalign{\kern 10pt}
&\equiv& 
- 3 H n 
- \langle \sigma_{\rm eff}^{\rm ND} v \rangle (n^2 - n_{\rm eq}^2),
\label{eq:mixedBoltzmann0}
\end{eqnarray}
where $n_{w0}$ and $n_{w+}$ ($w=h,l$) are the number densities of 
$F_{w,i}^{0}$ $F_{w,i}^+$ ($i=1,\cdots,\nlight$ for $w=l$, and $1,\cdots,\nheavy$ for $w=h$), respectively.
$\sigma_{w0}$, $\sigma_{w+}$ and $\sigma_{hc}$ are the cross section of 
$F_{w,i}^0 \bar{F}^0_{w,i}$, $F_{w,i}^+ F_{w,i}^-$ annihilations
and $F_{h}^+ F_{h}^0$ co-annihilation, respectively.
We also have used 
\begin{eqnarray}
\frac{n_{w0/+}}{n} \simeq \frac{n_{w0/+,\rm eq}}{n_{\rm eq}},
\quad n_{w0}^{({\rm eq})} \simeq n_{w+}^{({\rm eq})} \equiv n_{w}^{(\rm eq)},
\quad w= h,l.
\end{eqnarray}
The number densities in the thermal equilibrium are given by
\begin{eqnarray}
\frac{n_l^{\rm eq}}{n_{\rm eq}} &=& \frac{1}{4 \nlight + 4\nheavy (1+\eta)^{3/2}\exp(-\eta/x)},
\nonumber\\
\frac{n_h^{\rm eq}}{n_{\rm eq}} &=& \frac{(1+\eta)^{3/2}\exp(-\eta/x)}{4 \nlight + 4\nheavy (1+\eta)^{3/2}\exp(-\eta/x)},
\quad
\eta \equiv \frac{m_{F_h} - m_{F_l}}{m_{F_l}},
\label{eq:density-ratio-nondegenerate}
\end{eqnarray}
and $g_{\rm eff}$ in \eqref{eq:density-equilibrium} will be replaced with
\begin{eqnarray}
g_{\rm eff}^{\rm ND} = 2 \cdot 4 \nlight + 2 \cdot 4\nheavy (1+\eta)^{3/2} \exp(-\eta/x).
\end{eqnarray} 
When $\eta/x \gg 1$, the Boltzmann equation \eqref{eq:mixedBoltzmann0} with \eqref{eq:density-ratio-nondegenerate} can be approximated by
\begin{eqnarray}
\frac{dn}{dt} &=& -3 H n 
\langle \sigma_{\rm eff}^{\rm ND} v|_{\eta\to\infty} \rangle
 (n^2 - n_{\rm eq}^2),
\nonumber\\
\sigma_{\rm eff}^{\rm ND}v |_{\eta\to\infty}
&=& \frac{1}{8\nlight} \left[\sigma_{l0} v  + \sigma_{l+} v \right],
\label{eq:mixedBoltzmann1}
\end{eqnarray}
and $g_{\rm eff}^{\rm ND}|_{\eta\to\infty} = 2\cdot4\nlight$.
With this approximation, one can calculate the relic density of the dark fermion by following 
the procedure described in Sec.~\ref{sec:annihilation}.
Since the effective cross section, and therefore $J_f$ in (\ref{Jf_definition}), 
 is enhanced by a factor $\sigma_{\rm eff}^{\rm ND}v|_{\eta\to\infty}/\sigma_{\rm eff}v  \simeq n_F/\nlight$, 
which results in the reduction of the relic density by a factor $\nlight/n_F$ as seen from \eqref{eq:relic_density}.
If $\eta$ is not so large, the approximation \eqref{eq:mixedBoltzmann1} is not valid any more.
In particular, for $\eta\sim0$ the Bolzmann equation \eqref{eq:mixedBoltzmann0} become almost identical 
to \eqref{eq:Boltzmann}, and the relic density will be increased up to that  in the degenerate case.
Effects of small $\eta$ on $\Omega_{\rm DM} h^2$ \eqref{eq:relic_density} mainly appear in the change of the value 
of $J_f$ (or $\la \sigma_\eff v \ra$).  Numerically we find that $J_f$ determined from \eqref{eq:mixedBoltzmann1} 
well approximates $J_f$ determined from \eqref{eq:mixedBoltzmann0} with \eqref{eq:density-ratio-nondegenerate} 
at ${\cal O}(5\%)$ accuracy when $\eta \gtrsim 0.10$ 
for $x = x_f \simeq 1/30$ and $\sigma_{l0/+} \sim \sigma_{h0/+}$.

We note that in the cross section \eqref{eq:annihilation}, the total decay width of 
$Z_R^{(1)}$ \eqref{eq:ZR-decayrate} can be modified
so that it consists $\nlight$ $F_l$ and $\nheavy$ $F_h$ partial decay widths, 
as $Z_{R} F_{l}\bar{F}_{l}$  and $Z_{R} F_{h}\bar{F}_{h}$ couplings are not the same.
The total decay width of $\gamma^{(1)}$ does not change so much, since 
$\gamma^{(1)}F\bar{F}$ copings are invariant under the exchange $SU(2)_L \leftrightarrow SU(2)_R$.
Numerically we find the change of the cross section \eqref{eq:annihilation} induced from the change in 
decay widths amounts only to a few percents.

\begin{figure}[tbp]
\centerline{\includegraphics[width=8cm]{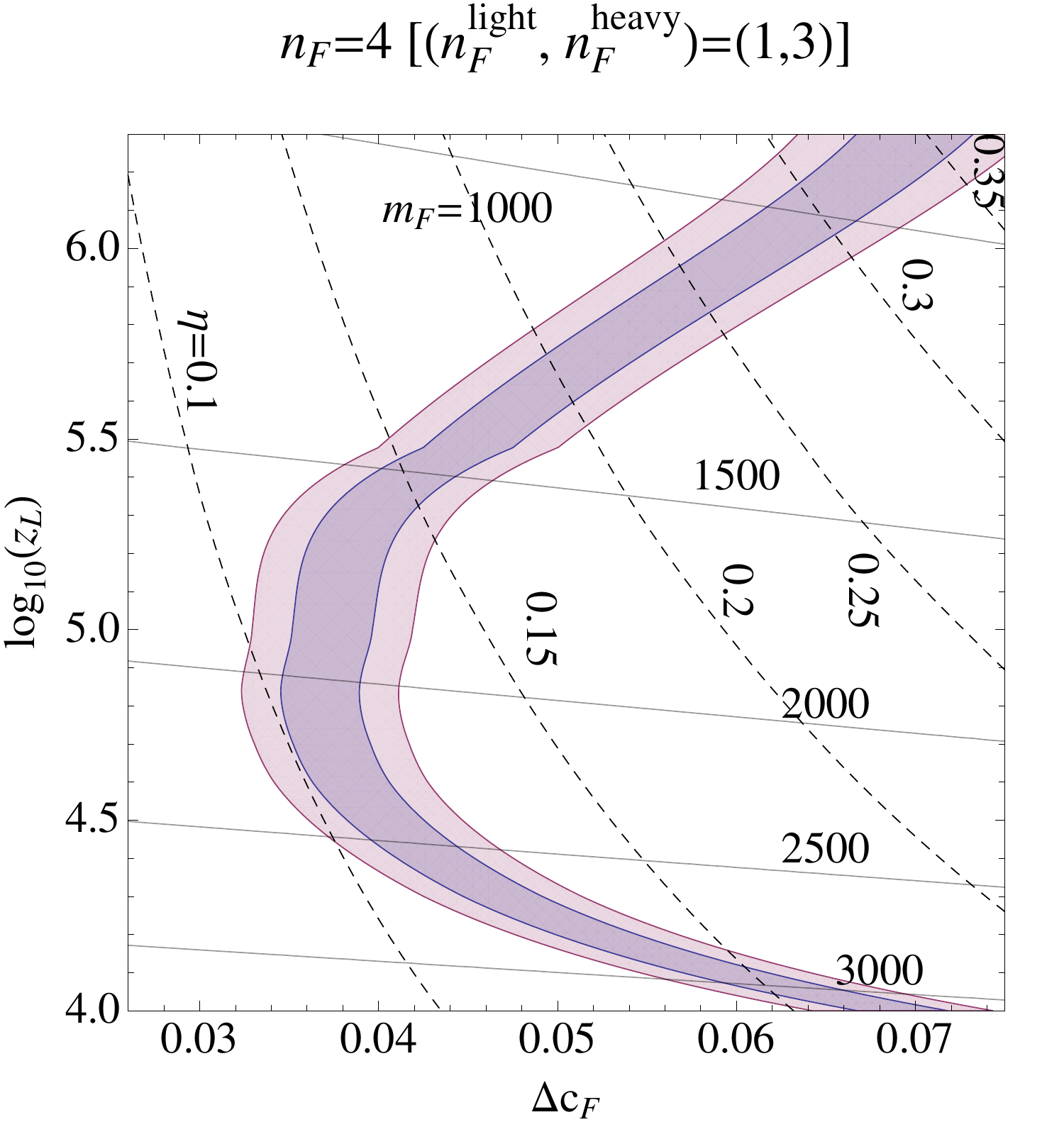}}
\caption{Parameter region $(\Delta c_F, z_L)$ allowed by the limits of relic density. Inner and outer colored regions are allowed with the 68\% CL limit and twice of the 68\% CL limit $\Omega_{\rm DM}h^2 \subset [0.1186 \pm2\times0.0031]$, respectively. Mass of the dark fermion $m_{F_l}$ and a mass ratio
 $\eta\equiv (m_{F_h}-m_{F_l})/m_{F_l}$ are also indicated
 as solid and dashed lines, respectively.}\label{fig:allowed}
\end{figure}

From Table.~\ref{tbl:diff_nf4}, we see that for $\Delta c_F \gtrsim 0.04$ the condition $\eta\gtrsim 0.1$ is satisfied and the cross section formula \eqref{eq:mixedBoltzmann1} is valid. 
\begin{figure}[tbp]
\begin{center}
\includegraphics[width=9cm]{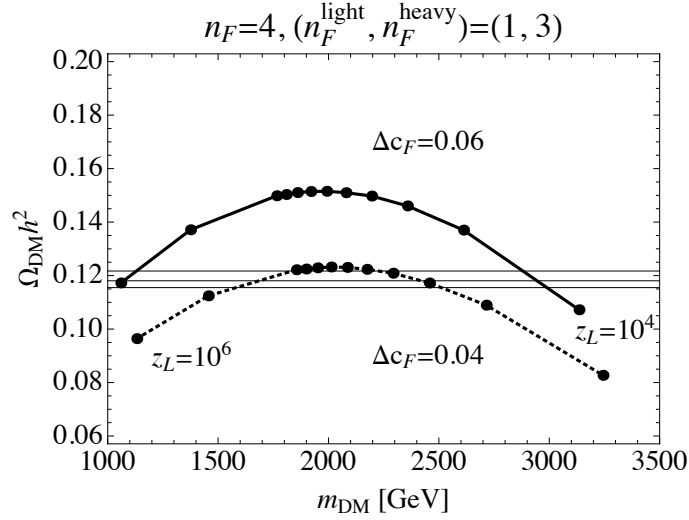}
\end{center}
\vskip -.5cm
\caption{
Relic density of the dark fermion versus $m_{\rm DM} = m_{F_l}$ for $n_F=4$ ($\nlight=1,\nheavy=3$).
Thick-solid and thick-dotted lines are $\Delta c_F \equiv c_{F_l} - c_{F_h} = 0.06$ and $0.04$, respectively.
Data points are, from right to left, $z_L=10^4$ to $10^5$ with an interval $10^4$,
$3\times10^5$ and $10^6$. 
Horizontal lines around $\Omega_{\rm DM} h^2 \sim 0.12$ show the observed 68\% confidence level (CL) limit of the relic density of the cold dark matter. 
}\label{fig:density-mix}
\end{figure}
In Fig.~\ref{fig:density-mix} we have plotted the relic density of the dark fermion determined from 
the Boltzmann equation \eqref{eq:mixedBoltzmann1} for $\Delta c_F=0.04$ and $0.06$ 
in the case of $n_F=4$ with $(\nlight,\nheavy) = (1,3)$.
For $\Delta c_F < 0.04$, the approximated formula \eqref{eq:mixedBoltzmann1} is no more valid,
and the relic-density can be much larger than those for $\Delta c_F \gtrsim 0.04$.
By inter-/extra-polating the $\Omega_{\rm DM} h^2$ with respect to $\Delta c_F$ and $z_L$,
we plot the parameter region $(\Delta c_F, z_L)$ allowed by 
the experimental limit on the current relic density in Fig.~\ref{fig:allowed}.
It is seen  that the observed current relic density is obtained when $10^4 \lesssim z_L\lesssim 10^6$ ($0.07 \lesssim \theta_H \lesssim 0.17$) in the range $0.04 \lesssim \Delta c_F \lesssim 0.07$.
The mass of the dark fermion $m_{DM}$ varies within the range of $[1000,\,3100]\, \text{GeV}$.
For $n_F=5,6$ and $n_F=4$ with $(\nlight,\nheavy)=(2,2)$, $(3,1)$,
we find no parameter region which explains the current DM density.

In the numerical study we have used an approximation explained in Sec.~\ref{sec:annihilation}. In the case where the Breit-Wigner resonance enhance the DM relic density, a more rigorous treatment may be required.\cite{Griest} In the case under consideration, the effect of the enhancement is found to be mild.
Quantitatively, in the notation of Ref.~\cite{Griest}
we obtain $\epsilon = (\Gamma_V/m_V)^2 = {\cal O}(0.005)$ 
($V=Z^{(1)},Z_R^{(1)},\gamma^{(1)}$) and $\sqrt{u} = 2m_F/m_V \lesssim 0.8$.
In this parameter region the approximation can be justified.\cite{Griest}

Before closing this section, we make a few comments.
First we comment on the effect of dark fermions on the electroweak precision parameters \cite{Peskin-Takeuchi1}, in particular on the $S$-parameter.
Since the dark fermions have vector-like couplings to the $Z$ boson, the contribution to the $S$ parameter from an $SU(2)$ doublet $\{F^+,F^0 \}$ is estimated to be
\begin{eqnarray}
\Delta (\alpha_{\rm EM} S) &\simeq& 4 s_w^2 c_w^2 \Pi'(0) \sum_{F=F^+,F^0} \left(
(g^Z_{F V})^2 - \frac{c_w^2 - s_w^2}{c_w s_w} g^Z_{F V} Q_{F} e - Q_{F}^2 e^2
\right),
\nonumber\\&&
c_w \equiv \cos\theta_W,\quad s_w \equiv \sin\theta_W, \label{eq:S-param}
\end{eqnarray} 
where $g^Z_{F V}\equiv (g^Z_{F L} + g^Z_{F R})/2$  and $Q_F$ is the vector coupling to $Z$ and the electric charge of $F$, respectively.
$\Pi(p^2)$ is the vacuum polarization function which is induced by the one-loop fermion with vector-type coupling.
Numerically we find that in both cases of $F_l$ ($\eta_{F_l}=+1$) and $F_h$ ($\eta_{F_h}=-1$) the sum of the right-hand side in \eqref{eq:S-param} vanishes accurately.
Hence there are no sizable corrections of the $S$ parameter from  dark fermions.

Secondly as an stabilization mechanism of the branes one can introduce some dynamical model 
a la Goldberger-Wise\cite{Goldberger}. In such a case the phase transition of the radion field may alter 
the thermal history of the universe drastically\cite{Radion-T}. 
Here we have supposed that the critical temperature of the radion phase transition, $T_\phi$, is much higher than the freeze-out temperature of the dark fermions, e.g.,
$T_\phi \gg T_f \sim 100\,\text{GeV}$.

\section{Direct detection}
In this section, we analyse the elastic scattering of  the dark fermion ($F^0$) off a nucleus 
\cite{JKG, EFO, Goodman-Witten}
and examine the constraint coming from  direct detection experiments.\cite{XENON, LUX} 
The dominant process of the $F^0$-nucleus scattering turns out the $Z$ boson exchange, 
though the $Z$-$F^0$ coupling is very small.
%$F^0$ is almost $SU(2)_R$ component, therefore $Z$-$F^0$ coupling is very small.
The $Z_R^{(n)}$-$F^0$ coupling is larger, but  $Z_R^{(n)}$  is heavy.
Subdominant  are the processes of $Z_R^{(1)}$ and Higgs exchange.   
Contributions from other processes are negligible.
\begin{figure}[htbp]
\begin{tabular}{ccc}
 \begin{minipage}{0.33\hsize}
  \begin{center}
   \includegraphics[width=50mm]{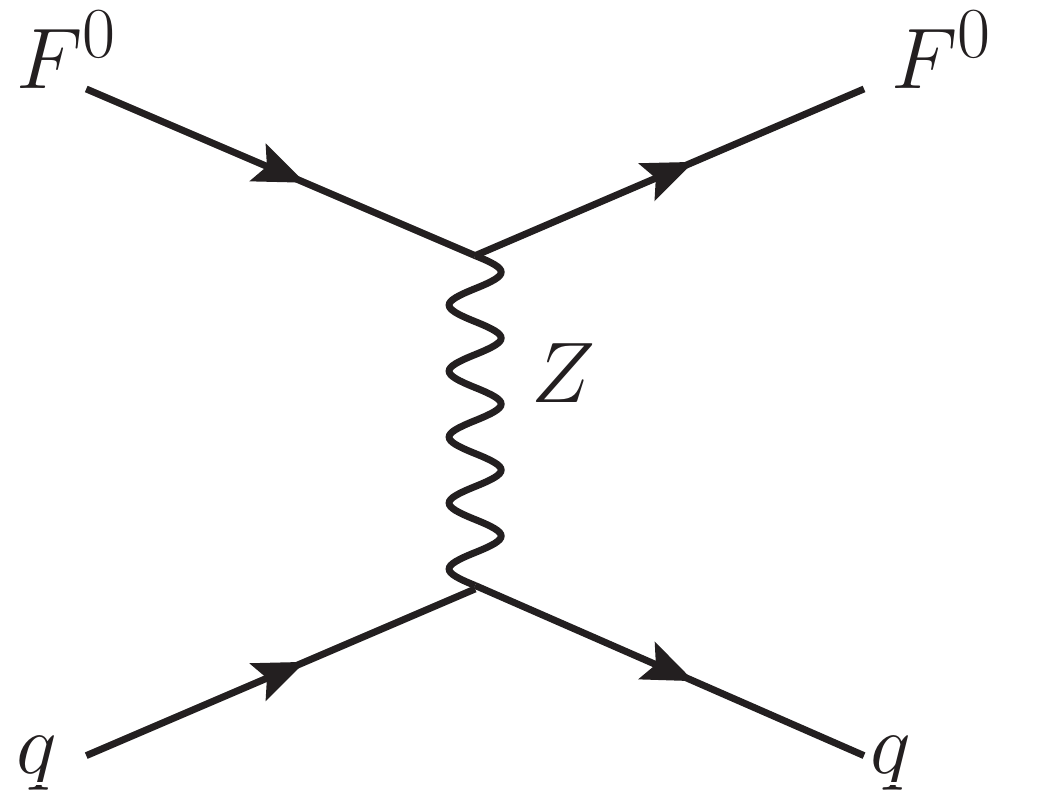}
  \end{center}
 \end{minipage}
 \begin{minipage}{0.33\hsize}
 \begin{center}
  \includegraphics[width=50mm]{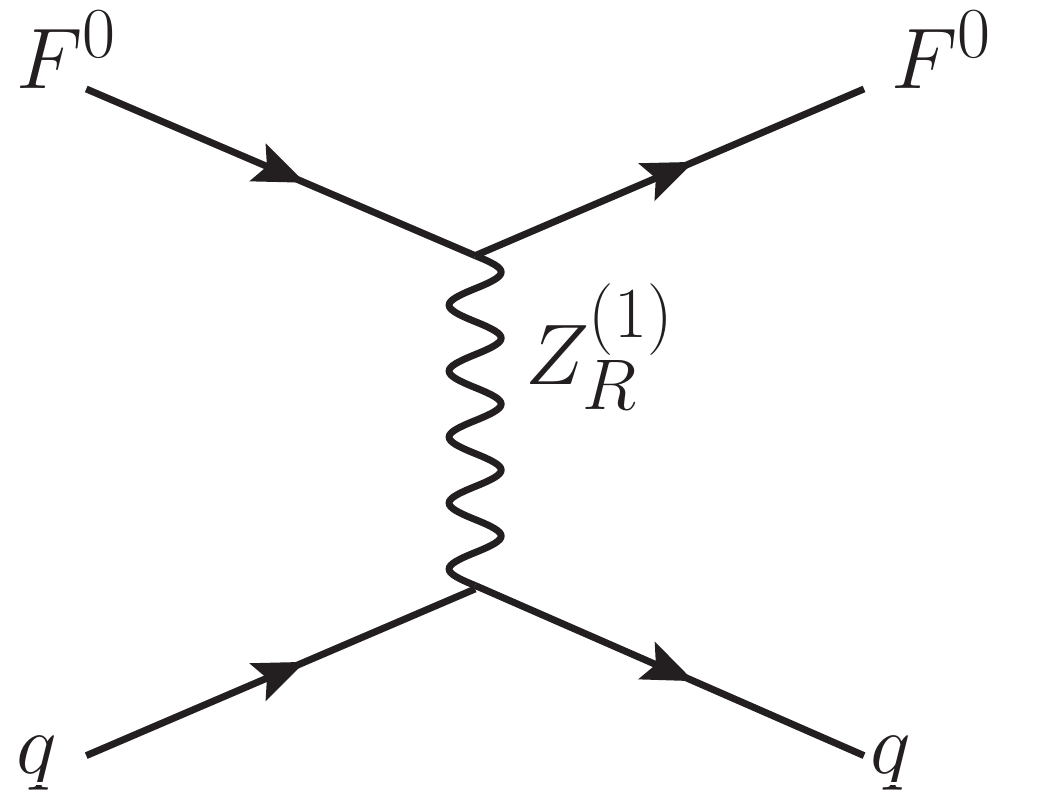}
 \end{center}
 \end{minipage}
 \begin{minipage}{0.33\hsize}
 \begin{center}
  \includegraphics[width=50mm]{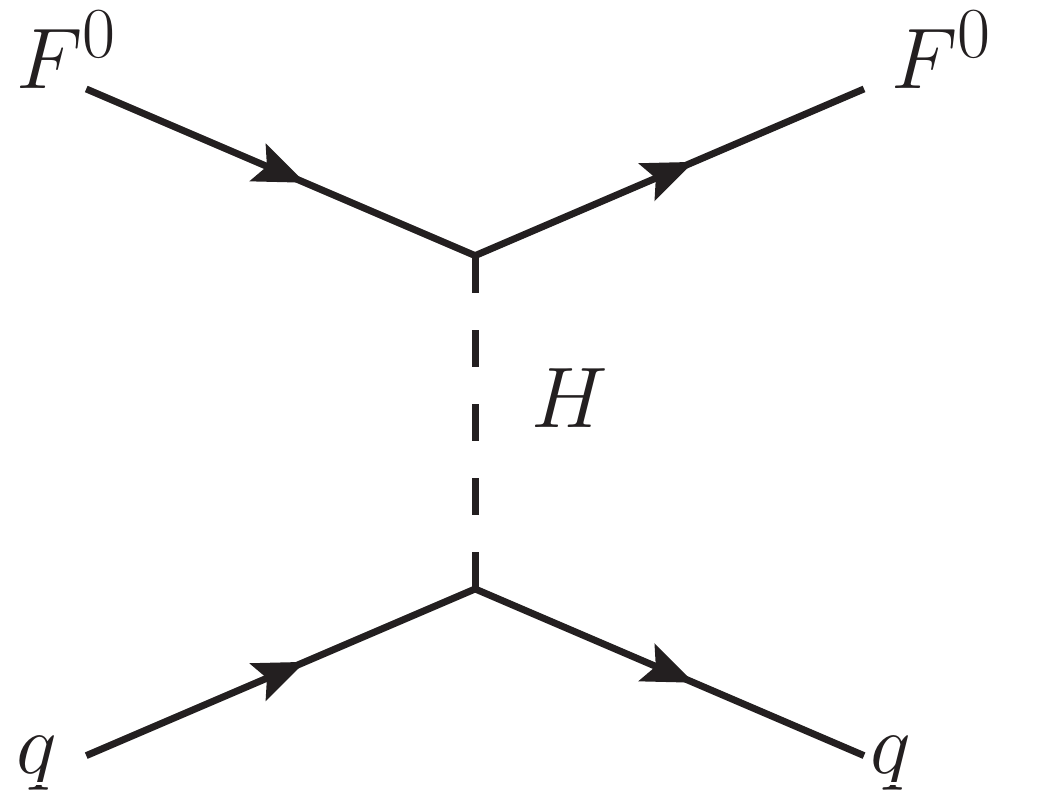}
 \end{center}
 \end{minipage}
\end{tabular}
\caption{Dominant and subdominant processes of the $F^0$-nucleus scattering}
\end{figure}

In the scattering of $F^0$ on nuclei with large mass number $A$, 
scalar and vector interactions  dominate
%add coherently, whereas pseudoscalar and axial vector  interactions are negligible 
for the spin-independent cross section.
Therefore the effective Lagrangian at low energies is  given by
\begin{equation}
\mathcal{L}_\text{int} \simeq \sum_q  %\sum_{q=u, d, c, s, t, b}
\Bigg\{-\bigg(  \frac{g_w^2 v_q}{m_Z^2\cos\theta_W^2}  V_F
+\frac{g_w^2 v^{(Z_R^{(1)})}_q}{m_{Z_R^{(1)}}^2}  V^{(Z_R^{(1)})}_F\bigg)
\bar{q}\gamma^0 q \; \bar{F^0}\gamma_0 F^0+\frac{y_qY_F}{m_H^2}\bar{q}q\bar{F^0}F^0\Bigg\} .
\label{eq:DM4couplings}
\end{equation}

To evaluate the scattering amplitude by the Higgs exchange,
we need estimate the nucleon matrix element
\begin{equation}
\langle N\vert m_q \bar{q}q\vert N\rangle=m_N f^{(N)}_{Tq},
\end{equation}
where $N=p, n$.
For  heavy quarks ($Q=c$, $b$, $t$) one has
\begin{equation}
f^{(N)}_{TQ}=\frac{2}{27} \bigg(1-\sum_{q=u, d, s}f^{(N)}_{Tq} \bigg).
\end{equation}
In the GHU model, quark couplings satisfy
$v_q\vert_{\text{GHU}}\simeq v_q\vert_{\text{SM}}$ and
\begin{equation}
y_q\vert_{\text{GHU}}\simeq y_q\vert_{\text{SM}}\cos\theta_H 
=\frac{g_w}{2m_W}m_q \cos\theta_H ,
\end{equation} 
to good accuracy.\cite{HK2008}
Therefore, by dropping the small momentum dependence of  the form factor, 
the spin-independent cross section of the $F^0$-nucleus elastic scattering becomes
\begin{align}
\sigma_0
&\equiv \int_0^{4M_r^2v^2}\frac{d\sigma}{d|\bm{q}|^2}\bigg|_{|\bm{q}|=0}d|\bm{q}|^2 \nonumber\\
&=\frac{M_r^2}{\pi}\Big\{Z\left(b_p+f_p\right)+(A-Z)\left(b_n+f_n\right)\Big\}^2 ,
\end{align}
where $M_r$ is the $F^0$-nucleus  reduced mass and $Z$ ($A$) is the atomic (mass) number of 
the nucleus.
$|\bm{q}|$ is the momentum transfer and 
\beqn
&&\hskip -1.cm
b_p=2b_u+b_d ~, ~~ b_n=b_u+2b_d ~, \cr
\noalign{\kern 10pt}
&&\hskip -1.cm
b_q=-4\sqrt{2} G_F
\bigg( v_qV_F+\frac{m_W^2}{m_{Z_R^{(1)}}^2}v^{(Z_R^{(1)})}_qV^{(Z_R^{(1)})}_F \bigg) ~, \cr
\noalign{\kern 10pt}
&&\hskip -1.cm
f_N=\frac{Y_F}{m_H^2}\sum_q %{q=u, d, c, s, t, b}
\langle N\vert y_q \bar{q}q\vert N\rangle
=\frac{Y_F}{m_H^2}\frac{g_w m_N}{2m_W}\cos\theta_H
\bigg(\frac{2}{9}+\frac{7}{9}\sum_{q=u, d, s}f^{(N)}_{Tq} \bigg) ~.
\eeqn
The spin-independent cross section of the $F^0$-nucleon elastic scattering $\sigma_N$ 
can be written as 
\begin{align}
\sigma_N\equiv \frac{1}{A^2} \frac{m_r^2}{M_r^2}\sigma_0 ~,
\end{align}
where $m_r$ is the $F^0$-nucleon reduced mass.

\begin{figure}[htb]
\centering
\includegraphics[bb=0 0 504 360, width=12cm]{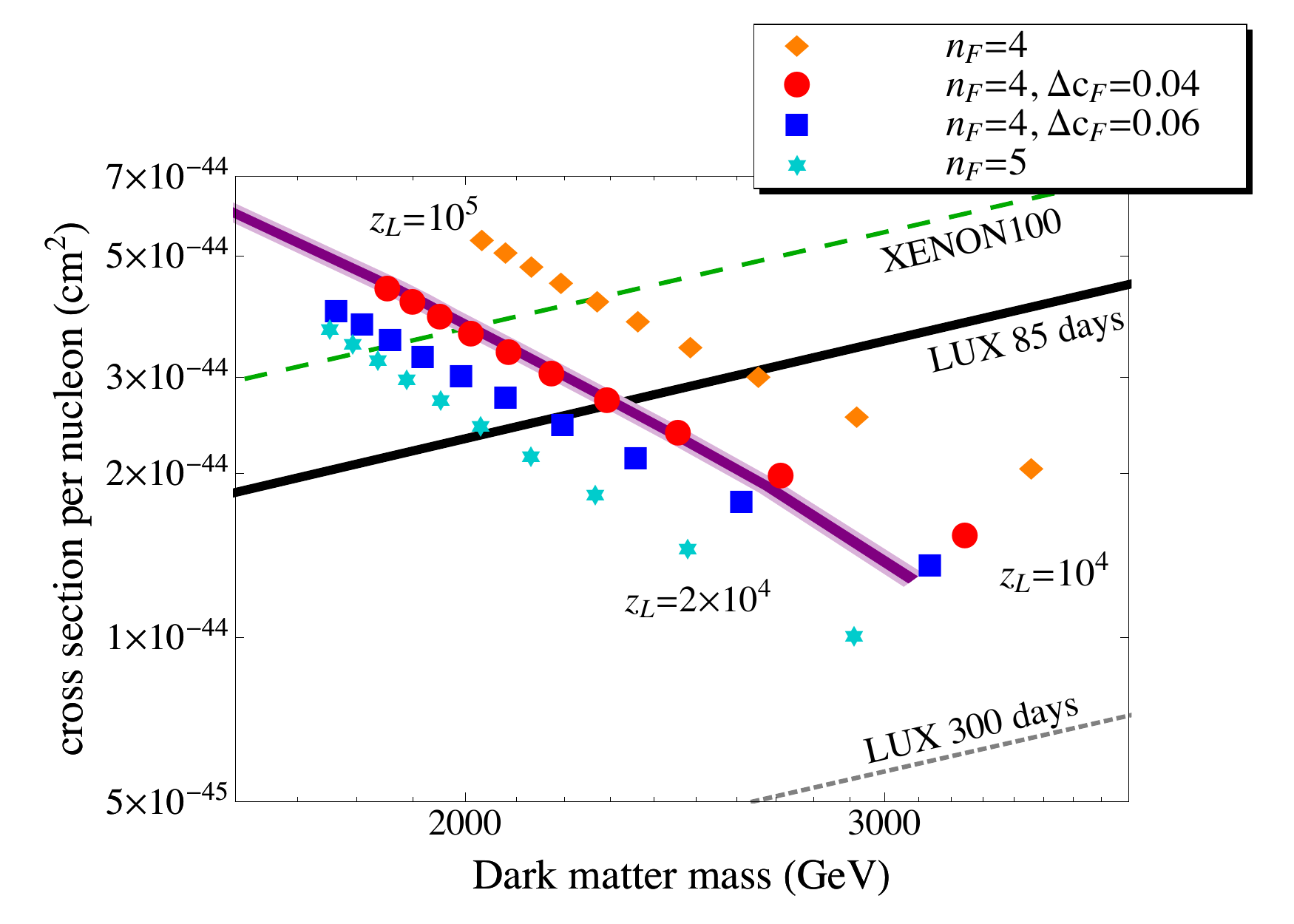}
\caption{The  spin-independent cross section of the $F^0$-nucleon elastic scattering 
for $10^4 \le z_L\le 10^5$. 
The orange diamonds and light blue stars represent the $n_F=4$ and $n_F=5$ cases of 
degenerate dark fermions with a step of $10^4$ in $z_L$, respectively. 
Red circles and blue squares represent the cases of non-degenerate dark fermions 
$(\nlight , \nheavy ) = (1,3)$ with $\Delta c_F = 0.04$ and 0.06, respectively.
The black solid line and green dashed line are the 90\% confidence limits set by 
the 85.3 live-days result of the LUX experiment\cite{LUX} and 
the 225 live-days result of the XENON100 experiment\cite{XENON}, respectively. 
For reference we have added the expected limit by the 300 live-days result of the LUX
experiment.  The XENON 1T experiment is expected to give a limit one order of magnitude
smaller than that of the LUX 300 live-days experiment in the cross-section.
The purple and light purple bands represent the regions allowed by the limit of the relic density 
of DM at the 68 \% CL depicted in Fig.~~\ref{fig:allowed} and by twice of that.
The model with dark fermions of $2.3\, {\rm TeV} < m_{F_l} <3.1\,$TeV ($4 \times 10^4 > z_L >10^4$)
gives a consistent scenario.  
}
\label{figure-cross-section}
\end{figure}

The $F^0$-nucleon cross sections $\sigma_N$ are shown in Table \ref{cross-section-table} 
and Figure \ref{figure-cross-section}.
In the numerical evaluation we have employed the values given by \cite{EFO}
%To calculate the cross sections numerically, the following values are used for the nucleon matrix elements 
\begin{align}
f^{(p)}_{Tu}&=0.020 ~,~~ f^{(p)}_{Td}=0.026 ~,~~ f^{(p)}_{Ts}=0.118 ~,\nonumber\\
f^{(n)}_{Tu}&=0.014 ~,~~ f^{(n)}_{Td}=0.036 ~,~~ f^{(n)}_{Ts}=0.118 ~.
\end{align}
Recent lattice simulations show smaller values for $f^{(N)}_{Ts}$ \cite{nucleon-matrix-element},
which yields slightly smaller cross sections than those described below.
%In the case that $f^{(N)}_{Ts}$ are small, 
%the cross sections are slightly smaller than the following results.

\begin{table}
\caption{$F^0$ mass $m_F$ and the spin-independent cross section $\sigma_N$ 
of the $F^0$-nucleon  scattering for $n_F=4, 5, 6$  degenerate dark fermions.}
\centering
  \begin{tabular}{|c|c|c|c|}
\noalign{\kern 5pt}
%  \multicolumn{4}{c}{$\:$}\\ 
  \multicolumn{4}{c}{$n_F=4$}\\ 
    \hline$z_L$ &
     $\;\quad \theta_H \quad\;$ & $\;m_F$ (TeV) &
      \quad\; $\sigma_N\;(\text{cm}^2$) \quad\; \\ 
    \hline$10^5$ & 0.115 & 2.03 & 5.33$\times10^{-44}$\\
    \hline$5\times10^4$ & 0.101 & 2.36 & 3.78$\times10^{-44}$\\
    \hline$3\times10^4$ & 0.092 & 2.66 & 2.99$\times10^{-44}$\\
    \hline$2\times10^4$ & 0.085 & 2.92 & 2.53$\times10^{-44}$\\
    \hline$10^4$ & 0.074 & 3.46 & 2.03$\times10^{-44}$\\ \hline
\noalign{\kern 5pt}
%  \multicolumn{4}{c}{$\:$}\\
  \multicolumn{4}{c}{$n_F=5$}\\ 
    \hline$\;\;\;\; z_L\;\;\;\;$ & 
    $\;\quad \theta_H \quad\;$ & $\;m_F$ (TeV) &
     $\quad\; \sigma_N\;(\text{cm}^2$) \quad\; \\
    \hline$10^5$ & 0.114 & 1.75 & 3.67$\times10^{-44}$\\
    \hline$10^4$ & 0.073 & 2.91 & 1.01$\times10^{-44}$\\  \hline
\noalign{\kern 5pt}
%   \multicolumn{4}{c}{$\:$}\\
  \multicolumn{4}{c}{$n_F=6$}\\
    \hline $\;\;\;\; z_L\;\;\;\;$ &
     $\;\quad \theta_H \quad\;$ & $\;m_F$ (TeV) &
      $\quad\; \sigma_N\;(\text{cm}^2$) \quad\;\\
    \hline$10^5$ & 0.113 & 1.57 & 2.96$\times10^{-44}$\\
    \hline$10^4$ & 0.072 & 2.56 & 0.72$\times10^{-44}$\\  \hline
   \end{tabular}\label{cross-section-table}
\end{table}
%%%%%%%%%%%%%%%%%%%%%%%%%%%%%%%%%%%%%%%%%%
\begin{table}
\caption{$m_{F_l}$, $m_{Z_R^{(1)}}$, the couplings of $F^0_l$ and the spin-independent cross section $\sigma_N$ of the $F^0_l$-nucleon scattering for $n_F=4$ and $(\nlight , \nheavy ) = (1,3)$. $V_F$, $v_q^{(Z_R^{(1)})}$ ($q=u,d$), $Y_F$ are defined in Eq.~\eqref{eq:DM4couplings}.}
\centering
  \begin{tabular}{|c|c|c|c|c|c|c|c|c|c|}
\noalign{\kern 5pt}
  \multicolumn{10}{c}{$\Delta c_F=0.04$}\\ 
    \hline$z_L$ & $ \theta_H$ & $m_{F_l}$& $m_{Z_R^{(1)}}$& 
     $V_F$ & $v_u^{(Z_R^{(1)})} $ & $v_d^{(Z_R^{(1)})} $ & $V_F^{(Z_R^{(1)})}$ & $Y_F$ & 
     $\sigma_N$ ($\text{cm}^2$)\\
     & & (TeV) & (TeV) &&&&&&\\ 
    \hline$4\times10^4$ & 0.097 & 2.29 & 6.47 & -0.00108 & 0.474 & -0.237 & 1.11 & -0.0299 & 2.69$\times10^{-44}$\\
    \hline$3\times10^4$ & 0.092 & 2.46 & 6.74 & -0.00100 & 0.469 & -0.234 & 1.11 & -0.0293 & 2.35$\times10^{-44}$\\
    \hline$2\times10^4$ & 0.085 & 2.72 &7.15 & -0.00092 & 0.461 & -0.231 & 1.10 & -0.0286 & 1.96$\times10^{-44}$\\
    		\hline$10^4$ & 0.074 & 3.24 & 7.92 & -0.00081 & 0.450 & -0.225 & 1.08 & -0.0280 & 1.53$\times10^{-44}$\\ \hline
		 \noalign{\kern 10pt}
  \multicolumn{10}{c}{$\Delta c_F=0.06$}\\ 
    \hline$z_L$ & $\theta_H$ & $m_{F_l}$& $m_{Z_R^{(1)}}$& 
     $V_F$ & $v_u^{(Z_R^{(1)})}$ & $v_d^{(Z_R^{(1)})}$ & $V_F^{(Z_R^{(1)})}$ & $Y_F$ &
     $\sigma_N$ ($\text{cm}^2$)\\
     & & (TeV) & (TeV) &&&&&&\\
    	\hline$2\times10^4$ & 0.085 & 2.61 & 7.15 & -0.00086 & 0.461 & -0.231 & 1.09 & -0.0266 & 1.76$\times10^{-44}$\\ 
    	\hline$10^4$ & 0.074 & 3.13 & 7.92 & -0.00075 & 0.450 & -0.225 & 1.07 & -0.0261 & 1.35$\times10^{-44}$\\ \hline
   \end{tabular}
\end{table}

In the previous section we have seen that when all $n_F$  dark fermions are degenerate, 
there are no parameter regions which reproduce the observed value of  the relic DM density.
It was shown that the observed  DM density can be obtained when there are $\nlight$ light 
dark fermions and  $\nheavy$ heavy dark fermions of opposite   $\eta_F$ in the boundary  conditions.  
In particular, for the parameter set of $(\nlight , \nheavy ) = (1,3)$, 
the region $0.04 \siml \Delta c_F \siml 0.07$, 
$z_L\siml  10^6$  successfully explains the relic abundance as shown in Fig.~\ref{fig:allowed}. 
The allowed band region in Fig.~\ref{fig:allowed} is mapped in Fig.~\ref{figure-cross-section}
for the spin-independent cross section for the $F^0$-nucleon elastic scattering.
The purple and light purple bands there represent the regions allowed by the limit of the relic abundance 
of DM at the 68 \% CL and by twice of that, respectively. It is seen that the band region from 
$z_L =  10^4$ to $4 \times10^4$ is allowed by the direct detection experiments of LUX \cite{LUX}
 and XENON100 \cite{XENON}.  In the allowed region the dark fermion mass ranges
 from 3.1$\,$TeV to 2.3$\,$TeV, whereas the AB phase $\theta_H$ ranges from 0.074
 to 0.097.   The mass of $Z'$ bosons (the lowest $Z_R$ boson and the first KK modes $Z^{(1)}$ 
 and $\gamma^{(1)}$) ranges from 8$\,$TeV to 6.5$\,$TeV.
For reference we have added, in Fig.~\ref{figure-cross-section}, 
 the expected limit by the 300 live-days result of the LUX
experiment.  The XENON 1T experiment is expected to give a limit one order of magnitude
smaller than that of the LUX 300 live-days experiment in the cross-section.

We remark that the $n_F=3$ case  predicts too small relic densities  as shown 
in Fig.~\ref{fig:density-mix}.
It implies that the dark fermions in the GHU model accounts for only a fraction of 
the dark matter of the universe, and the model is not excluded by the direct-detection
experiments.

\section{Conclusion and discussions}
In the present paper we have given a detailed analysis of DM in GHU.
In the $SO(5) \times U(1)$ GHU,  the observed unstable Higg boson 
is realized by introducing $SO(5)$-spinor fermions.   Spinor fermions do not directly 
interact with $SO(5)$-vector fermions which contain the SM quarks and leptons.
Therefore the total spinor-fermion number is conserved and the lightest one can remain 
as  dark matter in the current universe. 
Such fermions are referred to as ``dark fermions''.

In Sec.~\ref{sec:Abundance} we have evaluated the relic density of the dark fermions.
Although  charged and neutral dark fermions are degenerate at tree level,
charged fermions become heavier than neutral ones through loop effects so that the charged 
dark fermions decay into neutral ones much earlier than they cooled down at their freeze-out temperature. 
We found that among various annihilation processes of   dark fermions
dominant ones are those in which a dark fermion and its antiparticle
annihilate into the SM fermions mediated by the lowest KK $Z_R$ boson and the first KK photon.
We also have evaluated the annihilation cross section and obtained the relic densities of 
the dark fermions in the current universe for the various  values of $n_F$ and $z_L$.
The results depend sensitively on the number of dark fermions $n_F$.
When all neutral dark fermions are degenerate, no solution has been found which explains 
the observed value of the relic density of dark matter and is consistent with the limit from
the direct detection experiments.
For $n_F=3$ the relic density becomes much smaller than the bound, because 
 twice the mass of the dark fermion is close to the mass of the $Z_R$ boson and the annihilation is 
 enhanced by the resonance. 
For $n_F=4, 5$ and $6$ the relic densitiy becomes  larger than the bound.

We have considered the case in which $n_F$ dark fermions consist of $\nlight$ lighter fermions 
and $\nheavy$ heavier fermions. They are mixed with each other through the 
bulk  mass terms which can be introduced when lighter and heavier fermions have 
opposite signs of $\eta_F$ in the boundary conditions under reflections at the TeV and Planck branes.
When the mass difference of these fermions are sufficiently large (more than ${\cal O}(100\text{GeV})$), 
heavier ones decay quickly to lighter ones and the effective number of species of the dark fermions 
can be reduced from $n_F$ to $\nlight$.  Accordingly the relic density   reduces to $\nlight/n_F$ 
of that  in the degenerate case.
For $n_F=4$ it is  found that one can obtain the relic density consistent with the experimental bound for 
$10^4 \lesssim z_L \lesssim  10^6$ , $ 0.04 \lesssim \Delta c_F \lesssim 0.07$ 
when $(\nlight,\nheavy)=(1, 3)$.
In the cases of  $(\nlight,\nheavy)=(2,2),(3,1)$ 
and of  $n_F=5$ and $=6$  no solution has been found.
We comment that there are no sizable corrections to the $S$-parameter from the dark-fermion loops.

In Sec.~4, we calculated the scattering cross section of the dark fermions with nucleons.
The dark fermions have very small Higgs-Yukawa couplings and $Z$-boson couplings, 
both of which are suppressed by powers of $\sin\theta_H$.
We evaluated the spin-independent cross sections and compared with the 
experimental bound  obtained in the recent experiments of WIMP direct detection.\cite{XENON,LUX}
Combining with the constraint from the relic density,
we showed that  the region $10^4 \lesssim z_L \lesssim 4\times10^4$ 
for $(\nlight,\nheavy)=(1,3)$ is viable.
The corresponding mass of the dark matter candidate (dark fermions) ranges from
3.1$\,$TeV to 2.3$\,$TeV, whereas the AB phase $\theta_H$ ranges from 0.074 to 0.097.
The mass of $Z'$ bosons ranges from 8$\,$TeV to 6.5$\,$TeV.

The $n_F=4$ model   with one light and three heavy dark fermions with opposite boundary conditions
is consistent with the current direct detection experiments.
Such dark fermions should  be detected 
in the direct-detection experiments in near future.
For $n_F=3$, our model cannot explain the current DM density. In this case the current DM density 
should  be accounted for by dark matter generated by other mechanism such as axion DM \cite{Duffy}
and dynamical dark matter \cite{DynamicalDM}. 
In this case DM in the GHU model may or may not be detected, depending on the property 
of the dominant dark matter components. 

The gauge-Higgs unification scenario is viable and promising.  The $SO(5) \times U(1)$ GHU
predicts new $Z'$ bosons in the 6.5$\,$TeV$\sim$8$\,$TeV region and deviation of the self-couplings
of the Higgs boson from SM, which can be explored and checked at the upgraded LHC and
ILC experiments.  We stress again that the model naturally contains the dark matter 
candidate (dark fermions) in the mass range 2.3$\,$TeV$\sim$3.1$\,$TeV. 
The mass and cross section of the dark fermions are within the reach of the ongoing 
and future experiments, and the allowed parameter region of this model can be explored with 
future collider experiments \cite{LHCsignals}.
Pinning down its mass fixes the value of $\theta_H$, which further yields more predictions of GHU
in collider experiments.

\subsection*{Acknowledgements}
We thank Mitsuru Kakizaki and Minoru Tanaka for many valuable comments.
This work was supported in part  by  JSPS KAKENHI grants, 
No.\ 23104009 (YH\ and YO),  No.\ 21244036 (YH) and  No.\ 2518610 (TS),
and NRF Research Grant 2012R1A2A1A01006053 (HH), 
No.\ 2009-0083526 (YO) of the Republic of Korea.
\begin{appendix}
\section{$SO(5)$ generators and base functions\label{sec:so5generators}}

$SO(5)$ generators in the spinorial representation are defined as
\begin{eqnarray}
&& T_L^a = \frac{1}{2} \begin{pmatrix} \sigma^a & \\ &  \end{pmatrix},
\quad
T_R^a = \frac{1}{2} \begin{pmatrix} & \\ & \sigma^a \end{pmatrix},
\\
&&\hat{T}^a = \frac{1}{2\sqrt{2}} \begin{pmatrix} & i\sigma^a \\ -i\sigma^a & \end{pmatrix},
\quad
\hat{T}^4 = \frac{1}{2\sqrt{2}} \begin{pmatrix} & I \\ I & \end{pmatrix},
\end{eqnarray}
and $\Tr[T^\alpha, T^\beta]  = \delta^{\alpha\beta}$ holds.

Mode functions for KK towers are expressed in terms of Bessel functions.  
For gauge fields we define
\beqn
&&\hskip -1cm 
C(z;\lambda) =
      \frac{\pi}{2}\lambda z z_L F_{1,0}(\lambda z, \lambda z_L) ~, \quad
C'(z;\lambda) =
      \frac{\pi}{2}\lambda^2 z z_L F_{0,0}(\lambda z, \lambda z_L) ~ , \cr
\noalign{\kern 5pt}
&&\hskip -1cm 
S(z;\lambda) =
      -\frac{\pi}{2}\lambda z  F_{1,1}(\lambda z, \lambda z_L) ~ , \quad
S'(z;\lambda) =
      -\frac{\pi}{2}\lambda^2 z F_{0,1}(\lambda z,  \lambda z_L) ~,   \cr
\noalign{\kern 5pt}
&&\hskip -1cm 
\hat S(z;\lambda)  = \frac{C(1; \lambda)}{S(1; \lambda)} \, S(z; \lambda) ~, \cr
\noalign{\kern 5pt}
&&\hskip -1cm 
F_{\alpha,\beta}(u,v) =
    J_\alpha (u) Y_\beta(v)   -Y_\alpha(u) J_\beta(v) ~.
\label{BesselF1}
\eeqn
These functions satisfy
\beqn
&&\hskip -1cm 
C(z_L; \lambda) = z_L ~,~~ C' (z_L; \lambda) =0~, ~~ 
S(z_L; \lambda) = 0 ~,~~ S' (z_L; \lambda) =\lambda ~,  \cr
\noalign{\kern 5pt}
&&\hskip -1cm 
C S' - S C' = \lambda z ~. 
\label{BesselF2}
\eeqn
For fermions with a bulk mass parameter $c$ we define
\beqn
&&\hskip -1cm
\begin{pmatrix} C_L \cr S_L \end{pmatrix}  (z;\lambda, c)
= \pm \frac{\pi}{2} \lambda\sqrt{zz_L}
   F_{c+{1\over 2},c\mp{1\over 2}}  (\lambda z, \lambda z_L) ~, \cr
\noalign{\kern 10pt}
&&\hskip -1cm  
\begin{pmatrix} C_R \cr S_R \end{pmatrix}  (z;\lambda, c)
= \mp \frac{\pi}{2} \lambda\sqrt{zz_L}
F_{c-{1\over 2},c\pm {1\over 2}} 
 (\lambda z, \lambda z_L)~.
\label{BesselF3}
\eeqn
They satisfy
\beeq
D_+ (c)\begin{pmatrix} C_L \cr S_L \end{pmatrix} 
= \lambda \begin{pmatrix} S_R \cr C_R \end{pmatrix} ~~,~~
D_- (c) \begin{pmatrix} C_R \cr S_R \end{pmatrix} 
= \lambda \begin{pmatrix} S_L \cr C_L \end{pmatrix} ~~,~~
D_\pm (c)  = \pm \frac{d}{dz} + \frac{c}{z}~~,
\label{BesselF4}
\eneq
and
\beqn
&&\hskip -1cm 
C_R=C_L =1 ~,~~  S_R=S_L=0 ~,   ~~{\rm at~} z= z_L ~, \cr
\noalign{\kern 10pt}
&&\hskip -1cm  
C_L C_R - S_L S_R =1 ~.
\label{BesselF5}
\eeqn

\section{Wave functions of dark  fermions}

The dark fermion $\Psi_{F_i}$ is introduced  in the spinorial representation of 
$SO(5)$.  With the charge assignment of $Q_E = T^{3_L} + T^{3_R} + Q_X$ and $Q_X = \onehalf$,
$\Psi_{F_i}(x,z)$ is decomposed into KK modes $F_i^{+ (n)} (x)$ and $F_i^{0 (n)} (x)$
($n= 1, 2, 3, \cdots$) in the twisted gauge in which $\la  A_z \ra$ vanishes.
\beqn
&&\hskip -1.cm
\Psi_{F_i} = \Psi_{F_i ,R}  + \Psi_{F_i ,L} = \sum_n  \Psi_{F_i}^{(n)} ~~,~~
\Psi_{F_i}^{(n)}  =  \Psi_{F_i,R}^{(n)} +  \Psi_{F_i,L}^{(n)}~~, \cr
\noalign{\kern 10pt}
&&\hskip -1.cm
\gamma^5  \begin{pmatrix} \Psi_{F_i ,R} \cr  \Psi_{F_i ,L} \end{pmatrix}  = 
\begin{pmatrix} + \Psi_{F_i ,R} \cr  - \Psi_{F_i ,L} \end{pmatrix} ~, \cr
\noalign{\kern 10pt}
&&\hskip -1.cm 
\Psi_{F_i,R}^{(n)} (x, z) = \sqrt{k}z^2   
\left\{
\begin{pmatrix} f_{i,lR}^{(n)}(z) \cr 0 \cr f_{i,rR}^{(n)}(z) \cr 0 \end{pmatrix}   F^{+ (n)}_{i,R} (x)
+ \begin{pmatrix}0 \cr  f_{i,lR}^{(n)}(z) \cr 0 \cr f_{i,rR}^{(n)}(z) \end{pmatrix}   F^{0 (n)}_{i,R} (x)
\right\} ,  \cr
\noalign{\kern 10pt}
&&\hskip -1.cm 
\Psi_{F_i,L}^{(n)} (x, z) = \sqrt{k}z^2   
\left\{
\begin{pmatrix} f_{i,lL}^{(n)}(z) \cr 0 \cr f_{i,rL}^{(n)}(z) \cr 0 \end{pmatrix}   F^{+ (n)}_{i,L} (x)
+ \begin{pmatrix}0 \cr  f_{i,lL}^{(n)}(z) \cr 0 \cr f_{i,rL}^{(n)}(z) \end{pmatrix}   F^{0 (n)}_{i,L} (x)
\right\} .
\label{KKF1}
\eeqn
Here the suffixes $l$ and $r$ refer to two $SU(2)$'s of 
$SO(4) = SU(2)_L \times SU(2)_R  \subset SO(5)$.

$\Psi_{F_i}$ in the twisted gauge satisfies  a free Dirac equation.  
The left- and right-handed components of  $\tilde \Psi_{F_i} = z^{-2} \Psi_{F_i}$ satisfy
\beqn
&&\hskip -1.cm
\sigma \cdot \dd \, \tilde \Psi_{F_i,L} = k D_- (c) \tilde \Psi_{F_i,R} ~, \cr
\noalign{\kern 10pt}
&&\hskip -1.cm
\bar \sigma \cdot \dd \, \tilde \Psi_{F_i,R} = k D_+ (c) \tilde \Psi_{F_i,L} ~. 
\label{Feq1}
\eeqn
Let us denote  the $SU(2)_L$ ($SU(2)_R$) component of $\Psi_{F_i, L}$ by 
$\Psi_{F_i, lL}$ ($\Psi_{F_i, rL}$), etc.
The boundary condition for $\Psi_{F_i}$ with $\eta_{F_i}=+1$ in  (\ref{BC1}) 
is transformed in the twisted gauge  in the coformal  coordinates to
\beqn
&&\hskip -1.cm
\cos \onehalf \theta_H  \tilde \Psi_{F_i, lL} (1) 
- i \sin  \onehalf \theta_H \tilde \Psi_{F_i, rL} (1) =0 ~,\cr
\noalign{\kern 10pt}
&&\hskip -1.cm
- i \sin  \onehalf \theta_H \tilde \Psi_{F_i,lR} (1) 
+ \cos \onehalf \theta_H  \tilde \Psi_{F_i, rR} (1) =0 ~, \cr
\noalign{\kern 10pt}
&&\hskip -1.cm
\cos \onehalf \theta_H  \, D_- \tilde \Psi_{F_i, lR} (1) 
- i \sin  \onehalf \theta_H \, D_- \tilde \Psi_{F_i, rR} (1) =0 ~,\cr
\noalign{\kern 10pt}
&&\hskip -1.cm
- i \sin  \onehalf \theta_H \, D_+\tilde \Psi_{F_i, lL} (1)
+ \cos \onehalf \theta_H  \, D_+ \tilde \Psi_{F_i, rL} (1) = 0 ~,    \label{FBC1} \\
\noalign{\kern 10pt}
&&\hskip -1.cm
\tilde \Psi_{F_i, lR} (z_L) =0~, ~~ D_+ \tilde \Psi_{F_i, lL} (z_L) =0~, \cr
\noalign{\kern 10pt}
&&\hskip -1.cm
D_- \tilde \Psi_{F_i, rR} (z_L) =0~, ~~  \tilde \Psi_{F_i, rL} (z_L) =0~, 
\label{FBC2}
\eeqn
By making use of (\ref{FBC2}), eigenmodes can be written as
\beqn
&&\hskip -1.cm 
\begin{pmatrix}  \tilde \Psi_{F_i, lL} (z) \cr \tilde \Psi_{F_i, rL} (z) \end{pmatrix}
= \begin{pmatrix}A_1 C_L(z; \lambda, c_{F_i}) \cr B_1 S_L(z; \lambda, c_{F_i})\end{pmatrix} , \cr
\noalign{\kern 10pt}
&&\hskip -1.cm
\begin{pmatrix}  \tilde \Psi_{F_i, lR} (z) \cr \tilde \Psi_{F_i, rR} (z) \end{pmatrix}
= \begin{pmatrix}A_2 S_R(z; \lambda, c_{F_i}) \cr B_2 C_R(z; \lambda, c_{F_i})\end{pmatrix} .  
\label{Fwave1}
\eeqn
Then (\ref{FBC1}) leads to
\beqn
&&\hskip -1.cm 
M \begin{pmatrix} A_1 \cr B_1 \end{pmatrix} =
M \begin{pmatrix} A_2 \cr B_2 \end{pmatrix} = 0 ~, \cr
\noalign{\kern 10pt}
&&\hskip -1.cm
M = \begin{pmatrix} \cos \onehalf \theta_H C_L(1) &  - i \sin  \onehalf \theta_H  S_L(1) \cr
 - i \sin  \onehalf \theta_H  S_R(1) & \cos \onehalf \theta_H C_R(1) \end{pmatrix} ,
 \label{Fwave2}
 \eeqn
 where $C_L(z) = C_L(z; \lambda, c_{F_i})  ,~ S_R (z) = S_R(z;\lambda ,c_{F_i})$ etc.

The mass spectrum $\{ m_{F_i, n} = k \lambda_{i,n} \}$ is determined by $\det M =0$, or by
\beeq
C_L(1;\lambda_{i,n},c_{F_i}) C_R(1;\lambda_{i,n},c_{F_i}) - \sin^2 \frac{\theta_H}{2} = 0 ~.
\label{SpectrumF1}
\eneq
The corresponding wave functions  are given by
\begin{eqnarray}
&&\hskip -1.cm
\begin{pmatrix} f_{i, lL}^{(n)}(z) \\ f_{i, lR}^{(n)}(z) \end{pmatrix}
= \frac{i \sin \onehalf \theta_H S_L (1)}{\sqrt{r_i^{(n)}}}  \, 
\begin{pmatrix} C_L(z) \\ S_R (z) \end{pmatrix} 
=  \frac{\cos \onehalf \theta_H C_R(1)}{\sqrt{r_i'^{(n)}}}  \, 
\begin{pmatrix} C_L(z) \\ S_R (z) \end{pmatrix}, \cr
\noalign{\kern 10pt}
&&\hskip -1.cm
\begin{pmatrix} f_{i, rL}^{(n)}(z) \\ f_{i, rR}^{(n)}(z) \end{pmatrix}
=  \frac{\cos \onehalf \theta_H C_L(1)}{\sqrt{r_i^{(n)}}}  \, 
\begin{pmatrix} S_L(z) \\ C_R (z) \end{pmatrix} 
=  \frac{i \sin \onehalf \theta_H S_R (1)}{\sqrt{r_i'^{(n)}}}  \, 
\begin{pmatrix} S_L(z) \\ C_R (z) \end{pmatrix} ,
\label{WaveF1}
\end{eqnarray}
with $\lambda=\lambda_{i,n}$.
The normalization factors $r_i^{(n)}$ and $r_i'^{(n)}$ are determined by the condition
\begin{eqnarray}
\int_1^{z_L} dz \, \big\{ |f_{lL}^{(n)} |^2 + |f_{rL}^{(n)} |^2 \big\} = 
\int_1^{z_L} dz \, \big\{ |f_{lR}^{(n)} |^2 + |f_{rR}^{(n)} |^2 \big\} = 1
\label{eq:F-normalization}
\end{eqnarray} to be
\beqn
&&\hskip -1.cm
r_i^{(n)} = \int_1^{z_L} dz \,  \big\{ \sin^2 \onehalf \theta_H S_L(1)^2 C_L(z)^2 
+ \cos^2 \onehalf \theta_H  C_L(1)^2 S_L(z)^2 \big\}  \cr
\noalign{\kern 10pt}
&&\hskip -.6cm
= \int_1^{z_L} dz  \,  \big\{ \sin^2 \onehalf \theta_H S_L(1)^2 S_R(z)^2 
+  \cos^2 \onehalf \theta_H  C_L(1)^2 C_R(z)^2\big\} , \cr
\noalign{\kern 10pt}
&&\hskip -1.cm
r_i'^{(n)} = \int_1^{z_L} dz \,  \big\{ \cos^2 \onehalf \theta_H C_R(1)^2 C_L(z)^2 
+ \sin^2 \onehalf \theta_H  S_R(1)^2 S_L(z)^2 \big\}  \cr
\noalign{\kern 10pt}
&&\hskip -.6cm
= \int_1^{z_L} dz  \,  \big\{ \cos^2 \onehalf \theta_H C_R(1)^2 S_R(z)^2 
+  \sin^2 \onehalf \theta_H  S_R(1)^2 C_R(z)^2\big\} .
\label{normalizationF}
\eeqn

One comment is in order about the $\theta_H\go 0$ limit of the wave functions.
For $\theta_H = 0$ the spectrum (\ref{SpectrumF1}) is determined by
either $C_R(1)= C_R(1;\lambda_{i,2n-1},c_{F_i})=0$ or $C_L(1)=C_L(1;\lambda_{i,2n},c_{F_i})=0$ 
($n=1,2,3, \cdots$) where eigenvalues have been ordered as 
$0< \lambda_{i,1}  < \lambda_{i,2} <  \lambda_{i,3} < \cdots$.
The case $C_R(1)=0$ corresponds to excitations of the $SU(2)_R$ doublet component, 
whereas $C_L(1)=0$ to  excitations of the $SU(2)_L$ doublet component.
For $C_R(1)=0$ ($C_L(1)=0$), $r_i'^{(n)} / \sin^2 \onehalf \theta_H \not= 0$
( $r_i^{(n)} / \sin^2 \onehalf \theta_H \not= 0$) at $\theta_H =0$.

In  the boundary condition for $\Psi_{F_i}$, one could adopt $\eta_{F_i} = -1$  in (\ref{BC1}).  
In the case of non-degenerate dark fermions the heavy dark fermion
multiplet satisfies this flipped boundary condition.
In this case the corresponding wave functions and Kaluza-Klein masses are obtained 
from the above formulas by the replacement
\begin{eqnarray}
c_H \leftrightarrow is_H ~,
\quad
C_{L} \leftrightarrow S_{L} ~,
\quad
S_{R} \leftrightarrow C_{R} ~.
\end{eqnarray}
The spectrum is determined by the same equation as in (\ref{SpectrumF1}). 
The lowest mode mostly becomes an $SU(2)_L$ doublet for small $\theta_H$.

\section{Gauge and Higgs couplings of dark fermions}\label{sec:apdx-fermion-boson-couplings}

\subsection{Couplings to the Higgs boson}\label{higgs-coupling}
Couplings to the Higgs boson is read from the gauge interaction
\begin{eqnarray}
\int_1^{z_L} dz \sqrt{G} {e_4}^{z} \bar{\Psi}_F (g_A A_z + Q_X g_B B_z) i\gamma_5 \Psi_F,
\quad
\sqrt{G} {e_4}^z g_A = \frac{\sqrt{L}g_w}{k^4z^4},
\end{eqnarray}
where
\begin{eqnarray}
A_z(x,z) &=& \hat{H} + \sum_{a=1}^3 \hat{G}^a + \sum_{a=1}^3 \hat{D}^a,
\nonumber\\
\hat{H} &=& \sum_n H^{(n)}(x) u_{H^{(n)}} T^{\hat{4}},
\nonumber\\
\hat{G}^a &=& \sum_n G^{a(n)}(x) \left\{ 
u_{G^{(n)}} \frac{T^{a_L} +  T^{a_R}}{\sqrt{2}} \right\},
\nonumber\\
\hat{D}^a &=& \sum_n D^{a(n)}(x) \left\{ u_{D^{(n)}}^- \frac{T^{a_L} - T^{a_R}}{\sqrt{2}}
 + \hat{u}_{D^{(n)}} T^{\hat{a}}
\right\},
\nonumber\\
B_z(x,z) &=& \sum_n B^{(n)}(x) u_{B^{(n)}} (z),
\end{eqnarray}
$G^{a(n)}$, $D^{a(n)}$ and $B^{(n)}$ are NG-bosons and only the $\hat{H}$ is the tower of the physical scalar particles. Hereafter we consider only Higgs couplings.
The Higgs wave functions are given by
\begin{eqnarray}
u_{H^{(0)}}(z) &=& \sqrt{\frac{2}{k(z_L^2-1)}} \, z, 
\end{eqnarray}
for the zero-mode Higgs boson
and
\begin{eqnarray}
u_{H^{(n)}}(z) &=& \frac{1}{\sqrt{r_{H^{(n)}}}} S'(z;\lambda_{H^{(n)}}),
\quad
r_{H^{(n)}} = \int_1^{z_L} \frac{kdz}{z} S'(z;\lambda_{H^{(n)}})^2,
\end{eqnarray}
for KK excitations ($n\geq1$). Here $S(1;\lambda_{H^{(n)}}) = 0$ is satisfied.
The building-block for the $H\bar{F}^{(n)}F^{(n)}$ Yukawa coupling is given by
\begin{eqnarray}
\bar{\Psi}^{(n)}_{F_j} \gamma_5 T^{\hat{4}} \Psi^{(n)}_{F_j}
&=& i \frac{kz^4}{2\sqrt{2}} \frac{1}{r_j^{(n)}} \sin\frac{\theta_H}{2} \cos\frac{\theta_H}{2}S_L(1) C_L(1) [\bar{F}_{jL}^{(n)} F_{jR}^{(n)} - \bar{F}_{jR}^{(n)} F_{jL}^{(n)}],
\end{eqnarray}
where $C_L(z) C_R(z) - S_L(z) S_R(z) = 1 $ has been made use of.
Hence the Higgs Yukawa coupling in the 4D Lagrangian,
${\cal L}_{4D} \supset y_{F^{(n)}_i} H^{(0)} \bar{F}^{(n)}_i F^{(n)}_i$, is given by
\begin{eqnarray}
y_{F^{(n)}_i} = \frac{g_w}{4} \frac{1}{r_i^{(n)}} \sqrt{kL(z_L^2-1)} \sin\frac{\theta_H}{2} \cos\frac{\theta_H}{2} S_L(1;\lambda_{i,n},c_F) C_L(1;\lambda_{i,n},c_F),
\label{defHiggsYukawa}
\end{eqnarray}
In Tables~\ref{tbl:HYukawa} and \ref{tbl:HYukawa-ND}, we have summarized the Higgs Yukawa couplings of $F$.
%In Table~\ref{tbl:HYukawa}, we have also tabulated $\theta_H$ and $c_F$, which are obtained so that the Higgs mass is $125\,\text{GeV}$ \cite{LHCsignals}.
In Table~\ref{tbl:HYukawa-ND}  the  couplings in non-degenerate cases are summarized.

\begin{table}[tbp]
\caption{The Higgs-Yukawa couplings $y_{F_i^{(1)}}$ in (\ref{defHiggsYukawa}) in the case of 
degenerate dark fermions with the parameters  specified in Table \ref{tbl:Degenerate}.}\label{tbl:HYukawa}
\begin{center}
\begin{tabular}{c|c|c}
\hline\hline
$n_F$ & $z_L$ & $y_{F_i^{(1)}}$ 
\\
\hline
$3$ & $10^8$ &
     $-0.106$ 
\\
    & $10^6$ &
     $-0.071$ 
\\
    & $10^5$ &
     $-0.064$  
\\
    & $2\times10^4$ &
     $-0.089$ 
    \\
\hline
$4$ & $10^8$ & 
     $-0.082$ 
\\
    & $10^6$ &
     $-0.049$ 
\\
    & $10^5$ &
     $-0.038$ 
\\
    &$3\times10^4$ &
     $-0.034$ 
\\
    & $10^4$ &
     $-0.033$
\\
\hline
$6$ & $10^8$ & 
     $-0.060$ 
\\
    & $10^6$ &
     $-0.034$ 
\\
    & $10^5$ &
     $-0.024$ 
\\
    & $10^4$ &
     $-0.017$ 
\\
\hline\hline
\end{tabular}
\end{center}
\end{table}
\begin{table}[tbp]
\caption{The Higgs-Yukawa couplings $y_{F_l^{(1)}}$ of the light dark fermion in (\ref{defHiggsYukawa}) 
in the case of  non-degenerate $(\nlight,\nheavy)=(1,3)$ dark fermions with the parameters 
specified in Table \ref{tbl:diff_nf4}.}\label{tbl:HYukawa-ND}
\begin{center}
\begin{tabular}{cc|c}
\hline\hline
$\Delta c_F$ & $z_L$ & $y_{F_l^{(1)}}$
\\
\hline
$0.04$
&     $10^6$        
& $-0.042$ 
\\
&     $10^5$        
& $-0.033$ 
\\ 
&     $3\times10^4$ 
& $-0.029$ 
\\
&     $10^4$        
& $-0.028$ 
\\
\hline
$0.06$ 
&      $10^6$        
& $-0.038$ 
\\
&      $10^5$        
& $-0.030$ 
\\ 
&      $3\times10^4$ 
& $-0.027$ 
\\
&      $10^4$        
& $-0.026$ 
\\
\hline\hline
\end{tabular}
\end{center}
\end{table}

%%%%%%%%%%%%%%%%%%%%%%%%%%%%%%%%%%%%%%%%%%%%%%%%%%%%%%%%%%%%%%%%%%%%
\subsection{Couplings to vector bosons}\label{sec:vector-coupling}
Couplings to the vector bosons are read off from the gauge interaction in the 5D action
\begin{eqnarray}
\int_0^{z_L} dz \sqrt{G} {e_m}^\mu \bar{\Psi}_F\gamma^m( g_A A_\mu + Q_X g_B B_\mu)\Psi_F,
\quad
\sqrt{G} {e_m}^{\mu} g_A = \frac{g_w \sqrt{L}}{k z^4} \delta^\mu_m,
\label{apdx:FFV-5D}
\end{eqnarray}
where $A_\mu(x,z)$ and $B_\mu(x,z)$ decompose to the   Kaluza-Klein towers
\begin{eqnarray}
A_\mu(x,z)
&=& 
\hat{W}^{-}_\mu +
\hat{W}^{+}_\mu +
\hat{Z}^{(A)}_\mu +
\hat{A}_\mu^{\gamma(A)} + 
\hat{W}^{-}_{R\mu} +
\hat{W}^{+}_{R\mu} +
\hat{Z}^{(A)}_{R\mu} +
\hat{A}^{\hat{4}}_\mu ~,
\nonumber\\
B_\mu(x,z) 
&=& 
\hat{Z}^{(B)}_\mu +
\hat{A}^{\gamma(B)}_\mu + 
\hat{Z}^{(B)}_{R\mu} ~.
\end{eqnarray}
The gauge couplings in \eqref{apdx:FFV-5D} consist of
\begin{align}
&\bar{\Psi}_F \gamma^\mu (g_A \hat{V}_\mu) \Psi_F,
&&\text{for } V = W, \, W_R, \, A^{\hat{4}},
\nonumber \\
&\bar{\Psi}_F \gamma^\mu (g_A \hat{V}_\mu^{(A)} + Q_X g_B \hat{V}_\mu^{(B)}) \Psi_F,
&&\text{for } V = A^\gamma, \, Z, \, Z_R ~.
\end{align}
Each tower is decomposed to KK modes.
%$\hat{W}_\mu^{-} = \sum_n W_\mu^{-(n)}(x,z), \cdots$ or
For $W$, $W_R$ and $A^{\hat{4}}$ bosons
\begin{eqnarray}
\hat{W}^{\pm}_\mu &=& \sum_n W_\mu^{\pm(n)}(x) \left\{ 
h_{W^{(n)}}^L \frac{T^{1_L} \pm i T^{2_L}}{\sqrt{2}} +
h_{W^{(n)}}^R \frac{T^{1_R} \pm i T^{2_R}}{\sqrt{2}} +
\hat{h}_{W^{(n)}} \frac{T^{\hat{1}} \pm i T^{\hat{2}}}{\sqrt{2}}
\right\},
\nonumber\\
\hat{W}^{\pm}_{R\mu} &=& \sum_n W_{R\mu}^{\pm(n)}(x) \left\{ 
h_{W_R^{(n)}}^L \frac{T^{1_L} \pm i T^{2_L}}{\sqrt{2}} +
h_{W_R^{(n)}}^R \frac{T^{1_R} \pm i T^{2_R}}{\sqrt{2}} +
\hat{h}_{W_R^{(n)}} \frac{T^{\hat{1}} \pm i T^{\hat{2}}}{\sqrt{2}}
\right\},
\nonumber\\
\hat{A}^{\hat{4}}_\mu &=& \sum_n A_\mu^{\hat{4}(n)}(x) 
h_{A^{\hat{4}(n)}} T^{\hat{4}},
\end{eqnarray}
where $\hat{W}^{\pm}_{\mu} = (\hat{W}_{\mu}^1 \mp i\hat{W}_{\mu}^2)/\sqrt{2}$ etc., whereas
for $A^\gamma$, $Z$ and $Z_R$ bosons
\begin{eqnarray}
\left(\hat{A}^{\gamma(A)}_\mu ,\, \hat{A}^{\gamma(B)}_\mu \right) 
&=& \sum_n A^{\gamma(n)}_\mu(x) \left( 
h_{\gamma^{(n)}}^L T^{3_L} +
h_{\gamma^{(n)}}^R T^{3_R} ,\,
h_{\gamma^{(n)}}^B %T_B
\right),
\nonumber\\
\left( \hat{Z}^{(A)}_\mu ,\, \hat{Z}^{(B)}_\mu \right)
&=& \sum_n Z_\mu^{(n)}(x) \left(
h_{Z^{(n)}}^L T^{3_L} +
h_{Z^{(n)}}^R T^{3_R} +
\hat{h}_{Z^{(n)}} T^{\hat{3}} ,\,
h_{Z^{(n)}}^B %T_B
\right),
\nonumber\\
\left( \hat{Z}^{(A)}_{R\mu} ,\, \hat{Z}^{(B)}_{R\mu} \right)
&=& \sum_n Z_{R\mu}^{(n)}(x) \left(
h_{Z_R^{(n)}}^L T^{3_L} +
h_{Z_R^{(n)}}^R T^{3_R} ,\,
h_{Z_R^{(n)}}^B %T_B
\right).
\end{eqnarray}
Here $n=0,1,2,\cdots$ [$1,2,\cdots$] for $A^\gamma_\mu$, $W_\mu$ and $Z_\mu$ [$W_{R\mu}$, $Z_{R\mu}$ and $A^{\hat{4}}_\mu$].
$A^{\gamma(0)}_\mu$, $W^{(0)}_\mu$ and $Z^{(0)}_\mu$ correspond to the photon, $W$ and $Z$ bosons,
respectively.

%%%%%%%%%%%%%%%%%%%%%%%%%%%%%%%%%%%%%%%%%%%%%%%%%%%%%%%%
\subsubsection{Couplings to $\gamma^{(n)}$, $Z^{(n)}$, $Z_R^{(n)}$ and $A^{\hat{4}}$}
Here we summarize the dark fermion couplings to the neutral vector bosons.
It would be useful to collect the building blocks for the couplings.
For KK fermions $\Psi^{(n)}_F$ and $\Psi^{(m)}_F$,
we have
\begin{eqnarray}
\bar{\Psi}^{(n)}_F T^{3_L} \gamma^\mu \Psi^{(m)}_F
&=& \frac{k z^4}{2} f_{lL}^{(n)*} f_{lL}^{(m)} 
[\bar{F}^{+(n)}_{L} \gamma^\mu F^{+(m)}_{L} - \bar{F}^{0(n)}_{L}\gamma^\mu  F^{0(m)}_{L}] + (F_L,f_{lL}\to F_R,f_{lR}),
\nonumber\\
\bar{\Psi}^{(n)}_F T^{3_R} \gamma^\mu \Psi^{(m)}_F
&=& \frac{kz^4}{2} f_{rL}^{(n)*} f_{rL}^{(m)} 
[\bar{F}^{+(n)}_{L}\gamma^\mu  F^{+(m)}_{L,m} - \bar{F}^{0(n)}_{L}\gamma^\mu  F^{0(m)}_{L}] + (F_L,f_{rL}\to F_R,f_{rR}),
\nonumber\\
\bar{\Psi}^{(n)}_F T^{\hat{3}} \gamma^\mu \Psi^{(m)}_F
&=& \frac{kz^4}{2\sqrt{2}} i [f_{lL}^{(n)*} f_{rL}^{(m)} - f_{rL}^{(n)*} f_{rL}^{(m)}] 
[\bar{F}^{+(n)}_{L}\gamma^\mu  F^{+(m)}_{L} - \bar{F}^{0(n)}_{L}\gamma^\mu  F^{0(m)}_{L}]
\nonumber\\&& 
 + (F_L,f_{lL},f_{rL}\to F_R, f_{lR}, f_{rR}),
\nonumber\\
\bar{\Psi}^{(n)}_F \gamma^\mu \Psi^{(m)}_F
&=&kz^4 [f_{lL}^{(n)*} f_{lL}^{(m)} + f_{rL}^{(n)*} f_{rL}^{(m)}] 
[\bar{F}^{+(n)}_{L}\gamma^\mu  F^{+(m)}_{L} + \bar{F}^{0(n)}_{L}\gamma^\mu  F^{0(m)}_{L}]
\nonumber\\&&
 + (F_L,f_{lL},f_{rL} \to F_R,f_{lR},f_{rR}).
\end{eqnarray}
In the followings we summarize the couplings in the case of $n=m=1$.

\paragraph{Electromagnetic photon $\gamma = \gamma^{(0)}$}
For the photon $A_\mu^{\gamma(0)}$, wave functions are given by
\begin{eqnarray}
h_{\gamma^{(0)}}^L = h_{\gamma^{(0)}}^R = \frac{1}{\sqrt{(1+s_\phi^2)L}} s_\phi,
\quad
h_{\gamma^{(0)}}^B = \frac{1}{\sqrt{(1+s_\phi^2)L}} c_\phi,
\end{eqnarray}
where $c_\phi $ and $s_\phi$ given by
\begin{eqnarray}
c_\phi \equiv \cos\phi = \frac{g_A}{\sqrt{g_A^2 + g_B^2}},
\quad
s_\phi \equiv \sin\phi = \frac{g_B}{\sqrt{g_A^2 + g_B^2}},
\end{eqnarray}
parameterize the mixing of $A_M$ and $B_M$, and are related to the Weinberg angle 
$\theta_W$ by  $\sin\phi = \tan\theta_W$.
The couplings between dark fermions and the photon can be read from
\begin{eqnarray}
\lefteqn{
\int_1^{z_L} dz \sqrt{G} {e_l}^\mu \bar{\Psi}^{(n)}_F [g_A A_\mu^{\gamma(0)} + Q_X g_B A_\mu^{\gamma(0)}] \gamma^l \Psi^{(m)}_F}
\nonumber\\
&=& e A_\mu^{\gamma(0)}(x) \int_1^{z_L} dz
\biggl\{
  \left[ f_{lL}^{(n)*} f_{lL}^{(m)} + f_{rL}^{(n)*} f_{rL}^{(m)} \right] 
\nonumber\\&&  \phantom{MM} \times 
\biggl[ (Q_X + \tfrac{1}{2}) \bar{F}^{+(n)}_L \gamma^\mu F^{+(n)}_L
+ (Q_X - \tfrac{1}{2}) \bar{F}^{0(n)}_L \gamma^\mu F^{0(n)}_L
\biggr] \biggr\} + (L \to R) 
\nonumber\\
&=& e A_\mu^{\gamma(0)}(x) \delta_{n,m} 
\left\{
  (Q_X + \tfrac{1}{2}) \bar{F}^{+(n)} \gamma^\mu F^{+(n)}
+ (Q_X - \tfrac{1}{2}) \bar{F}^{0(n)} \gamma^\mu F^{0(n)}
\right\}, 
\end{eqnarray}
where the  orthonormality conditions \eqref{eq:F-normalization} has been used.
%Hence we observe that 
$F^+$ [$F^0$] has electric charge $Q_X+\frac{1}{2}$ [$Q_X - \frac{1}{2}$].
The Kaluza-Klein level for fermions is preserved.

\paragraph{KK photons}
Wave functions for the KK photons $\gamma^{(n)}$ ($n \ge 1$) are given by
\begin{eqnarray}
&&
\begin{pmatrix} h_{\gamma^{(n)}}^L = h_{\gamma^{(n)}}^R \\ h_{\gamma^{(n)}}^B \end{pmatrix}
= \frac{1}{\sqrt{1+s_\phi^2}} \frac{1}{\sqrt{r_{\gamma^{(n)}}}} 
\begin{pmatrix}  s_\phi \\ c_\phi \end{pmatrix} C(z),
\quad
 r_{\gamma^{(n)}} = \int_1^{z_L} \frac{dz}{kz} C(z)^2,
\end{eqnarray}
where $C(z) = C(z;\lambda_{\gamma^{(n)}})$ and $\lambda_{\gamma^{(n)}}$ satisfy 
$C'(1;\lambda_{\gamma^{(n)}}) = 0$.
Hence the couplings are given by
\begin{eqnarray}
\lefteqn{
\sum_{c=+,0}
\gamma^{(n)}_\mu
[g_{F^c L}^{\gamma^{(n)}} \bar{F}^{c}_L \gamma^\mu F^{c}_L
+g_{F^c R}^{\gamma^{(n)}} \bar{F}^{c}_R \gamma^\mu F^{c}_R]
}
\nonumber\\
&=&
\gamma_\mu^{(n)}(x)
\left\{
  (Q_X + \frac{1}{2}) \bar{F}^{+}_L \gamma^\mu F^{+}_L 
+ (Q_X - \frac{1}{2}) \bar{F}^{0}_L \gamma^\mu F^{0}_L
\right\}
\nonumber\\&&
\times
\frac{e\sqrt{L}}{\sqrt{r_{\gamma^{(n)}}} } 
\int_1^{z_L} dz \, C(z) [|f_{lL}|^2 + |f_{rL}|^2]
+ (L \to R) ~. \label{defHgamma}
\end{eqnarray}
Note that the couplings are left-right asymmetric, i.e.,
$
g_{F\gamma^{(n)}}^L \neq g_{F\gamma^{(n)}}^R
$
for $n \ge 1$.
In Tables~\ref{tbl:gamma(1)-coupling} and \ref{tbl:gamma(1)-coupling-ND}, $\gamma^{(1)}F^+F^-$ couplings are tabulated.
\begin{table}[tbp]
\caption{%
The  mass and left- and right-handed couplings to $F^+$ in (\ref{defHgamma}) in the unit of electromagnetic coupling $e$
of the first KK photon in the case of degenerate dark fermions with the parameters specified in Table \ref{tbl:Degenerate}.
}\label{tbl:gamma(1)-coupling}
\begin{center}
\begin{tabular}{c|c|ccc}
\hline\hline
$n_F$ & $z_L$ &
$m_{\gamma^{(1)}}$ &
$g_{F^+ L}^{\gamma^{(1)}}$ & 
$g_{F^+ R}^{\gamma^{(1)}}$ 
\\
& & [TeV] & & 
\\ 
\hline
$3$ & $10^8$ &
    $2.42$ & $0.19$ & $4.16$
\\
    & $10^6$ &
    $4.26$ & $0.28$ & $3.61$
\\
    & $10^5$ &
    $5.92$ & $0.38$ & $3.31$ 
\\
    & $2\times10^4$ &
    $7.55$ & $0.52$ & $3.09$
    \\
\hline
$4$ & $10^8$ & 
    $2.46$ & $0.06$ & $4.15$
\\
    & $10^6$ &
    $4.32$ & $0.11$ & $3.59$ 
\\
    & $10^5$ &
    $6.00$ & $0.15$ & $3.28$
\\
    &$3\times10^4$ &
    $7.19$ & $0.17$ & $3.10$
\\
    & $10^4$ &
    $8.52$ & $0.21$ & $2.93$
    \\
\hline
$6$ & $10^8$ & 
    $2.50$ & $-0.06$ & $4.14$ 
\\
    & $10^6$ &
    $4.40$ & $-0.05$ & $3.58$
\\
    & $10^5$ &
    $6.12$ & $-0.04$ & $3.26$
\\
    & $10^4$ &
    $8.68$ & $-0.03$ & $2.90$
\\
\hline\hline
\end{tabular}
\end{center}
\end{table}

\begin{table}[tbp]
\caption{
The  left- and right-handed couplings to the light $F_l^+$ in (\ref{defHgamma}) in the unit of electromagnetic 
coupling $e$ of the first KK photon in the case of non-degenerate  $(\nlight,\nheavy)=(1,3)$ dark fermions 
with the parameters specified in Table \ref{tbl:diff_nf4}.
}\label{tbl:gamma(1)-coupling-ND}
\begin{center}
\begin{tabular}{cc|ccccc}
\hline\hline
$\Delta c_F$ & 
$z_L$ &
$g_{F^+_L}^{\gamma^{(1)}}$ &
$g_{F^+_R}^{\gamma^{(1)}}$ 
\\
\hline
$0.04$ &
      $10^6$ & 
      $0.03$ & $3.58$
      \\
      &
      $10^5$ & 
      $0.08$ & $3.27$ 
      \\ 
      &
      $3\times10^4$ & 
      $0.11$ & $3.09$ 
      \\
      &
      $10^4$ & 
      $0.16$ & $2.92$ 
      \\
\hline
$0.06$ &
      $10^6$ &
      $0.01$ & $3.58$ 
      \\
      &
      $10^5$ & 
      $0.04$ & $3.26$ 
      \\ 
      &
      $3\times10^4$ & 
      $0.08$ & $3.09$ 
      \\
      &
      $10^4$ & 
      $0.13$ & $2.92$ 
      \\
\hline\hline
\end{tabular}
\end{center}
\end{table}

%%%%%%%%%%%%%%%%%%%%%%%

\paragraph{$Z$ boson}
Wave functions of $Z$ tower are given by
\begin{eqnarray}
&&\begin{pmatrix} h_{Z^{(n)}}^L \\ h_{Z^{(n)}}^R \\ \hat{h}_{Z^{(n)}} \\ (g_B/g_A) h_{Z^{(n)}}^B
\end{pmatrix}
= \frac{1}{\sqrt{1 + s_\phi^2}} \frac{1}{\sqrt{r_{Z^{(n)}}}}
\begin{pmatrix}
 \frac{c_\phi^2 + (1+s_\phi^2)\cos\theta_H}{\sqrt{2}} C(z) \\
 \frac{c_\phi^2 - (1+s_\phi^2)\cos\theta_H}{\sqrt{2}} C(z) \\
 -(1+s_\phi^2) \sin\theta_H \hat{S}(z) \\
 -\sqrt{2} s_\phi^2 C(z)
\end{pmatrix},
\nonumber\\
&&
r_{Z^{(n)}} = \int_1^{z_L} \frac{dz}{kz} 
\left\{
 c_\phi^2  C(z)^2 + (1+s_\phi^2)[\cos^2\theta_H C(z)^2 + \sin^2\theta_H \hat{S}(z)^2]
\right\},
\end{eqnarray}
where  $C(z) \equiv C(z;\lambda_{Z^{(n)}})$ , $\hat{S}(z) \equiv \hat{S}(z;\lambda_{Z^{(n)}})$
and $\lambda_{Z^{(n)}}$ satisfy
\begin{eqnarray}
2 S(z;\lambda_{Z^{(n)}}) C'(z;\lambda_{Z^{(n)}}) + (1+s_\phi^2) \lambda_{Z^{(n)}} \sin^2\theta_H
&=& 0 ~ .
\end{eqnarray}
The smallest positive root $\lambda_{Z^{(0)}}$ is related to the $Z$-boson mass 
by $m_Z = k\cdot\lambda_{Z^{(0)}}$. 
In terms of these the couplings of $F$ to the $Z^{(n)}$ boson are given by
\begin{eqnarray}
{\cal L}_{4D} 
&\supset&
 Z_\mu^{(n)} \sum_{c=+,-} 
[ g_{F^c L}^{Z^{(n)}} \bar{F}^{c}_L \gamma^\mu F^{c}_L 
+ g_{F^c L}^{Z^{(n)}} \bar{F}^{c}_R \gamma^\mu F^{c}_R]
\nonumber\\
&=& \frac{g_w \sqrt{L}}{\sqrt{2}\cos\theta_W\sqrt{r_{Z^{(n)}}}} 
Z_\mu^{(n)}\sum_{c=+,0} \bar{F}^{c}_L \gamma^\mu F^{c}_L
\int_1^{z_L} dz 
\biggl[
I_3^{(c)}
\biggl\{
C(z) [ |f_{lL}|^2 + |f_{rL}|^2]
\nonumber\\&&
 + \cos\theta_H C(z) [|f_{lL}|^2 - |f_{rL}|^2]
- i \sin\theta_H \hat{S}(z) [f_{lL}^* f_{rL} - f_{rL}^* f_{lL}]
\biggr\}
\nonumber\\&&
- (Q_X + I_3^{(c)}) \sin^2\theta_W \cdot 2 C(z)[ |f_{lL}|^2 + |f_{rL}|^2]
\biggl] + (L \to R), \label{FZcoupling}
\end{eqnarray}
where $I_3^{(c)} = \frac{1}{2}$ $[-\frac{1}{2}]$ for $c=+$ $[0]$. 
We note that if the $F$ obey the boundary condition $\etaF=+1$ the $Z^{(n)}$ coupling to a fermion $F^{0}$ with $Q_{EM} = Q_X + I_3^{(i)}=0$
is suppressed by $\sin^2(\theta_H/2)$, because $f_{lL} \propto \sin(\theta_H/2)$.
We have summarized the $ZF\bar{F}$ couplings in
Tables~\ref{tbl:ZFF_coupling}, \ref{tbl:ZFF_coupling-bc} and \ref{Z-ND},
and the $Z^{(1)}F\bar{F}$ couplings in 
Tables~\ref{tbl:Z(1)FF_coupling} and \ref{tbl:Z(1)FhFh_coupling}.
%In the numerical computation we adopted $\sin^2\theta_W=0.231$.

% Z couplings
\begin{table}[tbp]
\caption{%
The left- and right-handed couplings in the unit of $g_w$  of $F$ to the $Z$ boson in (\ref{FZcoupling}) 
with b.c. $\etaF=+1$ in the case of  degenerate dark fermions with the parameters specified  in
Table \ref{tbl:Degenerate}.}\label{tbl:ZFF_coupling}
\begin{center}
\begin{tabular}{c|c|cccc}
\hline\hline
$n_F$ &
$z_L$ &
$g_{F^+ L}^{Z}$  &
$g_{F^+ R}^{Z}$  &
$g_{F^0 L}^{Z}\times10^4$  & 
$g_{F^0 R}^{Z}\times10^4$  
\\
\hline
$3$ &
  $10^8$ & 
    $-0.260$ & $-0.242$ & $-40.1$ & $-227$ 
\\
  & $10^6$ &
    $-0.261$ & $-0.257$ & $-21.8$ & $-69.6$ 
\\
  & $10^5$ &
    $-0.262$ & $-0.260$ & $-19.7$ & $-42.7$ 
\\
  & $2\times10^4$ &
    $-0.259$ & $-0.258$ & $-41.3$ & $-58.4$ 
    \\
\hline
$4$ & $10^8$ & 
    $-0.261$ & $-0.244$ & $-25.2$ & $-204.9$ 
\\
    & $10^6$ &
    $-0.263$ & $-0.258$ & $-11.4$ & $-55.9$ 
\\
    & $10^5$ &
    $-0.263$ & $-0.261$ & $-7.6$ & $-27.7$ 
\\
    & $3\times10^4$ &
    $-0.263$ & $-0.262$ & $-6.7$ & $-19.7$
\\
    & $10^4$ &
    $-0.263$ & $-0.262$ & $-6.5$ & $-15.4$ 
\\
\hline
$6$ & $10^8$ & 
    $-0.263$ & $-0.246$ & $-14.2$ & $-186.0$ 
\\
    & $10^6$ &
    $-0.263$ & $-0.259$ & $-5.8$ & $-47.7$ 
\\
    & $10^5$ &
    $-0.263$ & $-0.262$ & $-3.4$ & $-21.9$ 
\\
    & $10^4$ &
    $-0.264$ & $-0.263$ & $-2.1$ & $-9.8$ 
\\
\hline\hline
\end{tabular}
\end{center}
\end{table}
\begin{table}[tbp]
\caption{The left- and right-handed couplings in the unit of $g_w$  of $F$ to the $Z$ boson in (\ref{FZcoupling}) 
with b.c. $\etaF=-1$ in the case of  degenerate dark fermions with the parameters specified  in 
Table \ref{tbl:Degenerate}.}\label{tbl:ZFF_coupling-bc}
\begin{center}
\begin{tabular}{c|c|cccc}
\hline\hline
$n_F$ & $z_L$ & 
$g_{F^+ L}^{Z}$ &
$g_{F^+ R}^{Z}$ &
$g_{F^0 L}^{Z}$ &
$g_{F^0 R}^{Z}$ 
\\
\hline
$4$ & $10^8$ &
    $0.304$ & $0.287$ & $-0.569$ & $-0.552$ 
    \\
    & $10^6$ &
    $0.306$ & $0.301$ & $-0.569$ & $-0.565$ 
    \\
    & $10^4$ &
    $0.306$ & $0.305$ & $-0.570$ & $-0.569$
    \\
\hline
\hline
\end{tabular}
\end{center}
\end{table}
\begin{table}[tbp]
\caption{The left- and right-handed couplings in the unit of $g_w$  of $F$ to the $Z$ boson in (\ref{FZcoupling}) 
with b.c. $\etaF=+1$ in the case of  non-degenerate $(\nlight,\nheavy)=(1,3)$ dark fermions 
with the parameters specified  in Table \ref{tbl:diff_nf4}.}
\label{Z-ND}
\begin{center}
\begin{tabular}{c|c|cccc}
\hline\hline
$\Delta c_F$ & $z_L$ & 
$g_{F^+ L}^{Z}$ &
$g_{F^+ R}^{Z}$ &
$g_{F^0 L}^{Z}\times10^4$ &
$g_{F^0 R}^{Z}\times10^4$ 
\\
\hline
$0.04$ 
    & $10^6$ &
    $-0.263$ & $-0.259$ & $-8.4$ & $-52.2$
    \\ 
    & $10^5$ &
    $-0.263$ & $-0.261$ & $-5.9$ & $-25.5$ 
    \\
    & $3\times10^4$ &
    $-0.263$ & $-0.262$ & $-5.1$ & $-17.8$ 
    \\
    & $10^4$ &
    $-0.263$ & $-0.262$ & $-5.0$ & $-13.5$
    \\
\hline
$0.06$ &
    $10^6$ &
    $-0.263$ & $-0.259$ & $-7.2$ & $-50.8$
    \\ 
    & $10^5$ &
    $-0.263$ & $-0.261$ & $-5.1$ & $-24.6$ 
    \\
    & $3\times10^4$ &
    $-0.263$ & $-0.262$ & $-4.5$ & $-17.0$ 
    \\
    & $10^4$ &
    $-0.263$ & $-0.263$ & $-4.4$ & $-12.8$
    \\
\hline
\hline
\end{tabular}
\end{center}
\end{table}

% 1st KK Z boson
\begin{table}[htbp]
\caption{%
The left- and right-handed couplings in the unit of $g_w$  of $F$ to the first KK $Z$ boson 
$Z^{(1)}$ in (\ref{FZcoupling}) 
with b.c. $\etaF=+1$ in the case of  degenerate dark fermions with the parameters specified  in
Table \ref{tbl:Degenerate}.}\label{tbl:Z(1)FF_coupling}
\begin{center}
\begin{tabular}{c|c|ccccc}
\hline\hline
$n_F$ & 
$z_L$ &
$m_{Z(1)}$ &
$g_{F^+ L}^{Z^{(1)}}$ & 
$g_{F^+ R}^{Z^{(1)}}$ & 
$g_{F^0 L}^{Z^{(1)}}$ & 
$g_{F^0 R}^{Z^{(1)}}$ 
\\ 
 & & [TeV] 
\\
\hline
$3$ &
  $10^8$ & 
    $2.42$ &
    $-0.02$ & $-1.07$ & $-0.04$ & $-0.08$      
\\
  & $10^6$ &
    $4.25$ &
    $-0.06$ & $-0.95$ & $-0.02$ & $-0.02$ 
\\
  & $10^5$ &
    $5.92$ &
    $-0.09$ & $-0.87$ & $-0.01$ & $-0.01$
\\
  & $2\times10^4$ &
    $7.54$ &
    $-0.12$ & $-0.81$ & $-0.02$ & $-0.00$
    \\
\hline
$4$ & $10^8$ & 
    $2.45$ &
    $0.00$ & $-1.06$ & $-0.02$ & $-0.08$
\\
    & $10^6$ &
    $4.32$ &
    $-0.02$ & $-0.94$ & $-0.01$ & $-0.02$ 
\\
    & $10^5$ &
    $6.00$ &
    $-0.03$ & $-0.86$ & $-0.01$ & $-0.01$ 
\\
    & $10^4$ &
    $8.52$ &
    $-0.05$ & $-0.77$ & $-0.00$ & $-0.00$
\\
\hline
$6$ & $10^8$ & 
    $2.50$ & 
    $0.02$ & $-1.06$ & $-0.01$ & $-0.07$
\\
    & $10^6$ &
    $4.40$ &
    $0.02$ & $-0.94$ & $-0.00$ & $-0.01$
\\
    & $10^5$ &
    $6.13$ &
    $0.01$ & $-0.86$ & $-0.00$ & $-0.01$
\\
    & $10^4$ &
    $8.68$ &
    $0.01$ & $-0.77$ & $-0.00$ & $-0.00$
\\
\hline\hline
\end{tabular}
\end{center}
\end{table}

\begin{table}[htbp]
\caption{The left- and right-handed couplings in the unit of $g_w$  of $F$ to the first KK $Z$ boson 
$Z^{(1)}$ in (\ref{FZcoupling}) 
with b.c. $\etaF=-1$ in the case of  degenerate dark fermions with the parameters specified  in
Table \ref{tbl:Degenerate}.}\label{tbl:Z(1)FhFh_coupling}
\begin{center}
\begin{tabular}{c|c|cccc}
\hline\hline
$n_F$ & $z_L$ & 
$g_{F^+ L}^{Z^{(1)}}$ &
$g_{F^+ R}^{Z^{(1)}}$ &
$g_{F^0 L}^{Z^{(1)}}$ &
$g_{F^0 R}^{Z^{(1)}}$ 
\\
\hline
$4$ & $10^8$ &
    $0.00$ & $1.25$ & $-0.02$ & $-2.39$ 
    \\
    & $10^6$ &
    $0.03$ & $1.10$ & $-0.06$ & $-2.05$ 
    \\
    & $10^5$ &
    $0.04$ & $1.00$ & $-0.08$ & $-1.87$ 
    \\
    & $10^4$ &
    $0.06$ & $0.90$ & $-0.12$ & $-1.67$ 
    \\
\hline
\hline
\end{tabular}
\end{center}
\end{table}

%%%%%%%%%%%%%%%%%%%%%%%%%%%%%%%
%\clearpage
\paragraph{$Z_R$ boson}
Wave functions of the $Z_R$-tower are given by
\begin{eqnarray}
\begin{pmatrix}
h_{Z_R^{(n)}}^L \\ h_{Z_R^{(n)}}^R \\ (g_B/g_A) h_{Z_R^{(n)}}^B
\end{pmatrix}
&=& \frac{1}{\sqrt{1 + (1 + 2t_\phi^2) \cos^2\theta_H}} \frac{1}{\sqrt{r_{Z_R^{(n)}}}}
\begin{pmatrix}
\frac{1-\cos\theta_H}{\sqrt{2}} \\ \frac{-1-\cos\theta_H}{\sqrt{2}} \\ \sqrt{2} t_\phi^2 \cos\theta_H 
\end{pmatrix} C(z),
\nonumber\\
r_{Z_R^{(n)}} &=& \int_1^{z_L}\frac{dz}{kz} C(z)^2,
\label{eq:Z_R-wavefunc}
\end{eqnarray}
where $C(z) = C(z;\lambda_{Z_R^{(n)}})$ and $\lambda_{Z_R^{(n)}}$ satisfy
$C(1;\lambda_{Z_R^{(n)}}) = 0$.
Hence the $Z_R^{(n)}\bar{F}F$ couplings are given by
\begin{eqnarray}
{\cal L}_{4D} &\supset& Z_{R\mu}^{(n)}
\sum_{c=+,0} 
[ g_{F^c L}^{Z_R^{(n)}} \bar{F}^{c}_L \gamma^\mu F^{c}_L 
+ g_{F^c R}^{Z_R^{(n)}} \bar{F}^{c}_R \gamma^\mu F^{c}_R]
\nonumber\\
&=& Z_{R\mu}^{(n)} \frac{g_w \sqrt{L}}{\sqrt{2}\sqrt{1 + \frac{\cos^2\theta_H}{\cos2\theta_W}}
\sqrt{r_{Z_R^{(n)}}}} 
\sum_{c=+,0} \bar{F}^{c}_L \gamma^\mu F^{c}_L
\int_1^{z_L} dz \, C(z) 
\nonumber\\&&
\qquad \times \biggl[  I_3^{(c)}
\biggl\{
-\cos\theta_H [ |f_{lL}|^2 + |f_{rL}|^2]
 + [|f_{lL}|^2 - |f_{rL}|^2]
\biggr\}
\nonumber\\&&
\qquad + 2Q_X \frac{\sin^2\theta_W}{\cos2\theta_W}  \cos\theta_H [ |f_{lL}|^2 + |f_{rL}|^2]
\biggl] + (L \to R). \label{FZRcoupling}
\end{eqnarray}
We note that unlike the case of the $Z$ boson the $Z_R F\bar{F}$ couplings, where $F$ obeys the b.c. $\etaF=+1$, are not suppressed even if  $\theta_H \to 0$.

In Tables~\ref{tbl:ZRFF-coupling}, \ref{tbl:Z_RFFcoupling-bc} and \ref{tbl:Z_RFFcoupling-ND},
we have summarized the $Z_R \bar{F}F$ couplings.
\begin{table}[tbp]
\caption{%
The left- and right-handed couplings in the unit of $g_w$  of $F$ to $Z_R^{(1)}$ in (\ref{FZRcoupling}) 
with b.c. $\etaF=+1$ in the case of  degenerate dark fermions with the parameters specified  in 
Table \ref{tbl:Degenerate}.}
\label{tbl:ZRFF-coupling}
\begin{center}
\begin{tabular}{c|c|ccccc}
\hline\hline
$n_F$ & $z_L$ &
$m_{Z_R(1)}$ &
$g_{F^+ L}^{Z_R^{(1)}}$ & 
$g_{F^+ R}^{Z_R^{(1)}}$ &
$g_{F^0 L}^{Z_R^{(1)}}$ &
$g_{F^0 R}^{Z_R^{(1)}}$ 
\\
&& [TeV] 
\\ 
\hline
$3$ & $10^8$ & 
    $2.34$ &
    $-0.09$ & $-1.05$ & $0.25$ & $2.55$ 
\\
    & $10^6$ &
    $4.06$ &
    $-0.13$ & $-0.90$ & $0.34$ & $2.23$
\\
    & $10^5$ & 
    $5.59$ &
    $-0.16$ & $-0.82$ & $0.42$ & $2.06$
\\
    & $2\times10^4$ &
    $7.05$ &
    $-0.20$ & $-0.77$ & $0.51$ & $1.93$
    \\
\hline
$4$ & $10^8$ & 
    $2.37$ &
    $-0.07$ & $-1.05$ & $0.18$ & $2.54$
\\
    & $10^6$ &
    $4.12$ &
    $-0.10$ & $-0.89$ & $0.24$ & $2.22$ 
\\
    & $10^5$ &
    $5.70$ &
    $-0.11$ & $-0.82$ & $0.29$ & $2.04$
\\
    & $3\times10^4$ &
    $6.74$ &
    $-0.12$ & $-0.78$ & $0.32$ & $1.94$
\\
    & $10^4$ &
    $7.92$ &
    $-0.14$ & $-0.73$ & $0.35$ & $1.84$
    \\
\hline
$6$ & $10^8$ & 
    $2.42$ &
    $-0.04$ & $-1.05$ & $0.12$ & $2.54$ 
\\
    & $10^6$ &
    $4.20$ &
    $-0.06$ & $-0.89$ & $0.16$ & $2.21$
\\
    & $10^5$ &
    $5.78$ &
    $-0.07$ & $-0.81$ & $0.18$ & $2.03$ 
\\
    & $10^4$ &
    $8.11$ &
    $-0.08$ & $-0.73$ & $0.21$ & $1.83$   
\\
\hline\hline
\end{tabular}
\end{center}
\end{table}
\begin{table}[tbp]
\caption{%
The left- and right-handed couplings in the unit of $g_w$  of $F$ to $Z_R^{(1)}$ in (\ref{FZRcoupling}) 
with b.c. $\etaF=-1$ in the case of  degenerate dark fermions with the parameters specified  in 
Table \ref{tbl:Degenerate}.}\label{tbl:Z_RFFcoupling-bc}
\begin{center}
\begin{tabular}{c|c|cccccccc}
\hline\hline
$n_F$ & $z_L$ &
$g_{F^+ L}^{Z_R^{(1)}}$ & 
$g_{F^+ R}^{Z_R^{(1)}}$ &
$g_{F^0 L}^{Z_R^{(1)}}$ & 
$g_{F^0 R}^{Z_R^{(1)}}$ &
\\ 
\hline
$4$ & $10^8$ & 
    $0.05$ & $0.80$ & $0.07$ & $0.69$
\\
    & $10^6$ &
    $0.07$ & $0.68$ & $0.08$ & $0.65$
\\
    & $10^5$ &
    $0.08$ & $0.62$ & $0.09$ & $0.61$
\\  
    & $3\times10^4$ &
    $0.09$ & $0.58$ & $0.10$ & $0.58$ &
\\
    & $10^4$ &
    $0.11$ & $0.55$ & $0.11$ & $0.55$
\\
\hline\hline
\end{tabular}
\end{center}
\end{table}
\begin{table}[tbp]
\caption{%
The left- and right-handed couplings in the unit of $g_w$  of $F$ to $Z_R^{(1)}$ in (\ref{FZRcoupling}) 
with b.c. $\etaF=+1$ in the case of  non-degenerate $(\nlight,\nheavy)=(1,3)$ dark fermions with the parameters 
specified  in Table \ref{tbl:diff_nf4}.}\label{tbl:Z_RFFcoupling-ND}
\begin{center}
\begin{tabular}{c|c|cccccccc}
\hline\hline
$\Delta c_F$ & $z_L$ &
$g_{F^+ L}^{Z_R^{(1)}}$ & 
$g_{F^+ R}^{Z_R^{(1)}}$ &
$g_{F^0 L}^{Z_R^{(1)}}$ & 
$g_{F^0 R}^{Z_R^{(1)}}$ &
\\ 
\hline
$0.04$ 
    & $10^6$ &
    $-0.08$ & $-0.89$ & $0.20$ & $2.22$
    \\
    & $10^5$ & 
    $-0.10$ & $-0.81$ & $0.25$ & $2.03$
\\
    & $3\times10^4$ &
    $-0.11$ & $-0.77$ & $0.28$ & $1.93$
\\
    & $10^4$ &
    $-0.13$ & $-0.73$ & $0.32$ & $1.84$    \\
\hline
$0.06$
    & $10^6$ &
    $-0.07$ & $-0.89$ & $0.18$ & $2.21$
    \\  
    & $10^5$ & 
    $-0.09$ & $-0.81$ & $0.23$ & $2.03$
\\
    & $3\times10^4$ &
    $-0.10$ & $-0.77$ & $0.26$ & $1.93$
\\
    & $10^4$ &
    $-0.12$ & $-0.73$ & $0.30$ & $1.83$\\
\hline\hline
\end{tabular}
\end{center}
\end{table}

\vskip 4.cm  % <--- may be deleted.
\paragraph{$A^{\hat{4}}$ boson}
Diagonal  $\bar{F}^{(n)}F^{(n)} A_\mu^{\hat{4}}$ couplings  vanish, 
because one finds for the left-hand couplings 
\begin{eqnarray}
\lefteqn{
\bar{\Psi}^{(m)}_{FL} \gamma^\mu T^{\hat{4}} \Psi^{(n)}_{FL}
}\nonumber\\
&=& kz^4 \frac{1}{2\sqrt{2}}
\bar{F}_{L}^{(m)}\gamma^\mu F_{L}^{(n)}
(f_{lL}^{(m)*} f_{rL}^{(m)*}) \begin{pmatrix} & 1_2 \\ 1_2 & \end{pmatrix}
 \begin{pmatrix} f_{lL}^{(n)} \\ f_{rL}^{(n)} \end{pmatrix},
\nonumber\\
&\propto&
 \biggl\{
   S_L(1,\lambda_m) C_L(z,\lambda_m)
   C_L(1,\lambda_n) S_L(z,\lambda_n)
 -
 ( \lambda_m \leftrightarrow \lambda_n)
 \biggr\},
\end{eqnarray}
and  a similar relation for right-handed couplings.

%\clearpage
%%%%%%%%%%%%%%%%%%%%%%%%%%%%%%%%%%%%%%%%%%%%%%%%%%%%%%%%%%%%%%%%
\subsubsection{Couplings to $W$ and $W_R$ bosons}\label{apdx-WFF-couplings}
The building-blocks for $W\bar{F}F$ and $W_{R}\bar{F}F$ couplings are 
\begin{eqnarray}
\bar{\Psi}^{(n)}_F T^{+_L} \gamma^\mu \Psi^{(m)}_F
&=& \frac{kz^4}{2} f_{lL}^{(n)*} f_{lL}^{(m)} 
[\bar{F}^{+(n)}_{L}\gamma^\mu F^{0(m)}_{L} - \bar{F}^{+(n)}_{L}\gamma^\mu  F^{0(m)}_{L}]
 + (F_L, f_{lL}\to F_R,f_{lR}),
\nonumber\\
\bar{\Psi}^{(n)}_F T^{+_R} \gamma^\mu \Psi^{(m)}_F
&=& \frac{kz^4}{2} f_{rL}^{(n)*} f_{rL}^{(m)} 
[\bar{F}^{+(n)}_{L}\gamma^\mu  F^{0(m)}_{L} - \bar{F}^{+(n)}_{L}\gamma^\mu  F^{0(m)}_{L}]
 + (F_L,f_{rL}\to F_R,f_{rR}),
\nonumber\\
\bar{\Psi}^{(n)}_F T^{\hat{+}} \gamma^\mu \Psi^{(m)}_F
&=& \frac{kz^4}{2} i [f_{lL}^{(n)*} f_{rL}^{(m)} - f_{rL}^{(n)*} f_{rL}^{(m)}] 
[\bar{F}^{+(n)}_{L}\gamma^\mu  F^{0(m)}_{L} - \bar{F}^{+(n)}_{L}\gamma^\mu  F^{0(m)}_{L}]
\nonumber\\&& 
 + (F_L,f_{lL}, f_{rL} \to F_R, f_{lR}, f_{rR}).
\end{eqnarray}
In the followings we summarize $W^- \bar{F}^{0(1)} F^{+(1)}$ and 
$W^-_{R} \bar{F}^{0(1)} F^{+(1)}$ ($m=n=1$)  couplings.

%%%%%%%%%%%%%%%%%%%%%%
\paragraph{$W$ boson}
Wave functions of the $W$-tower are
 \begin{eqnarray}
 \begin{pmatrix}
 h_{W^{(n)}}^L \\ h_{W^{(n)}}^R \\ \hat{h}_{W^{(n)}}
\end{pmatrix}
&=& \frac{1}{\sqrt{r_{W^{(n)}}}} \begin{pmatrix}
\frac{1+\cos\theta_H}{\sqrt{2}} C(z) \\
\frac{1-\cos\theta_H}{\sqrt{2}} C(z) \\
 -\sin\theta_H \hat{S}(z)\end{pmatrix},
\nonumber\\
r_{W^{(n)}} &=& \int_1^{z_L} \frac{dz}{kz} 
\left\{ (1 + \cos^2\theta_H) C(z)^2 + \sin^2\theta_H \hat{S}(z)^2  \right\},
\end{eqnarray}
where $C(z) = C(z;\lambda_{W^{(n)}})$, $\hat{S}(z) = \hat{S}(z;\lambda_{W^{(n)}})$
and $\lambda_{W^{(n)}}$ satisfies
\begin{eqnarray}
2 S(z;\lambda_{W^{(n)}}) C(1;\lambda_{W^{(n)}}) + \lambda_{W^{(n)}} \sin^2\theta_H &=& 0.
\end{eqnarray}
$W^{(0)}$ is the $W$-boson whose mass is given by $m_W = k\cdot\lambda_{W^{(0)}}$. 
The   couplings
\begin{eqnarray}
{\cal L}_{4D} &\supset& W_\mu^{-(n)} \left[
 g^{W^{(n)}}_{F L} \bar{F}^{0}_L \gamma^\mu F^{+}_L
+g^{W^{(n)}}_{F R} \bar{F}^{0}_R \gamma^\mu F^{+}_R \right] + (h.c.),
\nonumber
\end{eqnarray}
are given by
\begin{eqnarray}
g^{W^{(n)}}_{F L} &=& \frac{g_w}{2\sqrt{2}} \frac{\sqrt{L}}{\sqrt{r_{W^{(n)}}}}
\int_1^{z_L}dz 
\biggl\{  C(z) [ 
  (1+\cos\theta_H) |f_{lL}|^2
+ (1-\cos\theta_H) |f_{rL}|^2 ]
\nonumber\\&&
- \sin\theta_H \hat{S}(z) i [f_{lL}^* f_{rL} - f_{rL}^* f_{lL}] \biggr\}, \label{FWcoupling}
\end{eqnarray}
and $g^{W^{(n)}}_{F R}$ is obtained by replacements $f_{l(r)L} \to f_{l(r)R}$.
We note that for the dark fermion obeying b.c. $\etaF=+1$ these couplings are suppressed by $\sin^2(\theta_H/2)$, 
because $f_{lL} \propto \sin(\theta_H/2)$.
The $WF\bar{F}$ and $W^{(1)} F\bar{F}$ couplings are summarized in 
Tables~\ref{tbl:WFF-coupling} and \ref{tbl:WFhFh-coupling}.
%In Tables~\ref{tbl:WFF-coupling} and \ref{tbl:WFhFh-coupling}, we have summarized
%$VF\bar{F}$ where $V$ denotes $W$ boson and the first KK $W$ boson $W^{(1)}$. 

%%%%%%%%%%%%%%%%%%%%%%%%
\paragraph{$W_R$ boson}
Wave functions of $W_R$-tower are given by
\begin{eqnarray}
\begin{pmatrix} h_{W_R^{(n)}}^L  \\ h_{W_R^{(n)}}^R \end{pmatrix}
&=& 
\frac{1}{\sqrt{1 + \cos^2\theta_H}} 
\frac{1}{\sqrt{r_{W_R^{(n)}}}}
\begin{pmatrix} \frac{-\cos\theta_H+1}{\sqrt{2}} \\ \frac{-1-\cos\theta_H}{\sqrt{2}} \end{pmatrix} C(z),
\nonumber\\
r_{W_R^{(n)}} &=& \int_1^{z_L} \frac{dz}{kz} C(z)^2,
\end{eqnarray}
where $C(z) \equiv C(z;\lambda_{W_R^{(n)}})$ and $\lambda_{W_R^{(n)}}$
is defined by
$C(1;\lambda_{W_R^{(n)}}) = 0$.
In an analogous way to the $W$ boson, we obtain the couplings
\begin{eqnarray}
{\cal L}_{4D} &\supset& W_{R\mu}^{-(n)} \left[
  g^{W_R^{(n)}}_{F L} \bar{F}^{0}_L \gamma^\mu F^{+}_L
+ g^{W_R^{(n)}}_{F R} \bar{F}^{0}_R \gamma^\mu F^{+}_R \right]
+ (h.c.),
\nonumber\\
g^{W_R^{(n)}}_{F L} 
&=& \frac{g_w}{2\sqrt{2}} \frac{\sqrt{L}}{\sqrt{r_{W_R^{(n)}}}\sqrt{1+\cos^2\theta_H}}
\nonumber\\&& \times
\int_1^{z_L} dz
C(z)\biggl\{ (1- \cos\theta_H) |f_{lL}|^2 +
(-1-\cos\theta_H) |f_{rL}|^2] \biggr\}, \label{FWRcoupling}
\end{eqnarray}
and $g^{W_R^{(n)}}_{F R}$ is obtained by replacing $f_{l(r)L}$ with $f_{l(r)R}$.
The  $W_R^{(1)} F\bar{F}$ couplings are summarized in 
Tables~\ref{tbl:WFF-coupling} and \ref{tbl:WFhFh-coupling}.

\begin{table}[tbp]
\caption{%
The left- and right-handed couplings $\bar{F}^0 F^+ V^-$ (in the unit of $g_w/\sqrt{2}$)  
of $F$  to a charged vector boson $V^-$  ($V = W$, $W^{(1)}$ and $W_R^{(1)}$) in (\ref{FWcoupling})  
with b.c. $\etaF=+1$ in the case of  degenerate dark fermions with the parameters specified  in 
Table \ref{tbl:Degenerate}.}
\label{tbl:WFF-coupling}
\begin{center}
\begin{tabular}{c|c|cc|ccc|ccc}
\hline\hline
$n_F$ & $z_L$ & 
$g_{F L}^W$ & 
$g_{F R}^W$ & 
$m_{W^{(1)}}$ &
$g_{F L}^{W^{(1)}}$ & 
$g_{F R}^{W^{(1)}}$ & 
$m_{W_R^{(1)}}$ &
$g_{F L}^{W_R^{(1)}}$ & 
$g_{F R}^{W_R^{(1)}}$ 
\\
&& $\times10^3$ & $\times10^3$ & [TeV] &$\times10^3$ & $\times10^3$ & [TeV] && \\
\hline 
$3$ & $10^8$ &
    $7.0$ & $39.8$ &
    $2.42$ & $61.9$ & $136$ &
    $2.34$ & $-0.41$ & $-3.11$
\\
    & $10^6$ &
    $3.8$ & $12.2$ &
    $4.25$ & $26.4$ & $28.0$ &
    $4.06$ & $-0.57$ & $-2.66$ 
\\
    & $2\times10^4$ &
    $7.2$ & $10.2$ &
    $7.54$ & $31.5$ & $7.8$ &
    $7.05$ & $-0.84$ & $-2.28$
\\
\hline
$4$ & $10^8$ & 
    $4.4$ & $35.9$ &
    $2.45$ & $40.2$ & $132.2$ &
    $2.37$ & $-0.30$ & $-3.10$ 
\\
    & $10^6$ &
    $2.0$ & $9.7$ &
    $4.32$ & $15.0$ & $26.8$ &
    $4.12$ & $-0.41$ & $-2.65$
\\
    & $10^4$ &
    $1.1$ & $2.7$ & 
    $8.52$ & $6.1$ & $3.8$ &
    $7.92$ & $-0.59$ & $-2.18$
    \\
\hline
$6$ & $10^8$ & 
    $2.5$ & $32.6$ &
    $2.50$ & $23.4$ & $127.3$ &
    $2.42$ & $-0.19$ & $-3.10$
\\
    & $10^6$ &
    $1.0$ & $8.4$ &
    $4.40$ & $8.0$ & $25.9$ &
    $4.20$ & $-0.26$ & $-2.64$
\\
    & $10^4$ &
    $0.4$ & $1.7$ &
    $8.68$ & $2.3$ & $3.6$ &
    $8.07$ & $-0.36$ & $-2.16$
\\
\hline\hline
\end{tabular}
\end{center}
\end{table}
\begin{table}[tbp]
\caption{The left- and right-handed couplings $\bar{F}^0 F^+ V^-$ (in the unit of $g_w/\sqrt{2}$)  
of $F$  to a charged vector boson $V^-$  ($V = W$, $W^{(1)}$ and $W_R^{(1)}$) in (\ref{FWcoupling})  
with b.c. $\etaF=-1$ in the case of  degenerate dark fermions with the parameters specified  in 
Table \ref{tbl:Degenerate}.}
\label{tbl:WFhFh-coupling}
\begin{center}
\begin{tabular}{c|c|cccccc}
\hline\hline
$n_F$ & $z_L$ & 
$g^{W}_{F L}$ &
$g^{W}_{F R}$ &
$g^{W^{(1)}}_{F L}$ &
$g^{W^{(1)}}_{F R}$ &
$g^{W_R^{(1)}}_{F L}$ &
$g^{W_R^{(1)}}_{F R}$  
\\
\hline
$4$ & $10^8$ &
    $0.997$ & $0.966$ & 
    $0.04$ & $4.15$ &
    $-0.019$ & $0.099$ 
    \\
    & $10^6$ &
    $0.998$ & $0.991$ &
    $0.10$ & $3.59$  &
    $-0.008$ & $0.020$ 
    \\
    & $10^4$ &
    $0.999$ & $0.998$ &
    $0.21$ & $2.93$ &
    $-0.004$ & $0.003$
    \\
\hline
\end{tabular}
\end{center}
\end{table}

%\clearpage
%%%%%%%%%%%%%%%%%%%%%%%%%%%%%%%%%%%%%%%%%%%%%%%%%%%%%%%%%%%%%%%%%%%%%%%%%%%%%%%%%%
\section{$VW^+W^-$ vector-boson couplings}\label{sec:apdx-boson-boson-couplings}

In terms of the wave functions for the $W$ boson and other neutral vector bosons $V=Z,Z_R,A^\gamma,A^{\hat{4}}$, one can read the $VW^+ W^-$ couplings from the relation
\begin{eqnarray}
\lefteqn{
g_A \int_1^{z_L} \frac{dz}{kz} \tr \partial_\mu \hat{V}_\nu [\hat{W}_\rho, \hat{W}_\sigma](x,z)
}\nonumber\\
&=& \sum_{n,r,s} g_{V^{(n)}W^{+(r)}W^{-(s)}} (\partial_\mu V_\nu^{(n)}) W_\rho^{+(r)} W_\sigma^{-(s)}(x).
\end{eqnarray}
Hereafter we summarize the formulas for $V^{(n)}W^{+}W^{-}$ couplings.
Numerically computed values of the $VW^+W^-$ ($V=Z,Z^{(1)},Z_R^{(1)}$ and $\gamma^{(1)}$) couplings 
are summarized in Table.~\ref{tbl:ZWW-coupling}.
These couplings depend sensitively on $z_L$ and $\theta_H$, but very weakly on $n_F$, 
thanks to the universality relations in the model.\cite{FHHOS2013, LHCsignals}

\paragraph{$\gamma^{(n)}W^{+}W^{-}$ coupling}
The $\gamma^{(n)} W^+ W^-$ coupling is given by
\begin{eqnarray}
g_{\gamma^{(n)}WW} &=& g_w \sqrt{L} \int_1^{z_L} \frac{dz}{kz}
\biggl\{
 h_{\gamma^{(n)}}^L \left[ (h_{W}^L)^2 + \frac{(\hat{h}_{W})^2}{2} \right]
+h_{\gamma^{(n)}}^R \left[ (h_{W}^R)^2 + \frac{(\hat{h}_{W})^2}{2} \right]
\biggr\}.
\end{eqnarray}
In particular, for the photon $\gamma=\gamma^{(0)}$ we obtain
\begin{eqnarray}
g_{\gamma WW} = e  \quad \text{(electromagnetic coupling)},
\end{eqnarray}
and for KK excited photons ($n \ne 0$) we have
\begin{eqnarray}
g_{\gamma^{(n)}WW}
&=& e \sqrt{L} \int_1^{z_L} \frac{dz}{kz} \frac{C(z,\lambda_{\gamma^{(n)}})}{\sqrt{r_{\gamma^{(n)}}}} [ (h_{W}^L)^2 + (h_{W}^R)^2 + (\hat{h}_{W})^2].
\end{eqnarray}

\paragraph{$Z^{(n)}W^{+}W^{-}$ coupling}
\begin{eqnarray}
g_{Z^{(n)}WW} &=& g_w \sqrt{L} \int_1^{z_L} \frac{dz}{kz}
\biggl\{ 
 h_{Z^{(n)}}^L \left[ (h_{W}^L)^2 +  \frac{(\hat{h}_{W})^2}{2}\right] 
+h_{Z^{(n)}}^R \left[ (h_{W}^R)^2 +  \frac{(\hat{h}_{W})^2}{2}\right] 
\nonumber\\&& \phantom{MMMMMM}
 + \hat{h}_{Z^{(n)}} ( h_{W}^L + h_{W}^R) \hat{h}_{W}  \biggr\}. \label{WWZcoupling}
\end{eqnarray}

\paragraph{$Z_R^{(n)}W^{+} W^{-}$ coupling}
\begin{eqnarray}
g_{Z_R^{(n)}WW} &=& g_w \sqrt{L} \int_1^{z_L} \frac{dz}{kz} 
\left\{
 h_{Z_R^{(n)}}^L \left[(h_{W}^L)^2 + \frac{(\hat{h}_{W})^2}{2}\right]
+h_{Z_R^{(n)}}^R \left[(h_{W}^R)^2 + \frac{(\hat{h}_{W})^2}{2}\right]
\right\}.
\nonumber\\ \label{WWZRcoupling}
\end{eqnarray}
We note that this coupling is suppressed by $\sin^2\theta_H$ because
\begin{eqnarray}
h_{Z_R}^L,
h_{W}^R \propto \sin^2(\theta_H/2),
\quad
\hat{h}_{W} \propto \sin\theta_H. \nonumber
\end{eqnarray}

\paragraph{$A^{\hat{4}(n)}W^{+} W^{-}$ coupling}
$A^{\hat{4}(n)}W^{+(r)}W^{-(s)}$ coupling vanishes when $r=s$.
In particular, for $r=s=0$ we obtain
\begin{eqnarray}
g_{A^{\hat{4}(n)}WW} &=& 0.
\end{eqnarray}

\paragraph{$W_R^{+(n)} W^{-} Z$ coupling}
\begin{eqnarray}
g_{W_R^{(n)} W Z} &=& g_w \sqrt{L} \int_{1}^{z_L} \frac{dz}{kz}
\left[h_{W^{(n)}_R}^L h_{W}^L h_{Z}^L + h_{W^{(n)}_R}^R h_{W}^R h_{Z}^R
+ \frac{1}{2} (h_{W_R^{(n)}}^L + h_{W_R^{(n)}}^R) \hat{h}_{W} \hat{h}_{Z} 
\right].
\nonumber\\
\end{eqnarray}
This coupling is suppressed by $\sin^2(\theta_H/2)$ because
$h_{W_R^{(n)}}^L h_{W^{(n)}}^L,\, h_{W_R^{(n)}}^R h_{W^{(n)}}^R,\, \hat{h}_{W^{(n)}}\hat{h}_{Z^{(n)}} \propto \sin^2(\theta_H/2)$.

%%%%%%
\begin{table}[tbp]
\caption{%
Triple vector-boson couplings $VW^+W^-$ with 
$V = Z, Z^{(1)}, Z_R^{(1)}$ (\ref{WWZcoupling}), (\ref{WWZRcoupling}) in unit of $g_w$ 
and $\gamma^{(1)}W^+W^-$ 
in unit of the electric charge $e$.}\label{tbl:ZWW-coupling}
\begin{center}
\begin{tabular}{c|c|cccc}
\hline\hline
$n_F$ & $z_L$ & 
$g_{WWZ}$  & $g_{WWZ^{(1)}}\times10^2$ & $g_{WWZ_R^{(1)}}\times10^2$ & $g_{WW\gamma^{(1)}}\times10^2$ 
\\
\hline
$4$ & $10^8$ & %$0.355$ &
    $0.811$ & 
    $1.506$ &
    $0.391$ &
    $-0.417$
\\
    & $10^6$ & %$0.1742$ &
    $0.861$ &
    $0.459$ &
    $0.114$ &
    $-0.115$
\\
    & $10^5$ & %$0.1153$ &
    $0.870$ &
    $0.225$ &
    $0.055$ &
    $-0.054$
\\
    & $10^4$ & %$0.07372$ &
    $0.874$ &
    $0.105$ &
    $0.025$ &
    $-0.024$
    \\
\hline\hline
\end{tabular}
\end{center}
\end{table}

\end{appendix}

\vskip 1cm

\renewenvironment{thebibliography}[1]
         {\begin{list}{[$\,$\arabic{enumi}$\,$]}  % {\arabic{enumi}.}
         {\usecounter{enumi}\setlength{\parsep}{0pt}
          \setlength{\itemsep}{0pt}  \renewcommand{\baselinestretch}{1.2}
          \settowidth
         {\labelwidth}{#1 ~ ~}\sloppy}}{\end{list}}

% A useful Journal macro
%\def\jnl#1#2#3#4{{#1}{\bf #2} (#4) #3}
\def\jnl#1#2#3#4{{#1}{\bf #2},  #3 (#4)}

\def\Zphys{{\em Z.\ Phys.} }
\def\jssc{{\em J.\ Solid State Chem.\ }}
\def\jpsJ{{\em J.\ Phys.\ Soc.\ Japan }}
\def\ptps{{\em Prog.\ Theoret.\ Phys.\ Suppl.\ }}
\def\PTP{{\em Prog.\ Theoret.\ Phys.\  }}
\def\PTEP{{\em Prog.\ Theor.\ Exp.\ Phys.\  }}
\def\JMP{{\em J. Math.\ Phys.} }
\def\NPB{{\em Nucl.\ Phys.} B}
\def\NP{{\em Nucl.\ Phys.} }
\def\PLB{{\it Phys.\ Lett.} B}
\def\PL{{\em Phys.\ Lett.} }
\def\PRL{\em Phys.\ Rev.\ Lett. }
\def\PRB{{\em Phys.\ Rev.} B}
\def\PRD{{\em Phys.\ Rev.} D}
\def\PRe{{\em Phys.\ Rep.} }
\def\AP{{\em Ann.\ Phys.\ (N.Y.)} }
\def\RMP{{\em Rev.\ Mod.\ Phys.} }
\def\ZPC{{\em Z.\ Phys.} C}
\def\SCI{\em Science}
\def\CMP{\em Comm.\ Math.\ Phys. }
\def\MPLA{{\em Mod.\ Phys.\ Lett.} A}
\def\IJMPA{{\em Int.\ J.\ Mod.\ Phys.} A}
\def\IJMPB{{\em Int.\ J.\ Mod.\ Phys.} B}
\def\EPJC{{\em Eur.\ Phys.\ J.} C}
\def\PR{{\em Phys.\ Rev.} }
\def\JHEP{{\em JHEP} }
\def\JCAP{{\em JCAP} }
\def\cmp{{\em Com.\ Math.\ Phys.}}
\def\JPA{{\em J.\  Phys.} A}
\def\JPG{{\em J.\  Phys.} G}
\def\NJP{{\em New.\ J.\  Phys.} }
\def\CQG{\em Class.\ Quant.\ Grav. }
\def\ATMP{{\em Adv.\ Theoret.\ Math.\ Phys.} }
\def\ibid{{\em ibid.} }

\renewenvironment{thebibliography}[1]
         {\begin{list}{[$\,$\arabic{enumi}$\,$]}  % {\arabic{enumi}.}
         {\usecounter{enumi}\setlength{\parsep}{0pt}
          \setlength{\itemsep}{0pt}  \renewcommand{\baselinestretch}{1.2}
          \settowidth
         {\labelwidth}{#1 ~ ~}\sloppy}}{\end{list}}

%%%%  title of reference  %%%%
\def\reftitle#1{{\it ``#1'', }}    %to print.

\leftline{\bf References}

%%%%%%%%%%%%% BIBLIOGRAPHY (US) %%%%%%%%%%%%%%%%%%%%


\begin{thebibliography}{99}
%%%%%%%%%%%%%%%%%%%%%%%%%%%%%%%%%%%%%%%%%%%%%%%

\bibitem{Aad:2012tfa} 
G.~Aad {\it et al.}  [ATLAS Collaboration],
\reftitle{Observation of a new particle in the search for the Standard Model Higgs boson with the ATLAS detector at the LHC}
\jnl{\PLB}{716}{1}{2012}.
%  [arXiv:1207.7214 [hep-ex]].

%\cite{Chatrchyan:2012ufa}
\bibitem{Chatrchyan:2012ufa} 
S.~Chatrchyan {\it et al.}  [CMS Collaboration],
\reftitle{Observation of a new boson at a mass of 125 GeV with the CMS experiment at the LHC}
\jnl{\PLB}{716}{30}{2012}.
%  [arXiv:1207.7235 [hep-ex]].


%%%%%%%


\bibitem{YH1}
Y.~Hosotani,
\reftitle{Dynamical Mass Generation by Compact Extra Dimensions}
\jnl{\PLB}{126}{309}{1983};
\reftitle{Dynamics of Nonintegrable Phases and Gauge Symmetry Breaking}
\jnl{\AP}{190}{233}{1989}.


\bibitem{Davies1}
A.~T.~Davies and A.~McLachlan,
\reftitle{Gauge group breaking by Wilson loops}
\jnl{\PLB}{200}{305}{1988};
\reftitle{Congruency class effects in the Hosotani model}
\jnl{\NPB}{317}{237}{1989}.



\bibitem{Hatanaka1998}
H.\ Hatanaka, T.\ Inami and C.S.\ Lim,
\reftitle{The gauge hierarchy problem and higher dimensional gauge theories}
\jnl{\MPLA}{13}{2601}{1998}.
%[hep-th/9805067].

\bibitem{Nomura1}
G.\ Burdman and Y.\ Nomura, 
\reftitle{Unification of Higgs and Gauge Fields in Five Dimensions}
\jnl{\NPB}{656}{3}{2003}. 

\bibitem{Csaki2}
C.\ Csaki, C.\ Grojean and H.\ Murayama, 
\reftitle{Standard Model Higgs From Higher Dimensional Gauge Fields}
\jnl{\PRD}{67}{085012}{2003}.

\bibitem{Lim2013} 
C.~S.~Lim,
\reftitle{The Higgs Particle and Higher-Dimensional Theories}
\jnl{\PTEP}{2014}{02A101}{2014}.
%arXiv:1308.5579 [hep-ph].

%%%%%%%%%%%%%%%


\bibitem{ACP}
K.~Agashe, R.~Contino and A.~Pomarol,
\reftitle{The Minimal Composite Higgs Model}
\jnl{\NPB}{719}{165}{2005}.
%[hep-ph/0412089].

\bibitem{MSW}
A.~D.~Medina, N.~R.~Shah and C.~E.~M.~Wagner, 
\reftitle{Gauge-Higgs Unification and Radiative Electroweak Symmetry Breaking in
Warped Extra Dimensions}
\jnl{\PRD}{76}{095010}{2007}.
%[arXiv:0706.1281 [hep-ph]].

\bibitem{HOOS} 
Y.~Hosotani, K.~Oda, T.~Ohnuma and Y.~Sakamura,
\reftitle{Dynamical Electroweak Symmetry Breaking in $SO(5) \times U(1)$
Gauge-Higgs Unification with Top and Bottom Quarks}
\jnl{\PRD}{78}{096002}{2008}; 
{\it Erratum}-\jnl{\ibid}{79}{079902}{2009}.
% [arXiv:0806.0480 [hep-ph]].

\bibitem{HNU}
Y.\ Hosotani, S.\ Noda and N.\ Uekusa,
\reftitle{The Electroweak gauge couplings in $SO(5) \times U(1)$ gauge-Higgs unification}
\jnl{\PTP}{123}{757}{2010}.
%[arXiv:0912.1173 [hep-ph]].

\bibitem{HTU1}
Y.\ Hosotani, M.\ Tanaka and N.\ Uekusa,
\reftitle{H parity and the stable Higgs boson in the $SO(5)\times U(1)$ 
gauge-Higgs unification}
\jnl{\PRD}{82}{115024}{2011}.
% [arXiv:1010.6135 [hep-ph]].

\bibitem{HTU2}
Y.\ Hosotani, M.\ Tanaka, and N.\ Uekusa, 
\reftitle{Collider signatures of the $SO(5) \times U(1)$ gauge-Higgs unification}
\jnl{\PRD}{84}{075014}{2011}. 
%\archive{arXiv:1103.6076 [hep-ph]} 

\bibitem{FHHOS2013}
S.~Funatsu, H.~Hatanaka, Y.~Hosotani, Y.~Orikasa,  and T.~Shimotani,
\reftitle{Novel universality and Higgs decay $H \go \gamma \gamma, gg$ 
in the $SO(5) \times U(1)$ gauge-Higgs unification}
\jnl{\PLB}{722}{94}{2013}.
%\jnl{\PLB}{722}{94-99}{2013}.


\bibitem{LHCsignals}
S.~Funatsu, H.~Hatanaka, Y.~Hosotani, Y.~Orikasa and T.~Shimotani,
\reftitle{LHC signals of the $SO(5)\times U(1)$ gauge-Higgs unification}
\jnl{\PRD}{89}{095019}{2014}. 
  [arXiv:1404.2748 [hep-ph]].

%%%%%%%%%%%%%%%%%%%

\bibitem{SH1} 
Y.~Sakamura and Y.~Hosotani,
\reftitle{WWZ, WWH, and ZZH Couplings in the Dynamical Gauge-Higgs Unification in the Warped Spacetime}
\jnl{\PLB}{645}{442}{2007}.
% [arXiv:hep-ph/0607236].

\bibitem{HS2} 
Y.~Hosotani and Y.~Sakamura,
\reftitle{Anomalous Higgs Couplings in the $SO(5)\times U(1)_{B-L}$ Gauge-Higgs
Unification in Warped Spacetime}
\jnl{\PTP}{118}{935}{2007}.
% [arXiv:hep-ph/0703212].

\bibitem{Giudice2007}
G.F. Giudice,  C. Grojean,  A. Pomarol and  R. Rattazzi, 
\reftitle{The Strongly-Interacting Light Higgs}
\jnl{\JHEP}{0706}{045}{2007}.
% [arXiv: hep-ph/0703164].

\bibitem{Sakamura1}
Y.~Sakamura,
\reftitle{Effective theories of gauge-Higgs unification models in warped spacetime}
\jnl{\PRD}{76}{065002}{2007}.
% [arXiv:0705.1334 [hep-ph]].


\bibitem{HK2008} 
Y.~Hosotani and Y.~Kobayashi,
\reftitle{Yukawa Couplings and Effective Interactions in Gauge-Higgs Unification}
%  [arXiv:0812.4782 [hep-ph]].
\jnl{\PLB}{674}{192}{2009}.

\bibitem{Hasegawa:2012sy} 
K.~Hasegawa, N.~Kurahashi, C.~S.~Lim and K.~Tanabe,
\reftitle{Anomalous Higgs Interactions in Gauge-Higgs Unification}
%  [arXiv:1201.5001 [hep-ph]].
\jnl{\PRD}{87}{016011}{2013}.

%%%%%%%%%

\bibitem{Haba:2009a} 
N.~Haba, Y.~Sakamura and T.~Yamashita,
\reftitle{Weak boson scattering in Gauge-Higgs Unification}
%  [arXiv:0904.3177 [hep-ph]].
\jnl{\JHEP}{0907}{020}{2009}.



\bibitem{Lim2007b}
Y.\ Adachi, C.S.\ Lim and N.\ Maru,
\reftitle{Finite anomalous magnetic moment in the gauge-Higgs unification}
\jnl{\PRD}{76}{075009}{2007};
% [arXiv:0707.1735 [hep-ph]];  
% [arXiv:0901.2229 [hep-ph]].
\reftitle{More on the Finiteness of Anomalous Magnetic Moment in the Gauge-Higgs Unification}
\jnl{\PRD}{79}{075018}{2009}.

\bibitem{Wagner2}
M.~Carena, A.~D.~Medina, B.~Panes, N.~R.~Shah and C.~E.~M.~Wagner,
\reftitle{Collider Phenomenology of Gauge-Higgs Unification Scenarios in Warped Extra
Dimensions}
\jnl{\PRD}{77}{076003}{2008}.
% [arXiv:0712.0095 [hep-ph]].


\bibitem{Lim2009}
Y.\ Adachi, C.S.\ Lim and N.\ Maru,
\reftitle{Neutron Electric Dipole Moment in the Gauge-Higgs Unification}
\jnl{\PRD}{80}{055025}{2009}.
% [arXiv:0905.1022 [hep-ph]].


\bibitem{Agashe2010}
K.\ Agashe, A.\ Azatov, T.\ Han, Y.\ Li,  Z.G.\ Si, L.\ Zhu,
% [arXiv:0911.0059 [hep-ph]].
\reftitle{LHC Signals for Coset Electroweak Gauge Bosons in Warped/Composite  PGB Higgs Models}
\jnl{\PRD}{81}{096002}{2010}.

\bibitem{Adachi2010} 
Y.~Adachi, N.~Kurahashi, C.~S.~Lim and N.~Maru,
\reftitle{Flavor Mixing in Gauge-Higgs Unification}
%  [arXiv:1005.2455 [hep-ph]].
\jnl{\JHEP}{1011}{150}{2010}.

\bibitem{Adachi2012} 
Y.~Adachi, N.~Kurahashi, N.~Maru and K.~Tanabe,
\reftitle{$B^0$-$\bar B^0$ Mixing in Gauge-Higgs Unification}
\jnl{\PRD}{85}{096001}{2012}.
%  [arXiv:1112.6062 [hep-ph]].
%\reftitle{CP Violation due to Flavor Mixing in Gauge-Higgs Unification}
% arXiv:1201.2290 [hep-ph].

%\cite{Haba:2012bc}
\bibitem{Haba:2012bc} 
  N.~Haba, K.~Kaneta and S.~Tsuno,
\reftitle{QCD parity violation at LHC in warped extra dimension}
\jnl{\PRD}{87}{095002}{2013}.
%  [arXiv:1211.5411 [hep-ph]].

\bibitem{MaruOkada2013decay} 
N.~Maru and N.~Okada,
\reftitle{Diphoton decay excess and 125 GeV Higgs boson in gauge-Higgs unification}
%  [arXiv:1303.5810 [hep-ph]].
\jnl{\PRD}{87}{095019}{2013};
\reftitle{$H \to Z\gamma$ in gauge-Higgs unification}
\jnl{\PRD}{88}{037701}{2013};
\reftitle{125 GeV Higgs Boson and TeV Scale Colored Fermions in Gauge-Higgs Unification}
arXiv:1310.3348 [hep-ph].


\bibitem{Kakizaki2013} 
M.~Kakizaki, S.~Kanemura, H.~Taniguchi and T.~Yamashita,
\reftitle{Higgs sector as a Probe of Supersymmetric Grand Unification with the Hosotani Mechanism}
\jnl{\PRD}{89}{075013}{2014}.
[arXiv:1312.7575 [hep-ph]].

\bibitem{Adachi:2014wva} 
Y.~Adachi, N.~Kurahashi and N.~Maru,
\reftitle{$\mu \go 3e$ and $\mu \go e$ Conversion in Gauge-Higgs Unification}
  arXiv:1404.4281 [hep-ph].



%%%%%%%%%%%%%

\bibitem{Kolb} 
  E.~W.~Kolb and M.~S.~Turner,
  \reftitle{The Early Universe}
  {\it Front.\ Phys.}\  {\bf 69}, 1 (1990).


\bibitem{JKG} 
G.\ Jungman, M.\ Kamionkowski and K.\ Griest, 
\reftitle{Supersymmetric dark matter}   
\jnl{\PRe}{267}{195}{1996}. 

\bibitem{EFO}
J.\ Ellis, A.\ Ferstl and K.A.\ Olive, 
\reftitle{Re-evaluation of the elastic scattering of supersymmetric dark matter}   
\jnl{\PLB}{481}{304}{2000}. 
 


%\cite{Servant:2002aq}
\bibitem{Servant:2002aq} 
  G.~Servant and T.~M.~P.~Tait,
  \reftitle{Is the lightest Kaluza-Klein particle a viable dark matter candidate?}
  \jnl{\NPB}{650}{391}{2003}.
%  [hep-ph/0206071].

% Second KK in UED
\bibitem{UED-Kakizaki1} 
  M.~Kakizaki, S.~Matsumoto, Y.~Sato and M.~Senami,
  \reftitle{Relic abundance of LKP dark matter in UED model including effects of second KK resonances}
  \jnl{\NPB}{735}{84}{2006}.
  %[hep-ph/0508283].
\bibitem{UED-Kakizaki2} 
  M.~Kakizaki, S.~Matsumoto and M.~Senami,
  \reftitle{Relic abundance of dark matter in the minimal universal extra dimension model}
  \jnl{\PRD}{74}{023504}{2006}.
  %[hep-ph/0605280].
\bibitem{UED-Kakizaki3} 
  G.~Belanger, M.~Kakizaki, and A.~Pukhov,
  \reftitle{Dark matter in UED: The Role of the second KK level}
  \jnl{\JCAP}{1102}{009}{2011}.
  %[hep-ph/0605280].

% Split UED
\bibitem{Split-UED1} 
  S.~C.~Park and J.~Shu,
  \reftitle{Split Universal Extra Dimensions and Dark Matter}
  \jnl{\PRD}{79}{091702}{2009}.
  %[arXiv:0901.0720 [hep-ph]].
\bibitem{Split-UED2} 
  C.~-R.~Chen, M.~M.~Nojiri, S.~C.~Park, J.~Shu and M.~Takeuchi,
  \reftitle{Dark matter and collider phenomenology of split-UED}
  \jnl{\JHEP}{0909}{078}{2009}.
  %[arXiv:0903.1971 [hep-ph]].

%little higgs

%\cite{Perelstein:2006bq}
\bibitem{Perelstein:2006bq} 
  M.~Perelstein and A.~Spray,
  \reftitle{Indirect Detection of Little Higgs Dark Matter}
  \jnl{\PRD}{75}{083519}{2007}.
%  [hep-ph/0610357].
  %%CITATION = HEP-PH/0610357;%%
  %36 citations counted in INSPIRE as of 04 Jun 2014

%\cite{Hooper:2006xe}
\bibitem{Hooper:2006xe} 
  D.~Hooper and G.~Zaharijas,
  \reftitle{Distinguishing Supersymmetry From Universal Extra Dimensions or Little Higgs Models With Dark Matter Experiments}
  \jnl{\PRD}{75}{035010}{2007};
  %[hep-ph/0612137].
  %%CITATION = HEP-PH/0612137;%%
  %20 citations counted in INSPIRE as of 04 Jun 2014
%\cite{Birkedal:2006fz}
%\bibitem{Birkedal:2006fz} 
  A.~Birkedal, A.~Noble, M.~Perelstein and A.~Spray,
  \reftitle{Little Higgs dark matter}
  \jnl{\PRD}{74}{035002}{2006}.
  %[hep-ph/0603077].
  %%CITATION = HEP-PH/0603077;%%
  %134 citations counted in INSPIRE as of 04 Jun 2014

%composite higgs

%\cite{DiazCruz:2007be}
\bibitem{DiazCruz:2007be} 
  J.~L.~Diaz-Cruz,
  \reftitle{Holographic dark matter and Higgs}
  \jnl{\PRL}{100}{221802}{2008}.
  %[arXiv:0711.0488 [hep-ph]].
  %%CITATION = ARXIV:0711.0488;%%
  %10 citations counted in INSPIRE as of 04 Jun 2014\end{thebibliography}

%\cite{Khlopov:2008ty}
\bibitem{Khlopov:2008ty} 
  M.~Y.~.Khlopov and C.~Kouvaris,
  \reftitle{Composite dark matter from a model with composite Higgs boson}
  \jnl{\PRD}{78}{065040}{2008}.
  %[arXiv:0806.1191 [astro-ph]].
  %%CITATION = ARXIV:0806.1191;%%
  %84 citations counted in INSPIRE as of 04 Jun 2014  

%\cite{Chala:2012af}
\bibitem{Chala:2012af} 
  M.~Chala,
  \reftitle{$h \rightarrow \gamma\gamma$ excess and Dark Matter from Composite Higgs Models}
  \jnl{\JHEP}{1301}{122}{2013}.
  %[arXiv:1210.6208 [hep-ph]].
  %%CITATION = ARXIV:1210.6208;%%
  %37 citations counted in INSPIRE as of 04 Jun 2014
  
  % Axion DM
%\cite{Preskill:1982cy}
\bibitem{Preskill:1982cy} 
  J.~Preskill, M.~B.~Wise and F.~Wilczek,
  \reftitle{Cosmology of the Invisible Axion}
  \jnl{\PLB}{120}{127}{1983}.
  %%CITATION = PHLTA,B120,127;%%
  %905 citations counted in INSPIRE as of 19 Jun 2014

%\cite{Abbott:1982af}
\bibitem{Abbott:1982af} 
  L.~F.~Abbott and P.~Sikivie,
  \reftitle{A Cosmological Bound on the Invisible Axion}
  \jnl{\PLB}{120}{133}{1983}.
  %%CITATION = PHLTA,B120,133;%%
  %876 citations counted in INSPIRE as of 19 Jun 2014

%\cite{Dine:1982ah}
\bibitem{Dine:1982ah} 
  M.~Dine and W.~Fischler,
  \reftitle{The Not So Harmless Axion}
  \jnl{\PLB}{120}{137}{1983}.
  %%CITATION = PHLTA,B120,137;%%
  %859 citations counted in INSPIRE as of 19 Jun 2014

\bibitem{Duffy} 
  L.~D.~Duffy and K.~van Bibber,
  \reftitle{Axions as Dark Matter Particles}
  \jnl{\NJP}{11}{105008}{2009}.
  %[arXiv:0904.3346 [hep-ph]].


%axino
%\cite{Covi:1999ty}
\bibitem{Covi:1999ty} 
  L.~Covi, J.~E.~Kim and L.~Roszkowski,
  \reftitle{Axinos as cold dark matter}
  \jnl{\PRL}{82}{4180}{1999}.
  %[hep-ph/9905212].
  %%CITATION = HEP-PH/9905212;%%
  %217 citations counted in INSPIRE as of 04 Jun 2014

  
  
  
  
%Higgs-portal

%\cite{Silveira:1985rk}
\bibitem{Silveira:1985rk} 
  V.~Silveira and A.~Zee,
  \reftitle{Scalar Phantoms}
  \jnl{\PLB}{161}{136}{1985}.
  %%CITATION = PHLTA,B161,136;%%
  %238 citations counted in INSPIRE as of 12 Jun 2014

%\cite{Davoudiasl:2004be}
\bibitem{Davoudiasl:2004be} 
  H.~Davoudiasl, R.~Kitano, T.~Li and H.~Murayama,
  \reftitle{The New minimal standard model}
  \jnl{\PLB}{609}{117}{2005}.
  %[hep-ph/0405097].
  %%CITATION = HEP-PH/0405097;%%
  %176 citations counted in INSPIRE as of 12 Jun 2014

%\cite{Patt:2006fw}
\bibitem{Patt:2006fw} 
  B.~Patt and F.~Wilczek,
  \reftitle{Higgs-field portal into hidden sectors}
  hep-ph/0605188.
  %%CITATION = HEP-PH/0605188;%%
  %234 citations counted in INSPIRE as of 12 Jun 2014

%\cite{Baek:2014jga}
\bibitem{Baek:2014jga} 
  S.~Baek, P.~Ko and W.~-I.~Park,
  \reftitle{Invisible Higgs Decay Width vs. Dark Matter Direct Detection Cross Section in Higgs Portal Dark Matter Models}
  arXiv:1405.3530 [hep-ph].
  %%CITATION = ARXIV:1405.3530;%%
  %1 citations counted in INSPIRE as of 12 Jun 2014

\bibitem{Okada:2014oda} 
  H.~Okada and Y.~Orikasa,
  \reftitle{X-ray line in Radiative Neutrino Model with Global $U(1)$ Symmetry}
  arXiv:1407.2543 [hep-ph].


% Dynamical Dark Matter

\bibitem{DynamicalDM} 
  K.~R.~Dienes and B.~Thomas,
  \reftitle{Dynamical Dark Matter: I. Theoretical Overview}
  \jnl{\PRD}{85}{083523}{2012};
  %[arXiv:1106.4546 [hep-ph]].
%\bibitem{DynamicalDM2} 
%  K.~R.~Dienes and B.~Thomas,
  \reftitle{Dynamical Dark Matter: II. An Explicit Model}
  \jnl{\PRD}{85}{083524}{2012}.
  %[arXiv:1107.0721 [hep-ph]].

%%%%%%%%%%%%%
\bibitem{PPSS} 
  G.~Panico, E.~Ponton, J.~Santiago and M.~Serone,
  \reftitle{Dark Matter and Electroweak Symmetry Breaking in Models with Warped Extra Dimensions}
  \jnl{\PRD}{77}{115012}{2008}.
  %arXiv:0801.1645 [hep-ph].

\bibitem{Carena2009} 
M.~Carena, A.~D.~Medina, N.~R.~Shah and C.~E.~M.~Wagner,
\reftitle{Gauge-Higgs Unification, Neutrino Masses and Dark Matter in Warped Extra Dimensions}
%  [arXiv:0901.0609 [hep-ph]].
\jnl{\PRD}{79}{096010}{2009}.

\bibitem{HKT}
Y.~Hosotani, P.~Ko and M.~Tanaka,
\reftitle{Stable Higgs Bosons as Cold Dark Matter}
\jnl{\PLB}{680}{179}{2009}.
%[arXiv:0908.0212 [hep-ph]].


\bibitem{Haba2009} 
N.~Haba, S.~Matsumoto, N.~Okada and T.~Yamashita,
\reftitle{Gauge-Higgs Dark Matter}
%  [arXiv:0910.3741 [hep-ph]].
\jnl{\JHEP}{1003}{064}{2010}.


%%%%%%%

\bibitem{XENON}
E.~Aprile  {\it et al.} [XENON100 Collaboration], 
  \reftitle{Dark Matter Results from 225 Live Days of XENON100 Data}
\jnl{\PRL}{109}{181301}{2012}.

\bibitem{LUX}
D.S.~Akerib  {\it et al.} [LUX Collaboration],  
  \reftitle{First results from the LUX dark matter experiment at the Sanford Underground Research Facility}    
\jnl{\PRL}{112}{091303}{2014}.


%%%%%%%%%%



\bibitem{HHHK}
N.~Haba, M.~Harada, Y.~Hosotani and Y.~Kawamura,
\reftitle{Dynamical rearrangement of gauge symmetry on the orbifold $S^1/Z_2$}
\jnl{\NPB}{657}{169}{2003}; 
{\it Erratum}-\jnl{\ibid}{669}{381}{2003}.
%  [arXiv:hep-ph/0212035].

\bibitem{HHK}
N.~Haba, Y.~Hosotani and Y.~Kawamura,
\reftitle{Classification and dynamics of equivalence classes in $SU(N)$ gauge theory on the orbifold $S^1/Z_2$}
\jnl{\PTP}{111}{265}{2004}. 
%  [arXiv: hep-ph/0309088].

\bibitem{Yamamoto2014}
K.\ Yamamoto
\reftitle{The formulation of gauge-Higgs unification with dynamical boundary conditions}
\jnl{\NPB}{883}{45}{2014}.
[arXiv:1401.0466 [hep-th]].

\bibitem{Cossu} 
G.~Cossu, H.~Hatanaka, Y.~Hosotani and J.~-I.~Noaki,
\reftitle{Polyakov loops and the Hosotani mechanism on the lattice}
\jnl{\PRD}{89}{094509}{2014}.
[arXiv:1309.4198 [hep-lat]].

%%%%%%%%%

\bibitem{Pomarol2009}
B.~Gripaios, A.~Pomarol, F.~Riva and J.~Serra,
\reftitle{Beyond the Minimal Composite Higgs Model}
\jnl{\JHEP}{0904}{ 070}{2009}.
%  [arXiv:0902.1483 [hep-ph]].


\bibitem{Sakamura2014}
Y.~Matsumoto and Y.~Sakamura,
\reftitle{6D gauge-Higgs unification on $T^2/Z_N$ with custodial symmetry}
arXiv:1407.0133 [hep-ph].


%%%%%%%%

%\cite{Griest}
\bibitem{Griest} 
  K.~Griest and D.~Seckel,
  \reftitle{Three exceptions in the calculation of relic abundances}
  \jnl{\PRD}{43}{3191}{1991}.
  
  
%\cite{Cheng:2002iz}
\bibitem{Cheng:2002iz} 
H.~-C.~Cheng, K.~T.~Matchev and M.~Schmaltz,
\reftitle{Radiative corrections to Kaluza-Klein masses}
\jnl{\PRD}{66}{036005}{2002}.

%\cite{Randall2001}
\bibitem{Randall2001} 
  L.~Randall and M.~D.~Schwartz,
  \reftitle{Quantum field theory and unification in AdS5}
  JHEP {\bf 0111}, 003 (2001);
  %[hep-th/0108114].
  \reftitle{Unification and the hierarchy from AdS5}
  Phys.\ Rev.\ Lett.\  {\bf 88}, 081801 (2002).
  %[hep-th/0108115].



%\cite{Srednicki:1988ce}
\bibitem{Srednicki:1988ce} 
  M.~Srednicki, R.~Watkins and K.~A.~Olive,
  \reftitle{Calculations of Relic Densities in the Early Universe}
  \jnl{\NPB}{310}{693}{1988}.
 
 \bibitem{PDG2012} 
  J.~Beringer {\it et al.}  [Particle Data Group Collaboration],
  \reftitle{Review of Particle Physics (RPP)}
  Phys.\ Rev.\ D {\bf 86}, 010001 (2012).

 
%\cite{Ade:2013zuv}
\bibitem{Ade:2013zuv} 
  P.~A.~R.~Ade {\it et al.}  [Planck Collaboration],
  \reftitle{Planck 2013 results. XVI. Cosmological parameters}
  arXiv:1303.5076 [astro-ph.CO].
  

\bibitem{Peskin-Takeuchi1} 
M.~E.~Peskin and T.~Takeuchi,
\reftitle{A new constraint on a strongly interacting Higgs sector}
\jnl{\PRL}{65}{964}{1990};
\reftitle{Estimation of oblique electroweak corrections}
\jnl{\PRD}{46}{381}{1992}.

  

% Goldberger-Wise 
\bibitem{Goldberger} 
  W.~D.~Goldberger and M.~B.~Wise,
  \reftitle{Modulus stabilization with bulk fields}
  \jnl{\PRL}{83}{4922}{1999}.
  %[hep-ph/9907447].

\bibitem{Radion-T} 
  P.~Creminelli, A.~Nicolis and R.~Rattazzi,
  \reftitle{Holography and the electroweak phase transition}
%  JHEP {\bf 0203}, 051 (2002)
 \jnl{\JHEP}{0203}{051}{2002}.
   %[hep-th/0107141].

%%%%%%%

\bibitem{Goodman-Witten} 
M.W.\ Goodman and E.\ Witten, 
\reftitle{Detectability of certain dark-matter candidates}   
\jnl{\PRD}{31}{3059}{1986}. 

%%%%%%
   
    
\bibitem{nucleon-matrix-element}
  S. D\"{u}rr {\it et al.}, 
% Z. Fodor, T. Hemmert, C. Hoelbling, J. Frison, S. D. Katz, S. Krieg, 
%  T. Kurth, L. Lellouch, T. Lippert, A. Portelli, A. Ramos, A. Schafer, K. K. Szabo,
  \reftitle{Sigma term and strangeness content of octet baryons}
  \jnl{\PRD}{85}{014509}{2012};
  
  G. S. Bali {\it et al.}, 
  % S. Collins, M. G\"{o}ckeler, R. Horsley, Y. Nakamura,
%  A. Nobile, D. Pleiter, P. E. L. Rakow, A. Sch\"{a}fer, G. Schierholz,
%  A. Sternbeck, J. M. Zanotti 
  [QCDSF Collaboration],
  \reftitle{The strange and light quark contributions to the nucleon mass from Lattice QCD}
   \jnl{\PRD}{85}{054502}{2012};
 
  S. Dinter {\it et al.}, 
  %V. Drach, R. Frezzotti, G. Herdoiza, K. Jansen, G. Rossi
  \reftitle{Sigma terms and strangeness content of the nucleon with 
  $N_f = 2 + 1 + 1$ twisted mass fermions}
   \jnl{\JHEP}{08}{037}{2012};
 
  A. Semke, M. F. M. Lutz,
   \reftitle{Strangeness in the baryon ground states}
   \jnl{\PLB}{717}{242}{2012};

  M. Engelhardt, 
  \reftitle{Strange quark contributions to nucleon mass and spin from lattice QCD}
    \jnl{\PRD}{86}{114510}{2012};

  H. Ohki {\it et al.}, 
  %K. Takeda, S. Aoki, S. Hashimoto, T. Kaneko, H. Matsufuru,  J. Noaki, T. Onogi 
  [JLQCD Collaboration],
  \reftitle{Nucleon strange quark content from $N_f=2+1$ lattice QCD with exact chiral symmetry}
 \jnl{\PRD}{87}{034509}{2013}; 
   
  P. E. Shanahan, A. W. Thomas, and R. D. Young,
  \reftitle{Sigma terms from an $SU(3)$ chiral extrapolation}
   \jnl{\PRD}{87}{074503}{2013};
 
  P. Junnarkar, A. Walker-Loud   
 \reftitle{The Scalar Strange Content of the Nucleon from Lattice QCD}
  \jnl{\PRD}{87}{114510}{2013};
   
  M. Gong {\it et al.}, 
  % A. Alexandru, Y. Chen, T. Doi, S. J. Dong, T. Draper, W. Freeman,   M. Glatzmaier, A. Li, K. F. Liu, Z. Liu
   \reftitle{Strangeness and charmness content of nucleon from overlap fermions on 
  $2+1$-flavor domain-wall fermion configurations}
   \jnl{\PRD}{88}{014503}{2013};

  W. Freeman, D. Toussaint,  [MILC Collaboration],
  \reftitle{The intrinsic strangeness and charm of the nucleon using improved staggered fermions}
    \jnl{\PRD}{88}{054503}{2013}.




%%%    %%%%%%%
%\input{bib_funatsu}
%\input{bib_hatanaka}

\end{thebibliography}
\end{document}